\documentclass[12pt,subeqn,a4paper]{article}

\usepackage{amssymb,amsmath,amsfonts,amsthm, amscd, mathrsfs,helvet}
\usepackage[usenames,dvipsnames]{color}
\usepackage{bm}
\usepackage{graphicx,verbatim}
\usepackage{psfrag}
\usepackage{wrapfig}
\usepackage[all]{xy}
\theoremstyle{plain}

\definecolor{LightBlue}{rgb}{0.02,0.47,0.7}

\newcommand{\black}[0]{\color{Black}}
\newcommand{\red}[0]{\color{Red}}
\newcommand{\blue}[0]{\color{Blue}}
\newcommand{\lightblue}[0]{\color{LightBlue}}
\newcommand{\green}[0]{\color{Green}}
\newcommand{\purple}[0]{\color{Purple}}

\numberwithin{equation}{section}

\begin{document}

\begin{titlepage}
\begin{flushright}
DAMTP-2008-21\\
\end{flushright}
\begin{centering}
\vspace{.2in}
{\large {\bf Semiclassical Quantisation\\ of Finite-Gap Strings}}\\

\vspace{.3in}

Beno\^{\i}t Vicedo\\
\vspace{.1 in}
DAMTP, Centre for Mathematical Sciences \\
University of Cambridge, Wilberforce Road \\
Cambridge CB3 0WA, UK \\
\vspace{.2in}
\vspace{.4in}
{\bf Abstract} \\
\vspace{.2in}
\end{centering}

We perform a first principle semiclassical quantisation of the
general finite-gap solution to the equations of a string moving on
$\mathbb{R} \times S^3$. The derivation is only formal as we do
not regularise divergent sums over stability angles. Moreover, with
regards to the AdS/CFT correspondence the result is incomplete as
the fluctuations orthogonal to this subspace in $AdS_5 \times S^5$ are
not taken into account. Nevertheless, the calculation serves the
purpose of understanding how the moduli of the algebraic curve gets
quantised semiclassically, purely from the point of view of finite-gap
integration and with no input from the gauge theory side. Our result
is expressed in a very compact and simple formula which encodes the
infinite sum over stability angles in a succinct way and reproduces
exactly what one expects from knowledge of the dual gauge theory. Namely,
at tree level the filling fractions of the algebraic curve get
quantised in large integer multiples of $\hbar = 1/\sqrt{\lambda}$. At
1-loop order the filling fractions receive Maslov index corrections of
$\frac{1}{2} \hbar$ and all the singular points of the spectral curve
become filled with small half-integer multiples of $\hbar$. For the
subsector in question this is in agreement with the previously
obtained results for the semiclassical energy spectrum of the string
using the method proposed in hep-th/0703191.

Along the way we derive the complete hierarchy of commuting flows
for the string in the $\mathbb{R} \times S^3$ subsector which are
generated by the Taylor coefficients of the quasi-momentum $p(x)$
through Hamilton's equation. Moreover, we also derive a very
general and simple formula for the stability angles around a
generic finite-gap solution which may be used in the study of
stability properties of solutions in the $\mathbb{R} \times S^3$
subsector. We also stress the issue of quantum operator orderings
and whether or not a given ordering preserves integrability since
this problem already crops up at 1-loop in the form of the
subprincipal symbol.

%\vspace{.05in}
%\baselineskip=.3in
\end{titlepage}

\tableofcontents

\input{epsf}

\setcounter{section}{-1}

\newpage

\section{Introduction}

The method of semiclassical quantisation in field theory has been
extensively developed by many authors in the 70's using different
approaches \cite{DHN1, DHN2, DHN3, Korepin1, Korepin2,
BerryTabor1, BerryTabor2} (see also the books
\cite{Coleman:1975qj, Rajaraman:1982is} for a more or less
complete survey and list of references). The aim of all these
methods is to give a quantum mechanical meaning to extended
classical solutions of the field equations which already
classically exhibit particle like properties. The role played by
such non-trivial classical solutions in the leading order
quantisation of any field theory is evident from the path integral
which is dominated by classical solutions in the $\hbar
\rightarrow 0$ limit. It follows then that the applicability of
semiclassical methods crucially relies on an explicit knowledge of
classical solutions. Yet for a generic field theory, very little
can be said about explicit solutions to the field equations and in
most cases a general solution does not exist. When the field
theory is classically integrable however, essentially everything
is known about the classical theory and the most general solution
can be constructed explicitly in terms of standard functions and
finitely many algebraic operations. In this case the complete
semiclassical spectrum of the theory can then be obtained by
applying the methods of semiclassical quantisation to the general
solution.

It is now very well established that the Metsaev-Tseytlin action
\cite{Metsaev:1998it} describing superstrings on $AdS_5 \times
S^5$ is classically integrable \cite{Bena:2003wd}, in the sense that
the theory possesses an infinite number of integrals of motion. This
fact has been thoroughly exploited in the literature \cite{KMMZ}
to completely classify the full set of classical solutions on
$AdS_5 \times S^5$ by assigning to every solution a finite genus
algebraic curve which encodes its integrals of motion $I_1,
\ldots, I_n$. However, the algebraic curve is not enough to
uniquely specify the solution, which can be seen as follows. Since
a given solution carries only finitely many non-zero integrals of
motion $I_1, \ldots, I_n$ it will be invariant under the action of
all the other integrals of the theory. Moreover, the solution
breaks all the symmetries generated by $I_1, \ldots, I_n$ and the
action of these integrals on the solution will generate new
solutions with the same integrals. Indeed, in the theory of
finite-gap integration \cite{Belokolos, Babelon, Krichever0,
Krichever1, Krichever2, Krichever3, Krichever4}, (finite-gap)
solutions are shown to be in one-to-one correspondence with sets
of algebro-geometric data which essentially consist of a finite
genus algebraic curve equipped with a finite set of points called
a \textit{divisor}. The action of the moduli $I_1, \ldots, I_n$ of
a solution on the solution itself will act non-trivially on the
divisor, thereby generating a new solution with different divisor.
The divisor therefore encodes the different zero-modes of a given
solution.

The treatment of the zero-modes is an important part of any
approach to semiclassical quantisation \cite{DHN1, DHN2,
Coleman:1975qj, Rajaraman:1982is}. Indeed, if a classical solution
has zero-modes then a naive semiclassical quantisation of the
solution will fail. Consider a solution $\phi_{\text{cl}}$ of a
field equation derived from an action $S[\phi]$, \textit{i.e.}
$S'[\phi_{\text{cl}}] = 0$, where $'$ denotes $\delta/\delta
\phi$. If $v$ denotes an infinitesimal symmetry of the equations
of motion, \textit{i.e.} $v(S'[\phi]) = S''[\phi] (v \phi)$, and
suppose that $\phi_{\text{cl}}$ is not invariant under the
symmetry then it follows immediately that $(v \phi_{\text{cl}})
\neq 0$ is in the kernel of the operator $S''[\phi_{\text{cl}}]$
which is therefore not invertible and so the propagator of the
theory in the background $\phi_{\text{cl}}$ cannot be defined. The
standard way around this difficulty is to treat the zero-mode
directions separately using the method of `collective
coordinates'. In short, collective coordinates parametrise the
zero-mode directions, namely the flat directions in field space,
along which the wave function will tend to spread out in the form
of a plane wave as a result of which the quantum counterpart of
the solution $\phi_{\text{cl}}$ will acquire dynamics along the
collective coordinates. Generally one has to perform a change of
variables in field space to include the collective coordinates
among the set of field variables and this can often only be done
implicitly. A nice feature of the finite-gap construction is that
it naturally lends itself to the separation of zero-modes since
the divisor, which plays the role of the collective coordinates,
already appears explicitly in the finite-gap solution \---\ no
change of variables was required.

The divisor of a finite-gap solution therefore plays a central
role in determining its semiclassical spectrum. But although the
algebraic curve is known in full generality for the $AdS_5 \times
S^5$ superstring, the divisor has only been identified so far in
the subsector $\mathbb{R} \times S^3$ for which the explicit
reconstruction of finite-gap solutions from the algebro-geometric
data has been studied \cite{Paper1, Paper2, Paper3}. The method of
semiclassical quantisation as stated above can therefore only be
applied directly in the subsector $\mathbb{R} \times S^3$. In this
paper we perform such a semiclassical analysis of bosonic string
theory on $\mathbb{R} \times S^3$ from first principles. We do not
attempt to include the fluctuations in the directions transverse
to the subspace $\mathbb{R} \times S^3 \subset AdS_5 \times S^5$
for clarity and because we believe that the method presented here
should carry over with few alterations to the full case of
superstrings on $AdS_5 \times S^5$ once the divisor is known. The
calculation therefore serves as a toy model for understanding from
the finite-gap perspective the origin of the discretisation of the
algebraic curve when leading order semiclassical corrections are
included. Nevertheless, our result agrees for fluctuations within
the $\mathbb{R} \times S^3$ subsector with the semiclassical
results\footnote{See also
\cite{Beisert:2003xu,Beisert:2005mq,Beisert:2005bv} for earlier work
on obtaining the fluctuation energies from the spectral curve. In
particular \cite{Beisert:2005bv} where the one-loop energy shift was
computed in the Landau-Lifshitz model.} of
\cite{Gromov+Vieira1, Gromov+Vieira2, Gromov+Vieira3}.

In the remainder of the introduction we start by recalling the
method of semiclassical quantisation \`a la Dashen, Hasslacher and
Neveu \cite{DHN1, DHN2, DHN3} when applied to the specific example
of the breather solution in Sine-Gordon theory. We reformulate
everything in a language that we hope will facilitate the
conceptual understanding of the method in the finite-gap setting
and in the last part of the introduction we give a sketch of the
ideas developed in the paper.

\subsection{Semiclassical Sine-Gordon breathers} \label{section: Breather}

Consider the example of the boosted breather solution in
Sine-Gordon theory \cite{DHN2, Coleman:1975qj, Rajaraman:1982is}
\begin{equation} \label{breather}
\phi_{\tau,v}(x,t) = \frac{4 m}{\sqrt{\lambda}} \tan^{-1}\left\{
\frac{ ((\tau m / 2 \pi)^2 - 1)^{\frac{1}{2}} \sin [ (2 \pi /
\tau) (t - v x)/(1 - v^2)^{\frac{1}{2}} ] }{\cosh [ ((\tau m / 2
\pi)^2 - 1)^{\frac{1}{2}} (2 \pi / \tau) (x - v t)/(1 -
v^2)^{\frac{1}{2}} ] } \right\}.
\end{equation}
This is really a two parameter family of solutions parametrised by
their proper period $\tau$ and their velocity $v$, or equivalently
by their energy $E$ and momentum $p$. To compute the (possibly
continuous) spectrum of the corresponding quantum states it is
always simpler at first to put the system in a very large but
finite box of length $L$ by identifying $x \sim x + L$ so as to
make the spectrum discrete, and then take the infinite volume
limit $L \rightarrow \infty$ at the end. In this closed-loop world
the breather solution \eqref{breather} is periodic in $t$ of
period $T$ provided $\tau$ and $v$ satisfy $T = l \tau /
(1-v^2)^{\frac{1}{2}} = m L / v$ with $l,m \in \mathbb{N}$.

If we were quantising the kink, we could move to its rest frame in
which it is static and study small fluctuations in terms of
eigenfrequencies. However, the breather is a little more
complicated since it is time dependent in its rest frame, and
because time dependent solutions are not point-like in field space, we
need a way to characterise perturbations of the orbit as a whole. As
we will describe in appendix \ref{section: BS isolated}, this is done
by considering the perturbation of a specific point on the orbit,
evolving that perturbation under the equations of motion for
roughly the period of the underlying solution, and comparing the
final perturbation with the original one. If the perturbation is
stable then it will have merely rotated and the angle of rotation
is called the \textit{stability angle}. If instead the
perturbation is unstable it will have grown exponentially in
magnitude, which corresponds to the case of a complex stability
angle. Finally, if the perturbation comes back exactly to itself,
this means it describes a nearby periodic solution, and in general
zero stability angles correspond to symmetries. In the case of the
Sine-Gordon breather we therefore need to look for generic nearby
solutions $\phi(x,t) = \phi_{\tau,v}(x,t) + \delta \phi$. This
perturbed solution won't be periodic in general, yet because the
linearised equation
\begin{equation} \label{linearised SG eq}
\Box \delta \phi = \left( \cos \phi_{\tau,v} \right) \delta \phi
\end{equation}
is invariant under time translation by $T$ we can always write its
solution as a superposition of eigenfunctions of time translation
$\delta \phi(x,t+T) = e^{-i \nu} \delta \phi(x,t)$, where $\nu$
are their stability angles. Notice that the Sine-Gordon equation
is invariant under arbitrary space and time translations, but the
breather solution $\phi_{\tau,v}$ is not. As a result, $\partial
\phi_{\tau,v}/\partial x$ and $\partial \phi_{\tau,v}/\partial t$
are both \textit{zero-modes}, \textit{i.e.} perturbations with
zero stability angles. In general, any symmetry of the action that
is not a symmetry of the classical solution will give rise to a
zero-mode.

\begin{figure}[h]
\centering
\begin{tabular}{ccc}
\includegraphics[height=15mm]{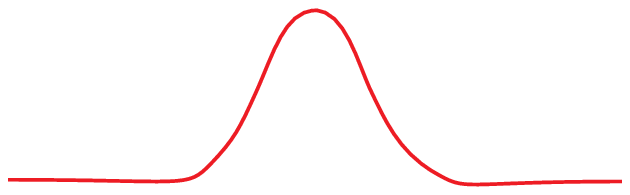} & \raisebox{5mm}{$\longrightarrow$} &
\includegraphics[height=15mm]{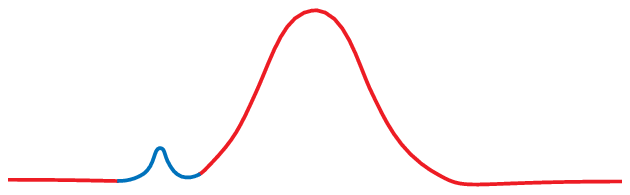}
\end{tabular}
\caption{Perturbing the breather by another small breather using
the B\"acklund transform} \label{figure: Backlund}
\end{figure}

The task of finding nearby solutions to the breather is greatly
facilitated by the fact that the Sine-Gordon equation is
integrable, since we can use the B\"acklund transform to get new
solutions from known solutions. In particular we can perturb our
breather by adding a little breather of small amplitude on top of
it (Figure \ref{figure: Backlund}).
Studying double breather solutions in the limit where the small
breather has vanishingly small amplitude corresponds to a
linearised study of the Sine-Gordon equation around the breather
solution. So integrability gives us a convenient way of writing
down explicit solutions to the linearised equation
\eqref{linearised SG eq} from which the stability angles of the
breather may be read off.

Identifying the space of classical solutions with phase-space, for
each $\tau, v$ (or equivalently $E,p$) the breather solution
\eqref{breather} is just a specific point in phase space. However,
the existence of two zero-modes $\partial \phi_{\tau,v}/\partial
x$ and $\partial \phi_{\tau,v}/\partial t$ for the breather
solution indicates that it really belongs to a two parameter
family of solutions with the same integrals of motion $E,p$. These
are the space and time translated breather solutions
\begin{equation} \label{breather family}
\phi_{\tau,v}(x + x_0,t + t_0).
\end{equation}
Since all the other stability angles of the breather are real,
when we include first order quantum corrections the wavefunction
will want to localise around not one breather, but around the
whole two parameter family \eqref{breather family} of breathers by
spreading along the flat directions, namely the $x_0$ and $t_0$
directions. Along these directions the wavefunction will therefore
be a plane wave, but since the $t_0$-direction is closed by
periodicity of the breather solution the plane wave along it must
have an integer number of peaks and troughs. In other words the
change of phase of the wavefunction around this closed direction
will have to be an integer multiple $n$ of $2 \pi$. Along all the
other non-zero stability angle directions the wavefunction will
decay rapidly and, intuitively, for states with higher excitation
number $n_i$ it will extend further in these directions. The
correct quantisation conditions encoding the semiclassical energy
spectrum of the wavefunction localised around the family of
breather solutions was first derived by Dashen, Hasslacher and
Neveu \cite{DHN1} and can be expressed as follows. If we define
the `action' of the breather solution as
\begin{subequations} \label{DHN}
\begin{equation} \label{DHN1}
W(E) = \int_0^T dt \int dx \pi_{\tau,v}(x,t) \partial_0
\phi_{\tau,v}(x,t),
\end{equation}
then the DHN quantisation conditions read
\begin{equation} \label{DHN2}
\frac{W(E)}{\hbar} = 2 \pi n + \sum_{\nu_i > 0} \left(n_i +
\frac{1}{2}\right) \nu_i + O(\hbar).
\end{equation}
\end{subequations}
Although the derivation of this formula is very complicated, it
intuitively makes a lot of sense. In general the phase of the
wavefunction in the semiclassial approximation is an action of the
form \eqref{DHN1} so the first term on the right hand side of
\eqref{DHN2} can be seen to comes from the single-valuedness of
the wavefunction along the compact $t_0$-direction whereas the
correction from the sum over stability angles is related to the
small fluctuations transverse to the $t_0$ and $x_0$ directions.

\begin{figure}[h]
\centering \psfrag{S}{\tiny $H^{-1}(E)$} \psfrag{g}{\tiny
$\gamma_E$} \psfrag{gp}{\tiny $\gamma_{E + \epsilon}$}
\psfrag{gm}{\tiny $\gamma_{E - \epsilon}$}
\includegraphics[height=3cm]{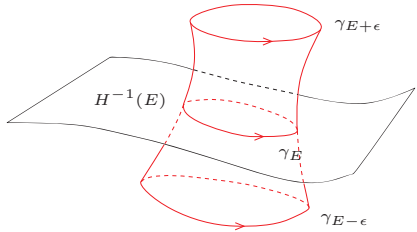}
\caption{Cylinder theorem: a periodic solution $\gamma_E$ on the
energy level $H^{-1}(E)$ is contained in a one parameter family of
periodic solutions of varying energy in the range $[E - \epsilon,
E + \epsilon]$.} \label{figure: cylinder}
\end{figure}

For the purpose of drawing the analogy between the Sine-Gordon
breather case here and that of finite-gap strings discussed later
it will be convenient to think of the conditions \eqref{DHN} in
more geometric terms in phase-space as follows. Since the breather
in \eqref{breather family} with $x_0 = 0$ is periodic, it can be
thought of as a closed orbit on the level set $\Sigma_{E,p}$ of
fixed $E,p$. The direction along the orbit, parametrised by $t_0$,
corresponds to the zero-mode $\partial \phi_{\tau,v}/\partial t$
of the breather. But since it has another zero-mode, namely
$\partial \phi_{\tau,v}/\partial x$, this orbit really belongs to
a continuous family of periodic orbits, parametrised by $x_0$, all
contained in $\Sigma_{E,p}$. However, because we are working in a
periodically identified finite box, this two parameter ($x_0,
t_0$) family of breathers is in fact a torus $\mathbb{T}^2_{E,p}$
lying within $\Sigma_{E,p}$. And since all the other stability
angles of the breather are non-zero, this means that
$\mathbb{T}^2_{E,p}$ is isolated on the level set $\Sigma_{E,p}$
in the sense that it does not belong to a larger continuous family
of periodic orbits within $\Sigma_{E,p}$. Yet if we leave the
level set $\Sigma_{E,p}$, one can show that in a neighbourhood of
$\Sigma_{E,p}$ the torus $\mathbb{T}^2_{E,p}$ persists, namely it
belongs to a two parameter family of torii parametrised by $E,p$.
This is the content of the `cylinder theorem', illustrated in
Figure \ref{figure: cylinder} for the case of a solution with a
single zero-mode, so that its zero-mode family in the level set
$H^{-1}(E)$ is just a circle $S^1_E$ that belongs to a cylinder
$S^1_E \times [E - \epsilon, E + \epsilon]$.
Looking back at the most general breather solution \eqref{breather
family} it contains four independent parameters: the two
parameters $x_0, t_0$ are parameters along the torus
$\mathbb{T}^2_{E,p}$ whereas $E,p$ parameterise the family of
torii of the cylinder theorem. Now the effect of the quantisation
condition \eqref{DHN} is to pick out a discrete set of breathers
from this `cylinder' of breathers \eqref{breather}, the energy and
momentum of which approximate to order $O(\hbar)$ the
semiclassical energy spectrum of the quantum states localised
around the breather solution. For instance, when applied to the
Sine-Gordon breather the quantisation conditions \eqref{DHN} yield
the following semiclassical spectrum \cite{DHN2}
\begin{equation*}
E_{k,n} = (p_k^2 + M_n^2)^{\frac{1}{2}}, \quad p_k = \frac{2 \pi
k}{L},
\end{equation*}
where $M_n = \frac{16 m}{\gamma'} \sin \frac{n \gamma'}{16}$ and
$\gamma' = \frac{\lambda}{m^2} \left( 1 - \frac{\lambda}{8 \pi
m^2} \right)^{-1}$, and in the infinite volume limit $L
\rightarrow \infty$ the momentum becomes continuous as expected.

\subsection{Sketch of semiclassical finite-gap strings} \label{section: sketch}

We would like to apply a similar kind of reasoning to the case of
superstring theory on $AdS_5 \times S^5$. However, since this
formalism requires the knowledge of explicit solutions we will
restrict attention to bosonic string theory on $\mathbb{R} \times
S^3$ for which the general finite-gap solution to the equations of
motion is known \cite{Paper1, Paper2}. In conformal static gauge
the string is given by an embedding $g(\sigma,\tau) \in SU(2)$ of
the worldsheet into $SU(2)$, and if we define the corresponding
Lie algebra current $j = -g^{-1}dg \in \mathfrak{su}(2)$ then the
equations of motion and Virasoro constraints take the following
form
\begin{equation} \label{eom + Vir}
d \ast j = 0, \quad dj - j \wedge j = 0, \quad \frac{1}{2} \,
\text{tr} j_{\pm}^2 = - \kappa^2.
\end{equation}
As is well know, the equations of motion are integrable and can be
rewritten in the form of a zero-curvature equation $dJ(x) - J(x)
\wedge J(x) = 0$. In this form one can make use of the powerful
methods of finite-gap integration to construct, at least
abstractly the general finite-gap solution to the equations of
motion. In fact, it is possible to incorporate the Virasoro and
static gauge constraints into the constructions \cite{Paper1,
Paper2} so as to get only physical motions of the string. The
general finite-gap solution is constructed from the following
piece of \textit{algebro-geometric data}:
\begin{itemize}
\item[$\bullet$] An algebraic curve \cite{KMMZ} of genus $g$.
\item[$\bullet$] A set of $g+1$ points \cite{Paper1} on this
curve.
\end{itemize}
Essentially, by the Riemann-Roch theorem there is an injective map
from this algebro-geometric data into the space of solutions to
\eqref{eom + Vir}. The idea of finite-gap integration is
illustrated in Figure \ref{FG integration}:
\begin{figure}[h]
\begin{gather*}
\begin{tabular}{c}
\psfrag{S}{\small $\Sigma$} \psfrag{g}{\small \red \small
$\hat{\gamma}$} \includegraphics[height=2cm]{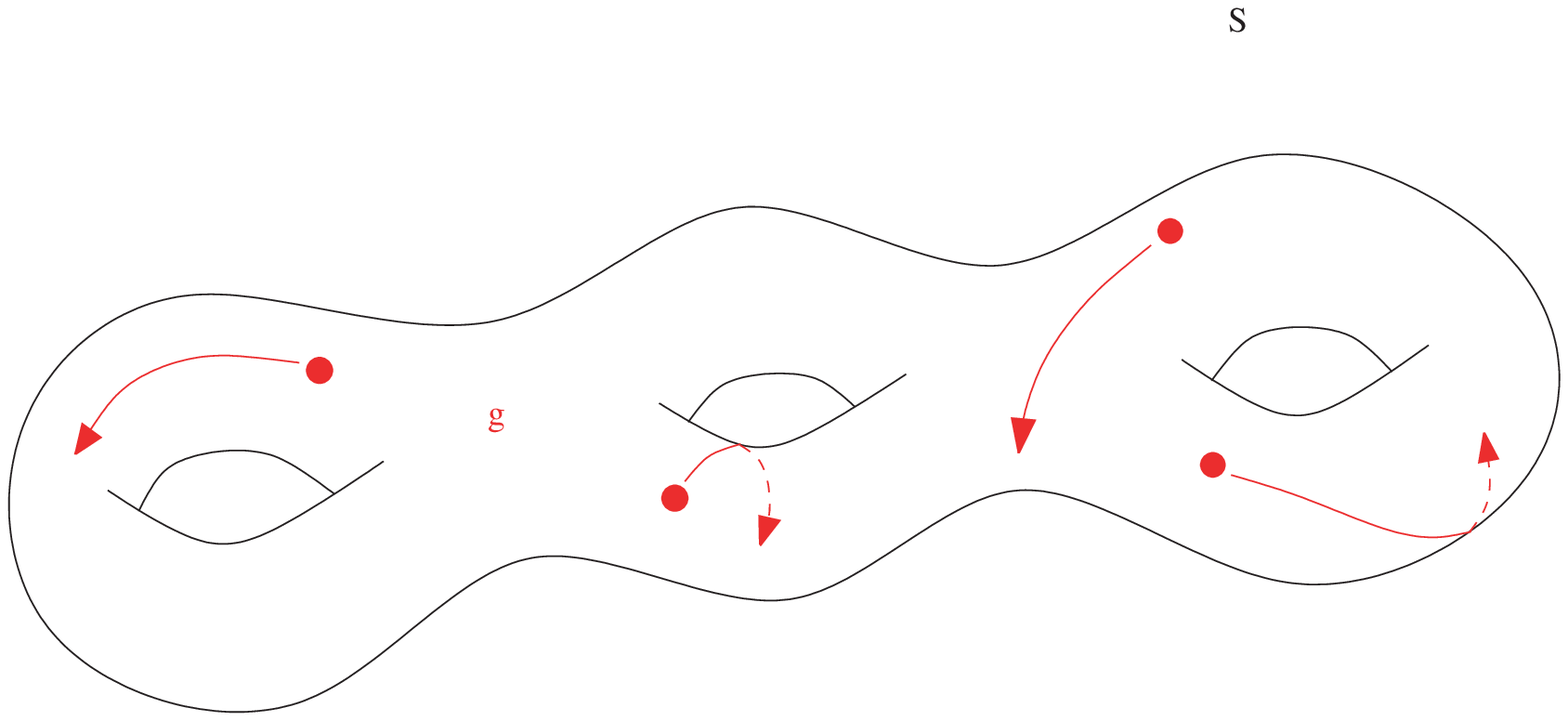}
\end{tabular} \quad {\blue \Leftrightarrow} \quad {\purple
\text{finite-gap solution to \eqref{eom + Vir}}}\\ \hspace{-70mm}
\qquad \qquad \qquad \psfrag{A}{\small \green $\vec{\mathcal{A}}:
\Sigma^{g+1}/S^{g+1} \rightarrow J(\Sigma) \times
\mathbb{C}^{\ast}$}
\includegraphics[height=1cm]{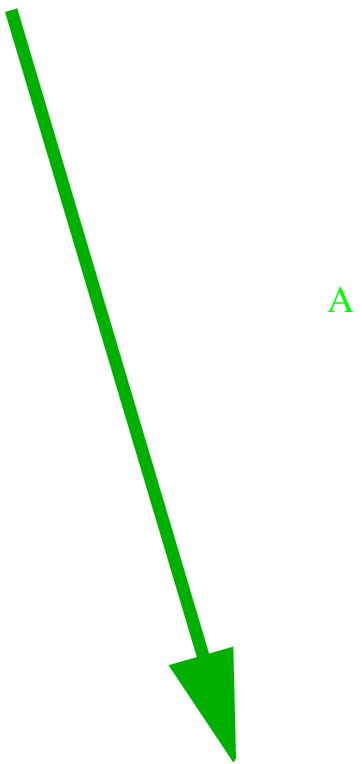}\\
\psfrag{J}{\small $J(\Sigma)$} \psfrag{C}{\small $\!\!\!\!\!
\times \mathbb{C}^{\ast}$}
\includegraphics[height=1.5cm]{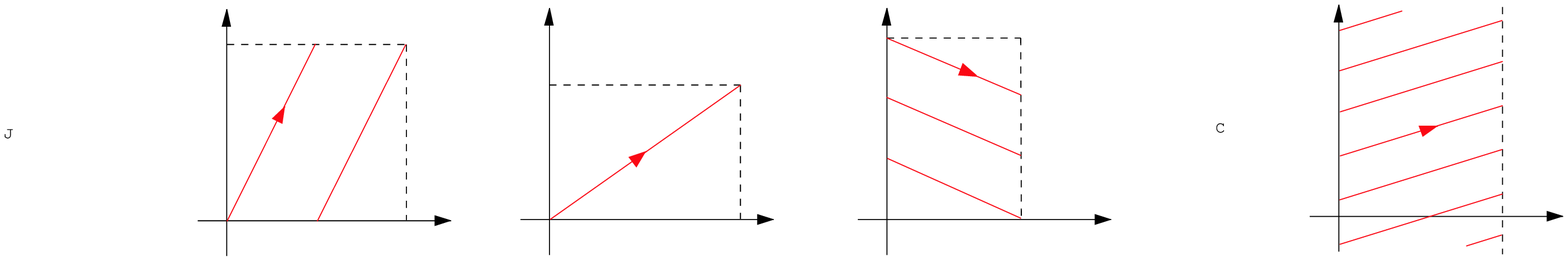}
\end{gather*}
\caption{Idea of finite-gap integration.} \label{FG integration}
\end{figure}
Every finite-gap solution to \eqref{eom + Vir} is in one-to-one
correspondence with an algebraic curve (of genus three in Figure
\ref{FG integration}) equipped with a set of marked points (four
of them in Figure \ref{FG integration}). The algebraic curve
encodes the integrals of motion of the solution, and these points
encode the dynamics. Their exact motion on the algebraic curve is
very complex, but what we find is that if we map the algebraic
curve to its (generalised) Jacobian, a $(g+1)$-torus, via the
(generalised) Abel map then the motion in $\sigma$ and $\tau$
becomes extremely simple, namely it linearises. The motion of the
string on this $(g+1)$-torus is like that of an infinitely rigid
string wrapping one cycle of the torus and moving linearly in time
along another direction.

An alternative way of picturing what a finite-gap solution looks
like that will be useful later is as follows. As we just saw, the
dynamics of a finite-gap solution corresponds to linear motion on
a $(g+1)$-torus, which is very reminiscent of a finite-dimensional
integrable system. In fact one can view the Jacobian as the
Liouville torus of a $(2g+2)$-dimensional dynamical system.
\begin{figure}[h]
\centering
\begin{tabular}{cc}
\psfrag{J}{\small $\mathbb{T}^{g+1}$} \psfrag{L}{\small
$\mathcal{L}$}
\includegraphics[height=2cm]{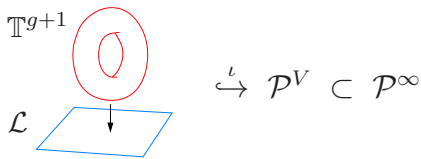} & \raisebox{.8cm}{$\;
\overset{\iota}\hookrightarrow \; \mathcal{P}^V \; \subset \;
\mathcal{P}^{\infty}$}
\end{tabular}
\caption{The algebro-geometric data as a $(2g + 2)$-dimensional
phase-space.} \label{finite-dim phase-space}
\end{figure}
The base space $\mathcal{L}$ of this $(2g+2)$-dimensional system
is the moduli space of the algebraic curve parametrised by the
\textit{filling fractions} $\{ \mathcal{S}_I =
\int_{\mathcal{A}_I} z dp \}_{I=1}^{g+1}$. But if the
algebro-geometric data is to be thought of as a finite-dimensional
phase-space it must be equipped with a natural symplectic
structure. This can be obtained as follows: the finite-gap
solution maps this algebro-geometric data to the space of
solutions to \eqref{eom + Vir}, see Figure \ref{finite-dim
phase-space}. Identifying the space of solutions to the equations
of motion with the phase-space $\mathcal{P}^{\infty}$, the
solutions to \eqref{eom + Vir} which also satisfy Virasoro and
static gauge define a second class constraint surface
$\mathcal{P}^V \subset \mathcal{P}^{\infty}$. This is equipped
with a Dirac bracket induced by the Poisson bracket on the
$\mathfrak{su}(2)$ current appropriately regularised \`a la
Maillet \cite{Maillet, Paper2}. If one then pulls back this Dirac bracket
to the algebro-geometric data using the finite-gap solution we
obtain a `natural' symplectic structure on the algebro-geometric
data which can be concisely written as (see \cite{Paper2} for details)
\begin{equation*}
\omega = \sum_{I=1}^{g+1} d \mathcal{S}_I \wedge d \varphi_I.
\end{equation*}
The upshot of this is that the filling fractions are precisely the
action variables of the finite-gap string. They are the analogues
of the period $\tau$ and velocity $v$ (or energy $E$ and momentum
$p$) of the generic breather \eqref{breather family} which defined
a four parameter family of solutions. A finite-gap solution
defines a whole $(2g+2)$ parameter family of solutions
parametrised by the algebro-geometric data and can be written as
follows
\begin{equation*}
g = g \Big( {\sum}_N t_N \bm{U}_N(\bm{S}) + \bm{D} \Big| \bm{S}
\Big),
\end{equation*}
where $t_N$ are a set of $g+1$ independent times (defined in
section \ref{section: hierarchy}), $\bm{S}$ is the vector of
action variables which plays the role of the parameters $(\tau,
v)$ here and $\bm{D} \in \mathbb{C}^{g+1}$ is the exact analogue
of the initial coordinates of the breather $(x_0,t_0)$. We
therefore expect a finite-gap solution constructed from a curve of
genus $g$ to have $g+1$ zero-modes corresponding to the $g+1$
components of the vector $\bm{D}$.

In view of applying a semiclassical quantisation formula like the
one in \eqref{DHN} we must first determine all the stability
angles of a given finite-gap solution. So just as in the case of
the Sine-Gordon breather, we would like to study perturbations of
finite-gap solutions described above. Once again integrability
will play a prominent role in solving the linearised equations. In
fact, finding solutions to the linearised problem is very simple
now that we have already fully exploited integrability to
construct the most general finite-gap solution. A perturbation of
a given finite-gap solution will simply be another `nearby'
finite-gap solution. Recall \cite{Paper1, Paper2} that in the
$SU(2)$ sector the algebraic curve is hyperelliptic and can be
represented by a set of $g+1$ cuts in the complex plane. How can
one describe perturbations of the $g$-gap solution corresponding
to this curve? Playing the same game as for the Sine-Gordon
breather where we used integrability to add another little
breather on it, here we can just take a solution corresponding to
a curve of genus one higher, but make the extra filling fraction
very small, which corresponds to making the cut very small, see
Figure \ref{figure: perturb}.
\begin{figure}[h]
\centering
\includegraphics[height=30mm]{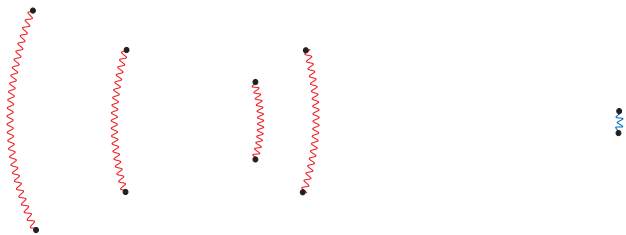}
\caption{Perturbation of a finite-gap solution.} \label{figure:
perturb}
\end{figure}
There is an obvious analogy here between breathers in Sine-Gordon
and cuts in bosonic strings on $\mathbb{R} \times S^3$ as one can
think of a finite-gap solution as a multi-breather solution
consisting of finitely many breathers. Cuts with small filling
fractions are analogous to breathers of small amplitude as both
describe perturbations. If we define the $a_i$-cycle ($i = 1,
\ldots, g$) as usual to encircle the $i^{\text{th}}$ cut
counterclockwise (on the upper sheet) then a perturbation of this
kind clearly corresponds to pinching an $a$-cycle of the algebraic
curve. So we want to take the difference between the solution
before pinching an $a$-cycle and the solution after pinching the
$a$-cycle; this will give us a perturbation of the latter and we
can then analyse its periodicity properties to extract the
corresponding stability angles. Notice however that any given
perturbation of a finite-gap string will have one stability angle
defined for each cycle on the Jacobian, or equivalently for each
macroscopic cut.

The semi-classical spectrum can be obtained by performing a WKB
analysis of the wavefunction that will localise around the
zero-mode directions of the solution, which in the case of the
finite-gap string is the Jacobian. Again, the leading term will
describe how many full waves fit on the compact Jacobian, and the
infinite sum corresponds to small fluctuations transversal to the
Jacobian. The result of such an analysis that will be sketched in
section \ref{section: BS} are the following set of Bohr-Sommerfeld
equations, the correct form of which involves Maslov
indices\footnote{Here and in the remainder of the paper, in the
string theory context we will always let $\hbar =
\frac{1}{\sqrt{\lambda}}$.}
\begin{equation} \label{BS0}
\frac{S_I}{\hbar} = N_I + \frac{\mu_I}{4} + \sum_{\alpha =
g+2}^{\infty} \left( n_{\alpha} + \frac{1}{2} \right)
\frac{\nu_{\alpha}^{(I)}}{2 \pi} + O(\hbar).
\end{equation}
Here $\mu_I = 2$ is the \textit{Maslov index} of the
$\mathcal{A}_I$-cycle ($I = 1, \ldots, g+1$) in the generalised
Jacobian $J(\Sigma,\infty^{\pm})$. Note that \eqref{BS0} is only
valid in the harmonic oscillator approximation $N_I \gg
n_{\alpha}$ where the perturbations are much smaller than the
background filling fractions. So the expression \eqref{BS0} really
contains two different orders, namely the tree level and 1-loop
level of order $O(1)$ and $O(\hbar)$ respectively (after
multiplying \eqref{BS0} throughout by $\hbar$). At tree level
\eqref{BS0} simply expresses the fact that the filling fractions
are quantised in integer multiples of $\hbar$, \textit{i.e.} $S_I
= N_I \hbar$, which is a straightforward consequence of the fact
that the $S_I$ are the action variables as was shown in
\cite{Paper2}. The non-trivial content of \eqref{BS0} is the 1-loop
correction which includes firstly the Maslov index correction
$\frac{\mu_I}{4} \hbar$ and secondly the infinite sum over stability
angles.

Obtaining the energy spectrum from \eqref{BS0} is relatively
straightforward since for a system to be semiclassically
integrable requires that $[\hat{S}_i,\hat{S}_j] = O(\hbar^3)$ and
so the energy eigenvalues are given to leading order in $\hbar$
simply by evaluating the classical energy $E_{\text{cl}}[S_1,
\ldots, S_{g+1}]$ on the eigenvalues of the action variables
\eqref{BS0}. As we show in section \ref{section: main result} this can
be expanded to order $O(\hbar)$, expressing the result as a sum of the
tree level term $E_{\text{cl}}[N_1 \hbar, \ldots, N_{g+1} \hbar]$ and
the 1-loop correction involving the sum over stability angles
\cite{Gromov+Vieira1, Gromov+Vieira2, Gromov+Vieira3}. But
moreover, in section \ref{section: main result} we also show,
using the result of section \ref{section: quasi-actions}, that the sum
of the tree level term and 1-loop correction term can be succinctly
rewritten in a compact form that captures the complete result at
1-loop in a unified way. Indeed, we show that the energy spectrum can
be formally obtained by evaluating the classical energy of an
\textit{infinite}-gap solution with all its infinite filling fractions
quantised to half-integer multiples of $\hbar$, namely
\begin{equation*}
E = E_{\text{cl}}\left[ \left( N_1 + \frac{1}{2} \right) \hbar,
\ldots \right].
\end{equation*}
This result is to be interpreted as a limit of expressions where a
finite but \textit{arbitrary} number of first entries are of order
$O(1)$ corresponding to the tree level order and the remaining
infinite number of entries encode the stability angle contribution
to the 1-loop corrections of order $O(\hbar)$, see \eqref{main
result}.

The paper is organised as follows: in section \ref{section:
semiclassical} we review some basic features of semiclassical
quantisation for finite-dimensional systems. In particular we
remind the reader how operator ordering enters in the
semiclassical regime: in the language of pseudo-differential
operators (appendix \ref{section: PsiDO}) the different operator
orderings are encoded in the subprincipal symbol \cite{Vu Ngoc 1}.
We also sketch the derivation of the Bohr-Sommerfeld quantisation
conditions \cite{Vu Ngoc 1, Voros, Voros1}. In section
\ref{section: hierarchy} we look back at the general construction
of finite-gap strings \cite{Paper1, Paper2} and derive the whole
hierarchy of commuting flows. That is, we show how the integrable
equations of motion for the embedding of the string in $\mathbb{R}
\times S^3$ are part of an infinite hierarchy of higher integrable
equations corresponding to the infinite set of conserved charges
of the string, as is usual in any integrable system. In section
\ref{section: perturbations} we discuss perturbations of a generic
finite-gap string through the pinching of $a$-cycles. This leads
to a general formula for the non-zero stability angles of a
generic finite-gap string. Using this result, in section
\ref{section: semiclassical 2} we come back to the issue of
semiclassical quantisation of finite-gap strings and apply the
formalism of section \ref{section: semiclassical} to obtain the
semiclassical spectrum of the string. Some appendices elaborate on
the discussion in each section.

\section{Semiclassical approximation generalities} \label{section: semiclassical}

Consider a classical Hamiltonian system described by a $2n$
dimensional phase-space $T^{\ast} X$ with Hamiltonian $H :
T^{\ast} X \rightarrow \mathbb{R}$. Given $E \in \mathbb{R}$ we
can consider the codimension one energy level set $\Sigma_E \equiv
H^{-1}(E) \subset T^{\ast} X$. Assume also that we have a desired
quantisation of the system, that is, we have a self-adjoint
operator $\hat{H}$ acting on $L^2(X)$ whose principal symbol is
the classical Hamiltonian $H$. If $H^{-1}([E-\epsilon,
E+\epsilon])$ is compact then the eigenvalues of $\hat{H}$ in the
range $[E-\epsilon, E+\epsilon]$ will be discrete since the
corresponding eigenfunctions are localised around this compact
set. The goal of semi-classical quantisation is to obtain the
spectrum of $\hat{H}$ in $[E-\epsilon, E+\epsilon]$ to leading
order in $\hbar$. One approach is to describe the spectrum using
what are known as trace formulae, the basic idea being to encode
the spectrum in terms of a single function $n(E) \equiv
\sum_{j=0}^{\infty} \delta(E - E_j^{\hbar}) = \text{tr} \;
\delta(E - \hat{H})$ where $E_j^{\hbar}$ denote the eigenvalues of
$\hat{H}$ and which can be rewritten as
\begin{equation} \label{trace formula idea}
n(E) = \text{Re} \frac{1}{\pi \hbar} \int_0^{\infty} dt \;
\text{tr} \, e^{\frac{i}{\hbar} (E-\hat{H}) t} = \text{Re}
\frac{1}{\pi \hbar} \int_0^{\infty} dt \; e^{\frac{i E t}{\hbar}}
\int_{{\tiny \begin{array}{c}\text{p.o. } \gamma\\
\text{period }t \end{array}}} [d \gamma] e^{- \frac{i}{\hbar}
\int_\gamma \mathcal{L}}.
\end{equation}
In the semiclassical limit $\hbar \rightarrow 0$ we perform a
stationary phase approximation of the integral on the right hand
side in order to obtain a semiclassical estimate of the spectrum
$\{ E_j^{\hbar} \}$ of $\hat{H}$. The presence of the trace means
that dominating contributions come from periodic orbits of the
classical system. This is a general feature of semiclassical trace
formulae which relate \textit{analytic} data of the operator
$\hat{H}$ (namely its eigenvalues) to \textit{geometric} data of
the corresponding classical Hamiltonian $H$ (namely its periodic
orbits). This is one advantage of trace formulae over other
semiclassical quantisation methods in that they elucidate the
relation between the semiclassical spectrum and the classical
periodic orbits.

On the downside however, despite the geometrical appeal of the
path integral approach to semiclassical quantisation, it is hard
to discuss the issues of operator ordering within this framework.
Indeed, thinking in terms of phase-space path integrals, since
everything in the integrand itself is classical, any information
about quantum ordering is neatly tucked away in the definition of
the regularisation used in the phase-space path integral measure
$[d \gamma]$. The standard choice of discretisation of the path
integral measure involves the mid-point prescription which
corresponds to the Weyl-ordering prescription in the operator
formalism. In particular the quantum Hamiltonian is the
Weyl-ordered classical Hamiltonian, i.e. $\hat{H} =
\text{Op}^W_{\hbar}(H)$. In order to deal with operator ordering
issues, it is therefore more convenient to work directly with
operators.

A convenient operator formalism for discussing semi-classical
quantisation involves pseudo-differential operators (referred to
as $\Psi$DOs for short). We refer to appendix \ref{section: PsiDO}
for a very brief introduction to $\Psi$DOs and their relevance for
treating semiclassical quantisation. The basic idea of this
approach is to associate with any operator $\hat{f}$ not a single
function on $T^{\ast} X$, which cannot by itself encode all the
information about operator ordering in $\hat{f}$, but a family of
functions $f_{\hbar} \in C(T^{\ast} X)$ called \textit{symbols}.
The leading function $f_0$ is exactly the classical function
corresponding to $\hat{f}$, whereas all the subleading functions
encode the operator ordering in $\hat{f}$. So instead of working
with operators one can work directly with their respective
symbols. Moreover, in the semiclassical approximation one only
needs to deal with the first two symbols of an operator, known as
the principal symbol (i.e. the classical function) and the
subprincipal symbol. We will turn to the formalism of $\Psi$DOs
and the issue of operator ordering in an integrable system in
section \ref{section: operator ordering}. In section \ref{section:
BS} we will show how the Bohr-Sommerfeld quantisation conditions
are modified by the presence of a subprincipal symbol which
reflects a choice of ordering.

But first, to get an intuitive idea of how operator ordering
ambiguities arise even at the semiclassical level to affect the
quantisation conditions, it is instructive to consider the simple
example of the harmonic oscillator for which the leading order
quantisation is exact. The classical harmonic oscillator
Hamiltonian is $H = \frac{p^2}{2 m} + \frac{1}{2} m \omega^2 x^2$,
and the action variable of the closed path of energy $E$ is given
by
\begin{equation*}
I = \frac{1}{2 \pi} \oint_{H = E} p dx = \frac{E}{\omega}.
\end{equation*}
By promoting the variables $x,p$ to operators $\hat{x},\hat{p}$ there
is only one reasonable choice of ordering in the Hamiltonian, namely
the Weyl-ordered Hamiltonian $\hat{H} = \frac{\hat{p}^2}{2 m} +
\frac{1}{2} m \omega^2 \hat{x}^2$. The spectrum of such an operator is
well known to be $E_n = \left(n + \frac{1}{2}\right) \hbar \omega, n \in
\mathbb{N}$ so that the spectrum of the Weyl-ordered action variable
$\hat{I} = \frac{1}{\omega} \hat{H}$ is simply given by the standard
Bohr-Sommerfeld quantisation condition,
\begin{equation*}
\text{Spec } (\hat{I}) \subset \left( \mathbb{Z} + \frac{1}{2} \right) \hbar,
\end{equation*}
where the index of $\frac{1}{2}$ by which the spectrum is shifted from
$\hbar \mathbb{Z}$ is known as the Maslov index in the context of
Bohr-Sommerfeld quantisation. Now since we are given at the outset
only the classical Hamiltonian, we could always choose to quantise it
with a more perverse choice of ordering. For instance, if we rewrite
the classical Hamiltonian as $H = \omega a a^{\ast}$ where $a \equiv
\sqrt{\frac{m \omega}{2 \hbar}} \left( x + \frac{ip}{2 m} \right)$ and
after promoting everything to operators request that in the quantum
Hamiltonian the $\hat{a}$ sits to the right of the $\hat{a}^{\dag}$
then we obtain the normal-ordered Hamiltonian $:\!\hat{H}\!: \; = \omega
\hbar \hat{a}^{\dag} \hat{a}$, where $[\hat{a}, \hat{a}^{\dag}] =
1$. The corresponding normal-ordered action operator is given by
$:\!\hat{I}\!: \; = \hbar \hat{a}^{\dag} \hat{a}$ whose spectrum is
easily seen to consists of integer multiples of $\hbar$,
\begin{equation*}
\text{Spec } (:\!\hat{I}\!:) \subset \mathbb{Z} \hbar.
\end{equation*}
We observe that the Maslov index is precisely cancelled by the shift
from Weyl-ordering to normal-ordering. Even though in the case of the
harmonic oscillator we know that the correct physical quantisation of
$H$ is the Weyl-ordered one $\hat{H}$ we would like to stress that in
general the choice of operator ordering in the quantisation of the
action or Hamiltonian may not be as obvious and their spectrum may
observe a shift from the standard Bohr-Sommerfeld spectrum
$\left(\mathbb{Z} +\frac{\mu}{4} \right) \hbar$, where $\mu \in
\mathbb{Z}_4$ is the Maslov index.

\subsection{Operator ordering issues} \label{section: operator ordering}

As explained in appendix \ref{section: PsiDO}, one can keep track
of operator orderings in the language of pseudo-differential
operators by retaining subleading terms beyond the principal
symbol in the full Weyl symbol of an operator. In most
applications of the theory of $\Psi$DOs the quantities of interest
are specified as $\Psi$DOs at the outset so that their full Weyl
symbol is known. In the present case however we start from a
classical system specified by its phase-space $(T^{\ast} X,
\omega)$ and the set of classical observables of interest are
$F_1,\ldots,F_n, H$. Quantising this classical system requires an
operator ordering prescription for obtaining operators from the
corresponding classical observables. At the semiclassical level
this boils down to the specification of an extra function, the
subprincipal symbol, for each classical observable. Specifically,
given a classical observable $f_0 \in C(T^{\ast} X)$, we construct
\begin{equation*}
\hat{f} = \text{Op}_{\hbar}^W (f_0 + f_1 \hbar),
\end{equation*}
where the presence of the subprincipal symbol $f_1 \in C(T^{\ast} X)$
reflects the operator ordering ambiguities already manifesting
themselves at the semiclassical level. Every possible choice of a
function $f_1 \in C(T^{\ast} X)$ corresponds to a different
prescription for the operator ordering in $\hat{f}$ at order
$O(\hbar)$. The principal symbol $f_0 = \sigma_0^W(\hat{f})$ is the
corresponding classical observable.

Recall the definition of an integrable system, which roughly speaking
is one which possesses the maximum possible number of independent
integrals of motion. Specifically, a Hamiltonian system $(T^{\ast}
X,H)$ is said to be classically integrable if there exists $n$
functions $F_1, \ldots, F_n \in C(T^{\ast} X)$ such that
\begin{itemize}
  \item[$(1')$] $d F_1 \wedge \ldots \wedge d F_n \neq 0$ almost
everywhere,
  \item[$(2')$] $\{ F_i, F_j \} = 0, \; \forall i,j =
1,\ldots,n$,
  \item[$(3')$] $H = H(F_1,\ldots,F_n)$.
\end{itemize}
Conditions $(2')$ and $(3')$ together imply that the $F_i$ are in
fact integrals of motion, $X_H F_i = 0$. In other words,
$T^{\ast} X$ admits a torus action with moment map
\begin{equation*}
\bm{F} \equiv (F_1, \ldots, F_n) : T^{\ast} X \rightarrow
\mathbb{R}^n.
\end{equation*}
At regular values $f$ of $\bm{F}$, the level sets $\bm{F}^{-1}(f)$
define $n$-torii (in the compact case) and foliate the phase-space
$T^{\ast} X$; namely $\mathbb{T}^n \hookrightarrow T^{\ast} X
\overset{\bm{F}}\rightarrow \mathbb{R}^n$. This foliation allows
one to define canonical action-angle coordinates with the action
variables $\{ I_i \}_{i=1}^n$ parametrising the base
$\mathbb{R}^n$ and the conjugate angle variables $\{ \theta_i
\}_{i=1}^n$, each taking values in $[0,2 \pi]$, parametrising the
independent cycles of the torus $\mathbb{T}^n$.

We will say that a $\Psi$DO $\hat{H}$ is \textit{semiclassically
integrable} if there exists $n$ $\Psi$DOs $\hat{F}_1, \ldots,
\hat{F}_n$ with principal symbols $F_i = \sigma_0^W(\hat{F}_i)$
such that
\begin{itemize}
  \item[$(1)$] $d F_1 \wedge \ldots \wedge d F_n \neq 0$ almost
everywhere,
  \item[$(2)$] $[ \hat{F}_i, \hat{F}_j ] = O(\hbar^3), \;
\forall i,j = 1,\ldots,n$,
  \item[$(3)$] $\hat{H} =
H(\hat{F}_1,\ldots, \hat{F}_n) + O(\hbar^2)$ for some function
$H$.
\end{itemize}
Notice that we only require commutativity modulo $O(\hbar^3)$ in
property $(2)$; it guarantees in particular that the operator
$H(\hat{F}_1,\ldots, \hat{F}_n)$ in $(3)$ is free of operator
ordering ambiguities certainly up to $O(\hbar^3)$, so that
property $(3)$ makes sense. Property $(2)$ is to be contrasted
with the definition of full quantum integrability which requires
exact commutativity $[ \hat{F}_i, \hat{F}_j ] = 0$. Now since
$\sigma_0^W([\hat{F}_i,\hat{F}_j]) = -i \hbar \{ F_i, F_j \}$ (see
appendix \ref{section: PsiDO}) and $\sigma_0^W(\hat{H}) =
H(F_1,\ldots,F_n)$, it follows that the principal symbols $F_i =
\sigma_0^W(\hat{F}_i)$ satisfy all three properties $(1')$-$(3')$
above for a classically integrable system with Hamiltonian $H =
\sigma_0^W(\hat{H})$. This means that any semiclassically
integrable system exhibits at leading order the full geometric
structure of the underlying classically integrable system given by
its principal symbols. In particular, the level set $\Lambda_f
\equiv \bm{F}^{-1}(f)$ of the moment map $\bm{F} =
(F_1,\ldots,F_n) : T^{\ast} X \rightarrow \mathbb{R}^n$ is a
Lagrangian $n$-torus and foliates phase-space $T^{\ast} X$ as we
let $f$ vary.

But the notion of semiclassical integrability contains more
information than that of its underlying classical integrable
structure \cite{Vu Ngoc 1, Vu Ngoc 2}. Property $(1)$ only
contributes at leading order since it is a statement about the
principal symbols $F_i$ alone, whereas property $(2)$ at
$O(\hbar^2)$ yields an equation for the subprincipal symbols
$F^s_i = \sigma_{\text{sub}}^W(\hat{F}_i)$ of the $\hat{F}_i$ (see
appendix \ref{section: PsiDO})
\begin{equation} \label{subprincipal symbol rel}
0 = \frac{i}{\hbar} \sigma_{\text{sub}}^W([\hat{F}_i,\hat{F}_j]) =
\left\{ F_i, F^s_j \right\} + \left\{ F^s_i, F_j \right\}.
\end{equation}
It is possible to interpret these equations geometrically so as to
supplement the geometrical structure already laid out by the
principal symbols with further geometrical data. For this we
define the \textit{subprincipal form} $\kappa$ on $\Lambda_f$ by
defining its action on the basis vectors $X_{F_i}$ at any point of
$\Lambda_f$ through \cite{Vu Ngoc 1}
\begin{equation} \label{subprincipal form}
\kappa(X_{F_i}) = - F^s_i, \quad i = 1,\ldots,n.
\end{equation}
It then follows immediately from \eqref{subprincipal symbol rel}
that $\kappa$ is closed since
\begin{align*}
d\kappa(X_{F_i},X_{F_j}) &= X_{F_i} \kappa(X_{F_j}) - X_{F_j}
\kappa(X_{F_i}) - \kappa([X_{F_i}, X_{F_j}])\\
&= - X_{F_i} F^s_j + X_{F_j} F^s_i - \kappa(X_{\{ F_i,F_j\}}) = -
\{F_i,F^s_j\} + \{F_j,F^s_i\} = 0.
\end{align*}
Hence the operator ordering in the $\hat{F}_i$ can be accounted
for at the semiclassical level by specifying a closed 1-form
$\kappa$ on the Liouville $n$-torus $\Lambda_f$. And in fact it is
clear from \eqref{subprincipal form} that every choice of a closed
1-form $\kappa \in \Omega^1(\Lambda_f)$ corresponds to a different
choice of operator ordering in the definition of the $\hat{F}_i$.

\subsection{Bohr-Sommerfeld conditions} \label{section: BS}

We are interested in the joint spectrum of the $\hat{F}_i$ up to
$O(\hbar)$ which requires solving the eigenvalue problem to that
order
\begin{equation} \label{eigenvalue problem}
(\hat{F}_i - f_i)\psi = O(\hbar^2).
\end{equation}
The Bohr-Sommerfeld conditions are conditions for the existence of
a solution to these coupled pseudo-differential equations. Their
rigourous derivation is rather involved but here we would just like to
outline how the subprincipal symbol comes about in these
conditions. To solve \eqref{eigenvalue problem} locally one considers a
local patch $V \subset \Lambda_f$ on which $\pi : T^{\ast} X
\rightarrow X$ is a diffeomorphism and uses the WKB
ansatz\footnote{The WKB ansatz isn't actually restrictive since one
can show that the space of solutions to \eqref{eigenvalue problem}
is one dimensional so that any other solution is proportional to the
WKB solution.}
\begin{equation*}
\psi_{WKB} = e^{\frac{i}{\hbar} \phi_{-1} + \phi_0} \rho + O(\hbar)
\end{equation*}
on $U = \pi(V) \subset X$ where the nature of $\rho$ will be specified
shortly. If we let $\iota_{d \phi_{-1}} : U \hookrightarrow T^{\ast}
X$ denote the 1-form $d\phi_{-1}$ viewed as a map then equation
\eqref{eigenvalue problem} implies to leading order in $\hbar$ that
\cite{Bates:1997kc}
\begin{equation} \label{leading evalue problem}
\text{im } \iota_{d \phi_{-1}} = V \subset \Lambda_f,
\end{equation}
or $\iota_{d\phi_{-1}} = \pi|_V^{-1}$. By a property of the
tautological 1-form $\alpha$, namely $d \phi_{-1} =
\iota_{d\phi_{-1}}^{\ast} \alpha$, we then have
\cite{Bates:1997kc}
\begin{equation}
d \pi|_V^{\ast} \phi_{-1} = \alpha,
\end{equation}
in other words, $\pi|_V^{\ast} \phi_{-1}$ is a local solution to
the classical integrability condition $\omega = d \alpha = 0$ on
$\Lambda_f$. If $\rho$ is a half-density\footnote{Since the
product of two half-densities is a density of weight one there is
a natural inner-product on half densities $\langle \rho_1, \rho_2
\rangle = \int_M \rho_1 \rho_2$ which makes the completion into a
Hilbert space.} on $U \subset X$ then the subleading order of
\eqref{eigenvalue problem} can be written invariantly as (Theorem
11.11 p126 of \cite{Sjostrand})
\begin{equation*}
\left(-i \mathcal{L}_{X_{F_i}} + F^s_i\right) \left(\pi|_V^{\ast} e^{i
\phi_0} \rho\right) = 0.
\end{equation*}
Now provided the subprincipal symbols are real this equation
implies on the one hand that $\pi|_V^{\ast} \rho$ is an invariant
half-density on $\Lambda_f$, i.e. $\mathcal{L}_{X_{F_i}} \pi|_V^{\ast}
\rho = 0$, and on the other hand that
\begin{equation}
d \pi|_V^{\ast} \phi_0 = \kappa,
\end{equation}
which says that $\pi|_V^{\ast} \phi_0$ is a local solution to the
subleading integrability condition that $d \kappa = 0$ on
$\Lambda_f$. What one would like to do is patch up the local WKB
solutions $\psi_{WKB}$ defined on local neighbourhoods of
$\Lambda_f$ for which the projection $\pi : T^{\ast} X \rightarrow
X$ is a diffeomorphism.

\begin{figure}[h]
\begin{center}
\psfrag{pi}{\tiny $\pi$} \psfrag{L}{\tiny $\Lambda_f \subset
T^{\ast} X$} \psfrag{M}{\tiny $X$} \psfrag{sing}{\tiny
singularity} \psfrag{caus}{\tiny caustic}
\includegraphics[height=40mm]{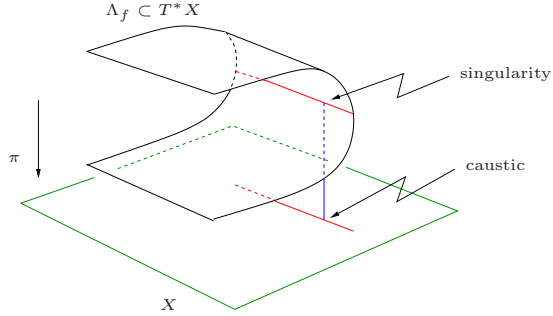}
\caption{Caustics of the Lagrangian submanifold $\Lambda_f$}
\label{Maslov}
\end{center}
\end{figure}
However, one runs into problems at caustic points where $\pi$ is
singular (see Figure \ref{Maslov}). A way around this problem was
proposed by Maslov which allows one to define a solution to
\eqref{eigenvalue problem} which is localised and defined
patchwise on $\Lambda_f$ (near caustics one uses the ``momentum''
projection $\pi_p$ of $T^{\ast} X$ onto a typical fibre of
$T^{\ast} X$ instead of $\pi$). The single valuedness of this
global solution requires its phase to be an integer multiple of
$2\pi$. The phase is essentially that of the local WKB solutions
$\psi_{WKB}$ introduced above but with additional Maslov index
corrections (coming from the caustics) so that single valuedness
conditions, known as the Bohr-Sommerfeld-Maslov conditions, read
\begin{equation} \label{BS3a}
\frac{1}{2 \pi \hbar} \int_{\gamma_i} \alpha + \frac{1}{2 \pi}
\int_{\gamma_i} \kappa = N_i + \frac{\mu_{\gamma_i}}{4} + O(\hbar),
\quad i=1,\ldots,n
\end{equation}
where $\gamma_i$ is a basis of $H_1(\Lambda_f, \mathbb{R})$ with
Maslov indices $\mu_{\gamma_i} \in \mathbb{Z}_4$ and integers $N_i
\in \mathbb{Z}$. Note in particular the presence of the
subprincipal form $\kappa$ which as we have argued is related to
operator ordering ambiguities in going from a classically
integrable system to its quantum (or just semiclassically)
integrable counterpart. It has the effect of shifting the spectrum
of the action variables similar to what happens in the case of the
harmonic oscillator when we change quantisation, from Weyl to
normal ordering say \cite{Voros2}. In the cases where all the
operators are chosen to be Weyl ordered, in particular the
$\hat{F}_i$, we have $\kappa = 0$ and \eqref{BS3} reduces to the
EBK quantisation conditions. In the remainder of the paper we
shall make the assumption that the cohomology class $[\kappa] \in
H^1(\Lambda_f)$ of the subprincipal form $\kappa$ vanishes. The
reason for this assumption is that the result is simpler to
express in this case and moreover it agrees with the results of
\cite{Gromov+Vieira1, Gromov+Vieira2, Gromov+Vieira3}. With this
assumption, the Bohr-Sommerfeld-Maslov conditions simplify
\begin{equation} \label{BS3}
\frac{1}{2 \pi \hbar} \int_{\gamma_i} \alpha = N_i +
\frac{\mu_{\gamma_i}}{4} + O(\hbar), \quad i=1,\ldots,n.
\end{equation}
We stress that this assumption does \textit{not} imply the choice
of Weyl ordering since it only corresponds to setting the
subprincipal symbol to zero, whereas Weyl ordering corresponds to
setting all the lower order Weyl symbols to zero as well.

\begin{figure}[h]
\begin{center}
\psfrag{p1}{\tiny $p_1$} \psfrag{x1}{\tiny $x_1$}
\psfrag{p2}{\tiny $p_2$} \psfrag{x2}{\tiny $\omega_2 x_2$}
\psfrag{T1}{\tiny $T_1 = \frac{2 \pi}{\omega_1}$}
\psfrag{T2}{\tiny $T_2 = \frac{2 \pi}{\omega_2}$}
\includegraphics[height=30mm]{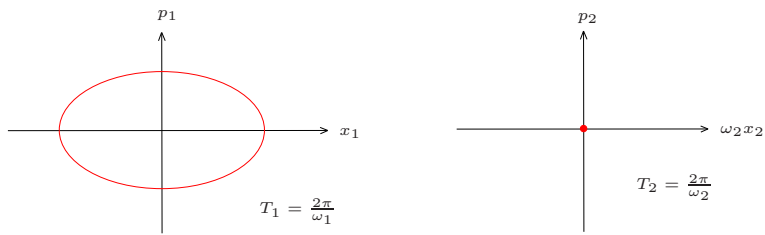}
\end{center}
\caption{Periodic orbit with $H_2 = 0$ of energy $H = H_1 = E$.}
\label{HOstab1}
\end{figure}

\begin{figure}[h]
\begin{center}
\psfrag{S1}{\tiny $E_1$} \psfrag{S2}{\tiny $E_2$}
\includegraphics[height=50mm]{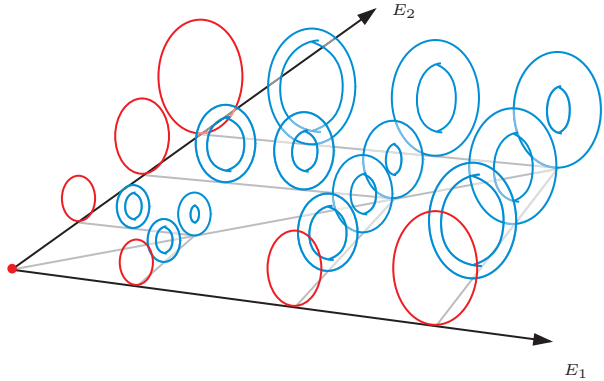}
\caption{The phase-space of the two-dimensional harmonic
oscillator.} \label{Phase_space}
\end{center}
\end{figure}

Now the derivation of the Bohr-Sommerfeld-Maslov conditions
\eqref{BS3a} or \eqref{BS3} essentially consisted in quantising a
Lagrangian $n$-torus $\Lambda_f$ by constructing a wave-function
localised around it. However, even though the level set $\Lambda_f
\equiv \bm{F}^{-1}(f)$ is indeed a Lagrangian $n$-torus for almost
every value of the integrals of motion $f_1, \ldots, f_n$ in an
integrable system, there exists interesting level sets
$\bm{F}^{-1}(f)$ in phase-space where this is not the case. This
happens at the (measure zero) set of critical values of the map
$\bm{F} = (F_1, \ldots, F_n)$. Consider for instance the
two-dimensional harmonic oscillator with different frequencies and
Hamiltonian
\begin{equation} \label{2d HO}
H = \frac{p_1^2}{2} + \frac{1}{2} \omega_1^2 x_1^2 +
\frac{p_2^2}{2} + \frac{1}{2} \omega_2^2 x_2^2 = H_1 + H_2,
\end{equation}
whose integrals of motion are given by $H_1, H_2$. For non-zero
values $E_1, E_2 \neq 0$ of $H_1, H_2$ the level sets
$\bm{H}^{-1}(E_1,E_2)$ consists of two ellipses, in other words a
Lagrangian 2-torus. However, if say $E_2 = 0$ the level set
$\bm{H}^{-1}(E_1,0)$ consists of just a single ellipse (Figure
\ref{HOstab1}).
The same thing is true when $E_1 = 0$ and at the point where $E_1
= E_2 = 0$ the level set consists of just a single point. One can
draw a picture of the phase-space in the region where $\mathcal{E}
\equiv \{ (E_1,E_2) : E_i \geq 0, i=1,2\}$ which is foliated by
2-torii in the interior of $\mathcal{E}$ but with the fibres over
the boundary $\partial \mathcal{E} \setminus \{ (0,0) \}$ being
ellipses and the fibre over the point $(0,0)$ being just a single
point, see Figure \ref{Phase_space}.
Note that the set of critical values $\partial \mathcal{E}$ is of
measure zero. However, if we are interested in the semiclassical
spectrum of the two-dimensional harmonic oscillator in the region
near $\partial \mathcal{E}$ then a modification of the
Bohr-Sommerfeld-Maslov quantisation conditions \eqref{BS3} is
required so that it applies to isotropic $p$-torii which are the
level sets of a limited number $p < n$ of integrals of motion
$F_1, \ldots, F_p$.

\begin{figure}[h]
\begin{center}
\psfrag{S}{\tiny $\Sigma_E$}
\includegraphics[height=40mm]{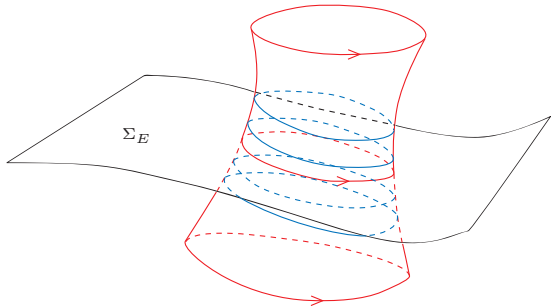}
\caption{Bohr-Sommerfeld semi-classical spectrum: the discrete set
of periodic orbits $\gamma(E^{\hbar}_{j_a})$ shown in blue have
energies $E^{\hbar}_{j_a}$ approximating the eigenvalues of
$\hat{H}$ to $O(\hbar^2)$.} \label{BS spectrum}
\end{center}
\end{figure}

It was pointed out by Voros \cite{Voros, Voros1} that the
Bohr-Sommerfeld conditions \eqref{BS3} for the apparently more
restrictive case of an integrable system may be used to obtain the
Bohr-Sommerfeld conditions in all other intermediate cases, namely
the partially integrable one (with $p < n$ integrals of motion)
and even the non-degenerate case $p=1$ (where $H$ is the only
integral). If the system has $p$ independent observables $\bm{F} =
(F_1, \ldots, F_p)$ in involution (with $H = H(\bm{F})$), then on
each codimension $p$ level set $\Sigma_f = \bm{F}^{-1}(f)$ the
system has a $p$-torus $\Lambda_f \subset \Sigma_f$ generated by
the vector fields $X_{F_i}$. Each of these $p$-torii is surrounded
by an $n$-torus of the linearised system to which the
Bohr-Sommerfeld-Maslov conditions \eqref{BS3} may be applied. This
results in a set of Bohr-Sommerfeld conditions for the cycles on
the $p$-torus which include stability angles for the small
fluctuations in the directions transverse to this $p$-torus. The
derivation of these Bohr-Sommerfeld conditions from those in the
integrable case \eqref{BS3} are a bit lengthy but the derivation
in the more general case $1 < p < n$ is conceptually the same as
the $p=1$ case \cite{Voros, Voros1}. For completeness and to
explain the appearance of the stability angles (which are related
to the eigenvalues of the Poincar\'e map) we repeat the details of
the derivation of \cite{Voros, Voros1} in appendix \ref{section:
BS isolated}. The result \eqref{BS4} is the following quantisation
condition for the \textit{isolated} orbit $\gamma$ \cite{Voros,
Voros1}
\begin{equation} \label{BS5}
\int_{\gamma} \alpha = \left[ 2 \pi \left( N +
\frac{\mu_{\gamma}}{4}\right) + \sum_{\alpha = 2}^n \left( n_{\alpha}
+ \frac{1}{2} \right) \nu_{\alpha} \right] \hbar + O(\hbar^2),
\end{equation}
where $N \in \mathbb{Z}$ and $n_{\alpha} \in \mathbb{N}$. Since
the periodic orbit $\gamma \subset \Sigma_E$ in fact belongs to a
continuous $1$-parameter family $\gamma(E)$ of periodic orbits
parametrised by the energy $E$ (`cylinder theorem'), what the
condition \eqref{BS5} does is pick out a discrete set of periodic
orbits $\gamma({E^{\hbar}_j})$, in a neighbourhood of $\Sigma_E$,
whose energies $E^{\hbar}_j$ approximate eigenvalues of $\hat{H}$
to leading order in $\hbar$, see Figure \ref{BS spectrum}.
The condition depends on the \textit{stability angles}
$\nu_{\alpha} \in \mathbb{R}$ (defined via the Poincar\'e map, see
appendix \ref{section: BS isolated}) of the stable isolated orbit
$\gamma$ and is valid only in the approximation where $0 <
n_{\alpha} \ll |N|$ which is required for the linear approximation
(used in deriving these condition) to hold.

The more general case of a system which has $p$ independent
observables $F_1, \ldots, F_p$ in involution (with $H = H(F_1,
\ldots, F_p)$), where $p$ lies in the range $1 < p < n$ is a
straightforward generalisation. In this case we get a set of $p$
quantisation conditions, one for each cycle $\gamma_k,
k=1,\ldots,p$ on the $p$-torus \cite{Voros, Voros1},
\begin{equation} \label{BS6}
\int_{\gamma_k} \alpha = \left[ 2 \pi \left( N_k +
\frac{\mu_{\gamma_k}}{4}\right) + \sum_{\alpha = p+1}^n \left( n_{\alpha}
+ \frac{1}{2} \right) \nu^k_{\alpha} \right] \hbar + O(\hbar^2).
\end{equation}
This time there are $p$ conditions on the $p$ parameters
$f_1,\ldots,f_p$ of the codimension $p$ level sets
$\Sigma_f = \bm{F}^{-1}(f)$.

\begin{figure}[h]
\begin{center}
\psfrag{p1}{\tiny $p_1$} \psfrag{x1}{\tiny $x_1$}
\psfrag{p2}{\tiny $p_2$} \psfrag{x2}{\tiny $\omega_2 x_2$}
\psfrag{T1}{\tiny $T_1 = \frac{2 \pi}{\omega_1}$}
\psfrag{T2}{\tiny $T_2 = \frac{2 \pi}{\omega_2}$} \psfrag{n}{\tiny
$\nu$}
\includegraphics[height=30mm]{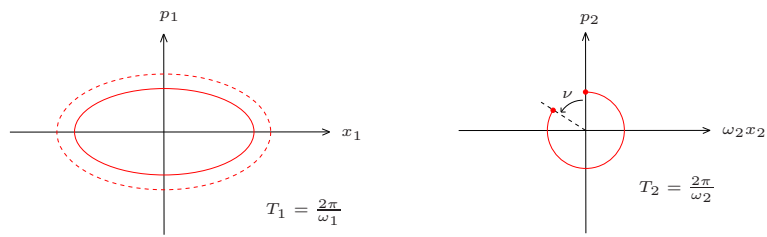}
\end{center}
\caption{Perturbed trajectory of energy $H = H_1 + H_2 = E$.}
\label{HOstab2}
\end{figure}

To illustrate the use of the modified Bohr-Sommerfeld conditions
\eqref{BS5} for an isolated orbit let use go back to the case of
the two-dimensional harmonic oscillators \eqref{2d HO}. This
system is obviously integrable and the exact spectrum of $H$ is
given by
\begin{equation*}
E_{n_1,n_2} = \left( n_1 + \frac{1}{2} \right) \hbar \omega_1 + \left(
n_2 + \frac{1}{2} \right) \hbar \omega_2.
\end{equation*}
However, suppose for the sake of argument that we can only solve
classically for the Hamiltonian $H_1$ and wish to obtain the
spectrum of $H = H_1 + H_2$ by perturbation as describe above.
Then consider a particular motion of the Hamiltonian $H_1$ of
total energy $H_1 = E$, through the point $(p_1,x_1,p_2,x_2) =
(p_0,0,0,0)$ say, see Figure \ref{HOstab1}. This defines a
1-parameter family of periodic orbits parametrised by their energy
$H = H_1 = E$. It is clear that the $(p_2,x_2)$-plane gives a
Poincar\'e section of the orbit through the point $(p_0,0,0,0)$
since all orbits of $H_1$ have the same period $T_1 = \frac{2
\pi}{\omega_1}$. The prescription for determining the stability
angles of this orbit is to consider small perturbations around it
within the same energy level $H = E$. If the periods of the two
harmonic oscillators are different, $T_1 \neq T_2$, then after a
length of time $T_1$, the motion in the $(p_2,x_2)$-plane does not
close and there is a deficit angle of $\nu = \omega_2 \cdot T_1$,
see Figure \ref{HOstab2}.
The tower of energy levels corresponding to the periodic motion in
Figure \ref{HOstab1} is therefore given by the Bohr-Sommerfeld
condition \eqref{BS5} which in this case reads
\begin{equation*}
I_1 = \left[ \left(n_1 + \frac{1}{2}\right) + \left( n_2 + \frac{1}{2}
\right) \frac{\nu}{2 \pi} \right] \hbar + O(\hbar^2)
\end{equation*}
and hence $E_{n_1,n_2} = \omega_1 \cdot I_1 = \left(n_1 +
\frac{1}{2}\right) \hbar \omega_1 + \left( n_2 + \frac{1}{2}
\right) \hbar \omega_2 + O(\hbar^2)$ so that the Bohr-Sommerfeld
condition is actually exact to first order in $\hbar$ on the
harmonic oscillator.

\section{The string hierarchy} \label{section: hierarchy}

In the general theory of finite-gap integration \cite{Belokolos,
Babelon, Krichever0, Krichever1, Krichever2, Krichever3,
Krichever4} the reconstruction of a solution requires an algebraic
curve $\Sigma$, which specifies the integrals of motion, as well
as a divisor (\textit{i.e.} a finite set of points)
$\hat{\gamma}_0$ on $\Sigma$ which specifies the initial
conditions for the dynamics. However, since the system is
integrable it possesses an infinite number of integrals of motion
$H_i$, each one generating a different Hamiltonian flow on
phase-space in the usual sense through the Hamilton equations
$\partial_{t_i} f = \{ H_i, f \}$. The dynamics of the divisor
$\hat{\gamma}(t_i)$ (with initial condition $\hat{\gamma}(0) =
\hat{\gamma}_0$) on $\Sigma$ with respect to some time $t_i$ is
then determined by the corresponding Hamiltonian $H_i$. In fact,
in this setup there is a natural correspondence between
Hamiltonian flows and meromorphic differentials on the algebraic
curve $\Sigma$. This is well known for instance in the case of the
worldsheet coordinate $\sigma$ which couples to the quasi-momentum
$dp$. Indeed, it was shown in \cite{Paper1} that the worldsheet
coordinates $(\sigma, \tau)$ enter the finite-gap solution only
through the meromorphic differential
\begin{equation*}
d \mathcal{Q} = \frac{1}{2 \pi} \left( \sigma dp + \tau dq
\right),
\end{equation*}
so that the coordinates $\sigma$ and $\tau$ are said to ``couple''
respectively to the quasi-momentum $dp$ and the quasi-energy $dq$.
The aim of this section is to similarly identify the dynamics
corresponding to all the higher conserved charges within the
finite-gap language in terms of meromorphic differentials on the
underlying curve $\Sigma$.

The reason for doing this is the following. As we have already
mentioned in the introduction and will recall again in section
\ref{section: semiclassical 2}, a finite-gap solution can be
understood as an injective map from a finite dimensional
phase-space to the full infinite dimensional phase-space of the
theory \cite{Paper2, Krichever0}. The Liouville torus of the
finite dimensional phase-space in question is the (generalised)
Jacobian $J(\Sigma, \infty^{\pm})$ of the algebraic curve $\Sigma$
which is a $(g+1)$-torus. We will show that the divisor moves
linearly on $J(\Sigma, \infty^{\pm})$ with respect to all the
higher flows. But since $J(\Sigma, \infty^{\pm})$ is $g+1$
dimensional one can use $g+1$ independent such flows to
parameterise it. This will give a nice coordinate system on the
Jacobian which will be useful when we come to consider
perturbations of this $(g+1)$-torus in section \ref{section:
perturbations} for computing stability angles. In particular, the
angle variables $\varphi_I$ which will couple to the quasi-actions
$dq^{(I)}$ defined in section \ref{section: quasi-actions} will
parameterise $g+1$ independent cycles $\mathcal{C}_I \equiv \{
\varphi_I \in [0,2 \pi) \}$ on the Jacobian along which the
Poincar\'e maps will be defined. Many of the techniques used in
this section can be found in the book \cite{Babelon}.

\subsection{Higher times and zero-curvature}

If one can rewrite the equations of motion of an integrable
two-dimensional field theory in the form of a zero-curvature
equation for a one-parameter family of $1$-forms $J(x)$, namely
\begin{equation*}
dJ(x) - J(x) \wedge J(x) = 0,
\end{equation*}
then this leads straight away to the construction of an infinite
set of conserved charges by considering the parallel transporter
$\Omega(x)$ with connection $J(x)$ along a loop winding once
around the worldsheet. The flatness of $J(x)$ immediately yields
\begin{equation} \label{conservation of charges}
\partial_{\sigma} \text{tr } \Omega(x)^n = \partial_{\tau} \text{tr }
\Omega(x)^n = 0, \quad \forall n \in \mathbb{N}.
\end{equation}
Moreover, as was shown in \cite{Paper2} the invariants $\text{tr }
\Omega(x)^n$ are in involution with respect to the Poisson bracket
\begin{equation} \label{involution of charges}
\left\{ \text{tr } \Omega(x)^n , \text{tr } \Omega(x')^m \right\}
= 0, \quad \forall n,m \in \mathbb{N}.
\end{equation}
This condition contains \eqref{conservation of charges} as a
special case since the worldsheet energy $\mathcal{E}$ and
momentum $\mathcal{P}$ are related to the leading order asymptotic
of $\Omega(x)$ near $x = \pm 1$. In fact \eqref{involution of
charges} is the statement of the invariance of $\text{tr }
\Omega(x')^m$ with respect to an infinite family of higher flows
generated by $\text{tr } \Omega(x)^n$. We will now show that the
Hamilton equations of motion corresponding to these higher
conserved charges $\text{tr} \, \Omega(x)^n$ also take the form of
a zero-curvature condition.

Let us start by determining the evolution of the space component
$J_1(x)$ of the lax connection under the higher flows, namely $\{
\text{tr } \Omega(x)^n , J_1(x') \}$. For this we first obtain the
following Poisson bracket with the transfer matrix
$T(\sigma_1,\sigma_2,x) = P \overleftarrow{\exp}
\int_{\sigma_2}^{\sigma_1} d\sigma J_1(\sigma,x)$,
\begin{equation} \label{TJ}
\{ T(\sigma_1, \sigma_2, x) \overset{\otimes}, J_1(\sigma_3, x')
\} = \int_{\sigma_2}^{\sigma_1} d\sigma (T(\sigma_1,\sigma,x)
\otimes {\bf 1}) \{ J_1(\sigma, x) \overset{\otimes},
J_1(\sigma_3, x') \} (T(\sigma,\sigma_2,x) \otimes {\bf 1}),
\end{equation}
which requires the Poisson bracket $\{J_1 \overset{\otimes},
J_1\}$ given in \cite{Paper2}, first obtained by J.-M. Maillet in
\cite{Maillet} in the context of the principal chiral model
\begin{equation} \label{Maillet bracket}
\begin{split}
\left\{ J_1(\sigma,x) \mathop{,}^{\otimes}
J_1(\sigma_3,x')\right\} &= \left[ r(x,x'), J_1(\sigma,x)\otimes
\mathbf{1} + \mathbf{1} \otimes J_1(\sigma_3,x') \right]
\delta(\sigma - \sigma_3)\\ &- \left[ s(x,x'),
J_1(\sigma,x)\otimes \mathbf{1} - \mathbf{1} \otimes
J_1(\sigma_3,x') \right] \delta(\sigma - \sigma_3)\\ &- 2 s(x,x')
\delta'(\sigma - \sigma_3),
\end{split}
\end{equation}
where
\begin{equation} \label{r,s-matrix}
r(x,x') = - \frac{2 \pi}{\sqrt{\lambda}} \frac{x^2 + {x'}^2 - 2
x^2 {x'}^2}{(x-x')(1 - x^2)(1 - {x'}^2)} \eta, \quad s(x,x') = -
\frac{2 \pi}{\sqrt{\lambda}} \frac{x+x'}{(1 - x^2)(1 - {x'}^2)}
\eta.
\end{equation}
Inserting \eqref{Maillet bracket} into \eqref{TJ}, integrating by
parts for the $\delta'$-term and using identities like
\begin{equation} \label{identities}
\left\{
\begin{split}
&\frac{\partial T}{\partial \sigma_1}(\sigma_1,\sigma_2,x) =
J_1(\sigma_1,x) T(\sigma_1,\sigma_2,x)\\
&\frac{\partial T}{\partial \sigma_2}(\sigma_1,\sigma_2,x) = -
T(\sigma_1,\sigma_2,x) J_1(\sigma_2,x),
\end{split}
\right.
\end{equation}
yields
\begin{multline} \label{PB T,J}
\{ T(\sigma_1, \sigma_2, x) \overset{\otimes}, J_1(\sigma_3, x')
\} \\
\begin{split} = &-2 (\delta(\sigma_3 - \sigma_1) -
\delta(\sigma_3 - \sigma_2)) (T(\sigma_1,\sigma_3,x) \otimes {\bf
1}) s(x,x') (T(\sigma_3,\sigma_2,x) \otimes {\bf 1})\\ &+
\epsilon(\sigma_1 - \sigma_2) \chi(\sigma_3;
\sigma_1,\sigma_2) (T(\sigma_1,\sigma_3,x) \otimes {\bf 1})\\
&\times [ (r + s)(x,x'), J_1(\sigma_3, x) \otimes {\bf 1} + {\bf
1} \otimes J_1(\sigma_3, x') ] (T(\sigma_3,\sigma_2,x) \otimes
{\bf 1}),
\end{split}
\end{multline}
where $\epsilon(\sigma) = \text{sign}(\sigma)$ is the usual sign
function and $\chi(\sigma; \sigma_1,\sigma_2)$ is the
characteristic function of the interval between $\sigma_1$ and
$\sigma_2$. If we are working on the circle, let $\sigma_1 =
\sigma + 2 \pi, \sigma_2 = \sigma, \sigma_3 = \sigma'$ and
identify the monodromy matrix as $\Omega(\sigma,x) = T(\sigma + 2
\pi, \sigma, x)$ then the previous equation reduces to
\begin{multline*}
\{ \Omega(\sigma,x) \overset{\otimes}, J_1(\sigma', x') \}\\ =
(T(\sigma  + 2 \pi,\sigma',x) \otimes {\bf 1}) [ (r + s)(x,x'),
J_1(\sigma', x) \otimes {\bf 1} + {\bf 1} \otimes J_1(\sigma', x')
] (T(\sigma',\sigma,x) \otimes {\bf 1}).
\end{multline*}
Making use of \eqref{identities} again, we can rewrite this as
\begin{equation} \label{OmegaJ}
\{ \Omega(\sigma,x) \overset{\otimes}, J_1(\sigma', x') \} =
\partial_{\sigma'} \mathcal{J}(\sigma, \sigma', x, x') +
[\mathcal{J}(\sigma, \sigma', x, x'), {\bf 1} \otimes J_1(\sigma',
x')],
\end{equation}
where
\begin{equation*}
\mathcal{J}(\sigma, \sigma', x, x') = (T(\sigma  + 2
\pi,\sigma',x) \otimes {\bf 1}) (r + s)(x,x') (T(\sigma',\sigma,x)
\otimes {\bf 1}).
\end{equation*}
Taking the trace over the first factor of the tensor product
yields
\begin{equation} \label{pre zero-curvature}
\{ \text{tr}\, \Omega(\sigma, x) , J_1(\sigma', x') \} =
\partial_{\sigma'} \mathcal{J}(\sigma, \sigma', x, x') +
[\mathcal{J}(\sigma, \sigma', x, x'), J_1(\sigma', x')],
\end{equation}
where $\mathcal{J}(\sigma, \sigma', x, x') = \text{tr}_{1} \left[
(T(\sigma',\sigma,x) T(\sigma  + 2 \pi,\sigma',x) \otimes {\bf 1})
(r + s)(x,x') \right]$. In fact, using the translation invariance
of the transfer matrix $T$ by $2 \pi$ and the definition of
$\Omega(x)$, we see that $\mathcal{J}(\sigma, \sigma', x, x')$
does not explicitly depend on $\sigma$ and can be written more
succinctly as
\begin{equation} \label{Lax connection}
\mathcal{J}(\sigma', x, x') = \text{tr}_{1} \left[
(\Omega(\sigma',x) \otimes {\bf 1}) (r + s)(x,x') \right].
\end{equation}
If we interpret the Poisson bracket $\{ \text{tr}\, \Omega(\sigma,
x) , J_1(\sigma', x') \}$ in \eqref{pre zero-curvature} as the
``time'' derivative of $J_1(\sigma', x')$ with respect to the time
generated by the Hamiltonian $\text{tr}\, \Omega(x)$ then
\eqref{pre zero-curvature} takes exactly the form of a
zero-curvature equation. This indicates that
\eqref{Lax connection} ought to be related to the Lax matrices
corresponding to all the higher order flows generated by the
Hamiltonians $\text{tr}\, \Omega(x)$, just as $J_0$ and $J_1$ were the
Lax matrices generating $\tau$ and $\sigma$ respectively. In fact, as
we will show, one should Taylor expand \eqref{pre zero-curvature} and
\eqref{Lax connection}  around $x = \pm 1$ thereby obtaining a
discrete set of independent times $t_{n,\pm}$.

An important remark is in order at this stage: since we are really
doing string theory in conformal static gauge by imposing the Virasoro
constraints and static gauge fixing conditions, which constitute a set
of second class constraints in the Hamiltonian formalism, one
should take care in imposing them consistently. This means that we
should define an appropriate Dirac Bracket corresponding to every
Poisson bracket and write everything in terms of those. Once this is
done, the Virasoro constraints and static gauge fixing conditions can
then be imposed without worry at any level of the
calculation. However, as we show in appendix \ref{section: Dirac
brackets}, for all the brackets of interest in the following, the
Dirac and Poisson brackets are identical. Thus in the remainder of
this section we shall denote brackets by $\{ \cdot, \cdot \}$ without
specifying whether they are Dirac brackets or Poisson brackets.

Let us first obtain the equations of motion for the monodromy
matrix with respect to the Hamiltonian $\text{tr}\, \Omega(x)$.
Starting from the Poisson algebra of the monodromies \cite{Maillet,
Paper2},
\begin{align} \label{fundamental Poisson bracket}
\left\{ \Omega(x) \mathop{,}^{\otimes} \Omega(x') \right\}
= &[r(x,x'), \Omega(x) \otimes \Omega(x')] \notag \\
+ &\left(\Omega(x) \otimes {\bf 1}\right) s(x,x') \left( {\bf 1}
\otimes \Omega(x') \right)
\\ - &\left( {\bf 1} \otimes \Omega(x') \right) s(x,x') \left(
\Omega(x) \otimes {\bf 1} \right), \notag
\end{align}
and taking the trace over the first factor of the tensor product
as above yields
\begin{equation} \label{pre monodromy evolution}
\{ \text{tr}\, \Omega(x) , \Omega(x') \} = [\mathcal{J}(x, x'),
\Omega(x')].
\end{equation}
Once again, if we interpret the Poisson bracket $\{ \text{tr}\,
\Omega(x) , \Omega(x') \}$ as a time derivative, this last
equation starts to take the form of the $(\sigma,\tau)$-evolution
equations
\begin{equation} \label{sigma tau evolution}
[\partial_{\tau} - J_0(x'), \Omega(x')] = 0, \quad
[\partial_{\sigma} - J_1(x'), \Omega(x')] = 0.
\end{equation}

The expression \eqref{Lax connection} for the Lax matrices can be
simplified further. Using \eqref{r,s-matrix} the sum of the
$(r,s)$-matrices entering in \eqref{Lax connection} is
\begin{equation*}
(r+s)(x,x') = - \frac{2 \pi}{\sqrt{\lambda}} \frac{2 x^2}{(x-x')(1
- x^2)} \eta.
\end{equation*}
Now by definition, $\eta = -t^a \otimes t^a$ where the
$\mathfrak{su}(2)$ generator $t^a$ is related to the Pauli
matrices as $t^a = \frac{i}{\sqrt{2}} \sigma_a$. Thus \eqref{Lax
connection} can be written as
\begin{equation} \label{Lax connection 2}
\mathcal{J}(\sigma', x, x') = - \frac{\pi}{\sqrt{\lambda}} \frac{2
x^2}{(x-x')(1 - x^2)} \text{tr} \left[ \Omega(\sigma',x) \,
\sigma_a \right] \sigma_a.
\end{equation}
Now it is straightforward to show that for any matrix $A \in
SL(2,\mathbb{C})$ the following is true
\begin{equation*}
V^{-1} \frac{\text{tr}\, [A \sigma_a] \sigma_a}{\lambda_+ -
\lambda_-} V = \sigma_3, \qquad \text{where} \quad V^{-1} A V =
\text{diag}\, (\lambda_+,\lambda_-),
\end{equation*}
\textit{i.e.} $V$ is the matrix of eigenvectors of $A$ and
$\lambda_{\pm}$ are the eigenvalues. Since the eigenvalues of
$\Omega(\sigma',x)$ are $e^{\pm i p(x)}$ and its matrix of
eigenvectors is $\Psi(x)$, this identity implies that the Lax
matrix \eqref{Lax connection 2} corresponding to the Hamiltonian
$\text{tr}\, \Omega(x)$ can be simplified as
\begin{equation} \label{Lax connection 3}
\text{tr}\, \Omega(x) \quad \longleftrightarrow \quad
\mathcal{J}(x, x') = \frac{4 \pi i}{\sqrt{\lambda}} \frac{\sin
p(x)}{1 - 1/x^2} \frac{\Psi(x) \sigma_3 \Psi(x)^{-1}}{x-x'}.
\end{equation}
But since $\text{tr}\, \Omega(x) = 2 \cos p(x)$, it follows that
the Lax matrix responsible for the flow of the Hamiltonian $p(x)$
is
\begin{equation} \label{Lax connection 4}
p(x) \quad \longleftrightarrow \quad J(x, x') = - \frac{2 \pi
i}{\sqrt{\lambda}} \frac{x^2}{x^2 - 1} \frac{\Psi(x) \sigma_3
\Psi(x)^{-1}}{x-x'}.
\end{equation}
Now we expand this around $x = \pm 1$ by extracting the Lax
matrices associated with the Taylor coefficients of the
quasi-momentum about $x = \pm 1$, namely
\begin{equation} \label{Lax connection 5}
\text{res}_{x = \pm 1} \, (x \mp 1)^{-n} p(x) \quad
\longleftrightarrow \quad \tilde{J}_{n,\pm}(x') = \text{res}_{x =
\pm 1} \, (x \mp 1)^{-n} J(x,x').
\end{equation}
Using the straightforward identity for a rational matrix $M(x)$
with singularities at $x = \pm 1$
\begin{equation} \label{res singular part identity}
\text{res}_{x = \pm 1} \, \frac{M(x)}{x - x'} = - \left( M(x')
\right)_{\pm 1},
\end{equation}
where $\left( M(x') \right)_{\pm 1}$ denotes the pole part of
$M(x')$ at $x' = \pm 1$, one can recast the Lax matrix \eqref{Lax
connection 5} in the much more useful form
\begin{equation} \label{Lax connection 6}
\tilde{J}_{n,\pm}(x') = \left( \frac{2 \pi i}{\sqrt{\lambda}}
\frac{x'^2}{x'^2 - 1} \frac{\Psi(x') \sigma_3 \Psi(x')^{-1}}{(x'
\mp 1)^n} \right)_{\pm 1}.
\end{equation}
At this point we can also define the corresponding
\textit{hierarchy of times} $\tilde{t}_{n,\pm}$ as the times
generated by the Hamiltonians $\text{res}_{x = \pm 1} \, (x \mp
1)^{-n} p(x)$ in \eqref{Lax connection 5}, namely we define
\begin{equation} \label{hierarchy of times}
\partial_{\tilde{t}_{n,\pm}} = \left\{ \text{res}_{x = \pm 1} \, (x \mp
1)^{-n} p(x), \cdot \right\}.
\end{equation}
Going back to equation \eqref{pre monodromy evolution}, if we
follow the prescription we just established to go from \eqref{Lax
connection 3} to \eqref{Lax connection 5}, namely of dividing
through by $- 2 \sin p(x)$ and taking the residue at $x = \pm 1$
one readily finds the equation governing the evolution of the
monodromy matrix under the hierarchy of times \eqref{hierarchy of
times}
\begin{equation} \label{monodromy evolution}
[\partial_{\tilde{t}_{n,\pm}} - \tilde{J}_{n,\pm}(x'), \Omega(x')]
= 0,
\end{equation}
which is exactly of the form \eqref{sigma tau evolution}. As an
application of equation \eqref{Lax connection 6} for the hierarchy
of Lax matrices we show that the first two of these matrices
$\tilde{J}_{0,\pm}$ are related to the original Lax connection
$J_{\pm} = J_0 \pm J_1$. Indeed, applying the asymptotics for the
quasi-momentum $p(x) \sim_{x \rightarrow \pm 1} - \frac{\pi
\kappa_{\pm}}{x \mp 1}$ to equation \eqref{Lax connection 5} with
$n=0$ we find
\begin{equation*}
- \pi \kappa_{\pm} \quad \longleftrightarrow \quad
\tilde{J}_{0,\pm}(x') = \pm \frac{\pi i}{\sqrt{\lambda}}
\frac{\Psi(\pm 1) \sigma_3 \Psi(\pm 1)^{-1}}{x' \mp 1}.
\end{equation*}
Now the components $J_{\pm}$ of the Lax connection are associated
to $\sigma^{\pm} = \frac{1}{2}(\tau \pm \sigma)$ translations
which are in turn generated by $\mathcal{E} \pm \mathcal{P} =
\frac{\sqrt{\lambda}}{2} \kappa_{\pm}^2$ and hence $J_{\pm}(x') =
- \frac{\sqrt{\lambda} \kappa_{\pm}}{\pi} \tilde{J}_{0,\pm}(x')$
since \cite{Paper1}
\begin{equation} \label{energy-momentum}
\mathcal{E} \pm \mathcal{P} \quad \longleftrightarrow \quad -
\frac{\sqrt{\lambda} \kappa_{\pm}}{\pi} \tilde{J}_{0,\pm}(x') =
\frac{i \kappa_{\pm}}{1 \mp x'} \Psi(\pm 1) \sigma_3 \Psi(\pm
1)^{-1} = J_{\pm}(x').
\end{equation}

Finally we derive the evolution equations for the Lax matrices
\eqref{Lax connection 6} under the hierarchy of times
\eqref{hierarchy of times} and show that they take the
zero-curvature form. We follow an argument given in \cite{Babelon} for
finite-dimensional systems which applies readily here. Writing the
monodromy matrix as $\Omega(x') = \Psi(x')\, \text{diag}(e^{i p(x)},
e^{-i p(x)}) \Psi(x')^{-1}$, equation \eqref{monodromy evolution}
implies that
\begin{equation} \label{Psi evolution}
\left[ \Psi(x')^{-1} \left(\partial_{\tilde{t}_{n,\pm}} \Psi(x')
\right) - \Psi(x')^{-1} \tilde{J}_{n,\pm}(x') \Psi(x'),
\text{diag}(e^{i p(x)}, e^{-i p(x)}) \right] = 0.
\end{equation}
But any $2 \times 2$ matrix commuting with a diagonal matrix must
itself be diagonal, and therefore we may write
\begin{equation} \label{Psi evolution 2}
\partial_{\tilde{t}_{n,\pm}} \Psi(x') = \tilde{J}_{n,\pm}(x') \Psi(x') + \Psi(x')
D(x'),
\end{equation}
for some unknown diagonal $2 \times 2$ matrix $D(x')$. Let us
denote the multi-indices labelling the hierarchy, such as $(n,+)$,
using capital letters, \textit{e.g.} $N = (n,s_n)$ where $n \in
\mathbb{N}$ and $s_n = \pm 1$. So let $N = (n,s_n)$ and $M =
(m,s_m)$, then we have for $\tilde{J}_N(x') =
\tilde{J}_{n,s_n}(x')$
\begin{equation} \label{J^N evolution}
\partial_{\tilde{t}_M} \tilde{J}_N(x') = \left[ \tilde{J}_M(x'), \frac{2 \pi
i}{\sqrt{\lambda}} \frac{x'^2}{x'^2 - 1} \frac{\Psi(x') \sigma_3
\Psi(x')^{-1}}{(x' - s_n)^n} \right]_{s_n},
\end{equation}
where we have made use of \eqref{Psi evolution 2} and the
subscript on the commutator means we take the pole part of the
whole commutator at $x' = s_n$. Let us start by assuming that $s_n
\neq s_m$, then $\tilde{J}_M(x')$ is regular at $x' = s_n$ and
only the pole part at $x' = s_n$ of the second term in the
commutator contributes which is just $\tilde{J}_N(x')$, so
\begin{equation*}
\partial_{\tilde{t}_M} \tilde{J}_N(x') = [\tilde{J}_M(x'), \tilde{J}_N(x')]_{s_n},
\end{equation*}
and likewise we also have $\partial_{\tilde{t}_N} \tilde{J}_M(x')
= [\tilde{J}_N(x'), \tilde{J}_M(x')]_{s_m}$. Since
$[\tilde{J}_M(x'), \tilde{J}_N(x')]$ is rational with poles only
at $x' = \pm 1$ and vanishes at $x' = \infty$ it can be written as
a sum over its pole parts, namely
\begin{equation*}
[\tilde{J}_M(x'), \tilde{J}_N(x')] = [\tilde{J}_M(x'),
\tilde{J}_N(x')]_{+1} + [\tilde{J}_M(x'), \tilde{J}_N(x')]_{-1}.
\end{equation*}
But because $s_n \neq s_m$ we have $\{ s_m, s_n\} = \{ \pm 1\}$
and the zero-curvature condition \eqref{zero-curvature} below follows.
If instead we assume that $s_n = s_m$, then we have
\begin{equation*}
\left[ \tilde{J}_N(x') -  \frac{2 \pi i}{\sqrt{\lambda}}
\frac{x'^2}{x'^2 - 1} \frac{\Psi(x') \sigma_3 \Psi(x')^{-1}}{(x' -
s_n)^n}, \tilde{J}_M(x') -  \frac{2 \pi i}{\sqrt{\lambda}}
\frac{x'^2}{x'^2 - 1} \frac{\Psi(x') \sigma_3 \Psi(x')^{-1}}{(x' -
s_n)^m} \right]_{s_n} = 0
\end{equation*}
since both arguments in the commutator are regular at $x' = s_n =
s_m$. The zero-curvature equation again readily follows from the
above equation and \eqref{J^N evolution}, \textit{i.e.}
\begin{equation} \label{zero-curvature}
\partial_{\tilde{t}_M} \tilde{J}_N(x') - \partial_{\tilde{t}_N} \tilde{J}_M(x') =
[\tilde{J}_M(x'), \tilde{J}_N(x')].
\end{equation}

Let us give an alternative basis $J_{n,\pm}$ for the string
hierarchy whose zeroth level $n=0$ corresponds exactly to the Lax
connection $J_{\pm}$. If we define $- \pi \kappa_{n,\pm} =
\text{res}_{x = \pm 1} (x \mp 1)^{-n} p(x)$ so that
$\kappa_{0,\pm} = \kappa_{\pm}$ then we have the following
correspondence between integral of motion and Lax connection
\begin{equation*}
\frac{\sqrt{\lambda}}{2} \kappa_{\pm} \kappa_{n,\pm} \quad
\longleftrightarrow \quad J_{n,\pm} = - \frac{\sqrt{\lambda}}{2
\pi} \left( \kappa_{\pm} \tilde{J}_{n,\pm} + \kappa_{n,\pm}
\tilde{J}_{0,\pm} \right).
\end{equation*}
In particular, from \eqref{energy-momentum} we see that the zeroth
level $n = 0$ of this hierarchy is precisely the Lax connection
$J_{\pm}$ associated with $\mathcal{E} \pm \mathcal{P} =
\frac{\sqrt{\lambda}}{2} \kappa_{\pm}^2$, so as desired $J_{0,\pm}
= J_{\pm}$. It is straightforward to see by the linearity of the
above expression for $J_{n,\pm}$ and the constancy of the
integrals of motion $\kappa_{n,\pm}$ that the new hierarchy is
also commuting, namely it also satisfies the zero-curvature
equation \eqref{zero-curvature}, with $\partial_{t_{n,\pm}} = \{
\frac{\sqrt{\lambda}}{2} \kappa_{\pm} \kappa_{n,\pm}, \cdot \}$
\begin{subequations} \label{zero-curvature + monodromy}
\begin{equation} \label{zero-curvature 2}
\partial_{t_M} J_N(x') - \partial_{t_N} J_M(x') = [J_M(x'), J_N(x')].
\end{equation}
Likewise, equation \eqref{monodromy evolution} also goes through
unaltered and reads
\begin{equation} \label{monodromy evolution 2}
[\partial_{t_M} - J_M(x'), \Omega(x')] = 0.
\end{equation}
\end{subequations}

\subsection{Baker-Akhiezer vector and linearization} \label{section: B-A and lin}

Equations \eqref{zero-curvature + monodromy} express the fact that
the operators $\partial_{t_M} - J_M(x')$ all commute among
themselves as well as individually with the monodromy matrix
$\Omega(x')$. This means they can all be simultaneously
diagonalised and there exists a solution $\bm{\psi}(P')$ to the
following equations, where $P' = (x',y') \in \Gamma$ and $\Gamma :
\text{det}\, (\Omega(x') - y' {\bf 1}) = 0$ is the spectral curve,
\begin{equation} \label{simultaneous}
\left\{
\begin{split}
\big(\partial_{t_M} - J_M(x')\big)
\bm{\psi}(P') &= 0, \quad \forall M\\
(\Omega(x') - y') \bm{\psi}(P') &= 0.
\end{split}
\right.
\end{equation}
In this section it will be important to keep track of the explicit
dependence of various functions on the hierarchy of times and so
we will use the notation $\{t\}$ for the complete set of times
$t_{0,\pm}, t_{1,\pm}, \ldots$ and write for instance $J_M(x',
\{t\})$, $\Omega(x', \{t\})$ and $\bm{\psi}(P', \{t\})$.

The idea of finite-gap integration (see \cite{Paper1, Paper2} and
references therein) is to identify the analytic properties of the
vector $\bm{\psi}(P', \{t\})$ which specify it uniquely. To this
aim we follow \cite{Paper2, Krichever4} and introduce the normalised
eigenvector $\bm{h}(P', \{t\})$ of $\Omega(x')$ which is
normalised by the condition $\bm{\alpha} \cdot \bm{h} = 1$ where
$\bm{\alpha} = (1,1)$. Using this vector we can look for solutions
to \eqref{simultaneous} in the form
\begin{equation} \label{psi}
\bm{\psi}(P', \{t\}) = \widehat{\Psi}(x', \{t\}) \bm{h}(P',
\{0\}),
\end{equation}
where $\widehat{\Psi}(x', \{t\})$ is a formal matrix solution to
$\big(\partial_{t_M} - J_M(x')\big) \widehat{\Psi}(x') = 0, \,
\forall M$ so that \eqref{psi} trivially satisfies
$\big(\partial_{t_M} - J_M(x')\big) \bm{\psi}(P') = 0, \, \forall
M$. If we fix the initial condition to be $\bm{\psi}(P',\{0\}) =
\bm{h}(P',\{0\})$ so that $\widehat{\Psi}(x', \{0\}) = {\bf 1}$
then by uniqueness of the solution with initial condition
$\widehat{\Psi}(x', \{0\}) = \Omega(x',\{0\})$ it follows that
$\widehat{\Psi}(x', \{t\}) \Omega(x',\{0\}) = \Omega(x',\{t\})
\widehat{\Psi}(x', \{t\})$ and therefore \eqref{psi} is indeed
also an eigenvector of $\Omega(x',\{t\})$.

We now analyse the analytic properties of the vector
$\bm{\psi}(P', \{t\})$ in the form \eqref{psi} by obtaining the
analytic properties of $\widehat{\Psi}(x')$ and $\bm{h}(P',
\{0\})$. First let us rewrite the hierarchy of Lax matrices in the
more transparent form
\begin{equation*}
J_{n,\pm}(x') = \left( \Psi(x') s_{n,\pm}(x') \sigma_3
\Psi(x')^{-1} \right)_{\pm 1},
\end{equation*}
where the \textit{singular parts} $s_{n,\pm}(x')$ are defined as
\begin{equation} \label{singular parts}
s_{n,\pm}(x') = \left( - i \frac{x'^2}{x'^2 - 1}
\left(\kappa_{n,\pm} + \frac{\kappa_{\pm}}{(x' \mp 1)^n}\right)
\right)_{\pm 1}.
\end{equation}
In the particular case of the zeroth level Lax matrix
$J_{0,\pm}(x')$ the singular parts are precisely those of the Lax
connection $J_{\pm}$ as defined in \cite{Paper1, Paper2}, namely
$s_{0,\pm}(x') = \frac{i \kappa_{\pm}}{1 \mp x'}$. Because
$J_{n,\pm}(x')$ only has poles at $x' = \pm 1$ it follows by
Poincar\'e's theorem on holomorphic differential equations that
$\widehat{\Psi}(x')$ is holomorphic outside $x' = \pm 1$. By
studying the asymptotics of the equation for $\widehat{\Psi}(x')$,
its behaviour near $x' = \pm 1$ is easily show to be
\begin{equation*}
\widehat{\Psi}(x',\{t\}) e^{- \sum_n s_{n,\pm} t_{n,\pm} \sigma_3}
= O(1) \qquad \text{as} \; x \rightarrow \pm 1,
\end{equation*}
where $O(1)$ denotes a matrix holomorphic in a neighbourhood of
$x' = \pm 1$. Moreover, using the fact that $J_{n,\pm}(\infty) =
0$ we observe that $\partial_{t_M} \widehat{\Psi}(\infty,\{t\}) =
0, \, \forall M$ and hence $\widehat{\Psi}(\infty,\{t\}) = {\bf
1}$ by the choice of initial conditions. Turning to the normalised
eigenvector $\bm{h}(P', \{t\})$, a standard analysis of its
analytic behaviour reveals that it is meromorphic in $P'$ and
uniquely specified by the following condition
\begin{equation*}
(h_1) \geq \hat{\gamma}(\{t\})^{-1} \infty^-, \quad h_1(\infty^+) = 1,
\quad \text{and} \quad (h_2) \geq \hat{\gamma}(\{t\})^{-1} \infty^+,
\quad h_1(\infty^-) = 1,
\end{equation*}
where the divisor $\hat{\gamma}(\{t\})$ of degree $g+1$ is called the
\textit{dynamical divisor}. The analytic data gathered above for
$\widehat{\Psi}(x')$ and $\bm{h}(P', \{0\})$ is sufficient to uniquely
characterise the components of $\bm{\psi}(P', \{t\})$ as
Baker-Akhiezer functions, namely
\begin{gather*}
(\psi_1) \geq \hat{\gamma}_0^{-1} \infty^-, \quad \psi_1(\infty^+) =
1, \quad \text{ and } \quad (\psi_2) \geq \hat{\gamma}_0^{-1}
\infty^+, \quad \psi_2(\infty^-) = 1,\\
\text{with } \quad \left\{
\begin{split}
&\psi_i(x'^{\pm},\{t\}) e^{\mp \sum_n s_{n,+} t_{n,+}} = O(1),
\quad \text{as } x' \rightarrow 1,\\
&\psi_i(x'^{\pm},\{t\}) e^{\mp \sum_n s_{n,-} t_{n,-}} = O(1),
\quad \text{as } x' \rightarrow -1,
\end{split}
\right. \notag
\end{gather*}
where $\hat{\gamma}_0 = \hat{\gamma}(\{0\})$ is the initial
divisor. Notice that the hierarchy of times enters linearly in the
definition of the Baker-Akhiezer vector $\bm{\psi}(P', \{t\})$
through the essential singularity. This is a very general feature
of finite-gap integration. When explicitly reconstructing the
Baker-Akhiezer vector satisfying the above conditions in terms of
Riemann $\theta$-functions on $\Sigma$, the singular parts give
rise to a unique normalised Abelian differential of the second
kind $d \mathcal{Q}$ with poles at $x' = \pm 1$ of the prescribed
form
\begin{equation*}
d \mathcal{Q} = i dS_{\pm}, \quad \text{as} \; x' \rightarrow \pm 1, \qquad
\text{where} \;
\left\{
\begin{array}{l}
S_+(x'^{\pm},\{t\}) = \pm \sum_n s_{n,+}(x') t_{n,+}, \\
S_-(x'^{\pm},\{t\}) = \pm \sum_n s_{n,-}(x') t_{n,-}.
\end{array}
\right.
\end{equation*}
All the time dependence of the Baker-Akhiezer vector, and hence of the
solution, is encoded in this meromorphic differential $d \mathcal{Q}$
which is linear in the hierarchy of times. In fact, we can define a
differential associated to each time of the hierarchy by writing
\begin{equation} \label{time-differential coupling}
d \mathcal{Q} = \sum_n t_{n,+} d \Omega_{n,+} + \sum_n t_{n,-} d
\Omega_{n,-} = \sum_N t_N d \Omega_N,
\end{equation}
using the multi-index notation, where the normalised Abelian
differentials of the second kind $d \Omega_{n,\pm}$ are defined
uniquely by their respective behaviours at the points $x' = \pm 1$,
namely
\begin{equation*}
d \Omega_{n,+}(x'^{\pm}) = \pm i d s_{n,+}(x') \quad \text{as} \;
x' \rightarrow +1, \qquad d \Omega_{n,-}(x'^{\pm}) = \pm i d
s_{n,-}(x') \quad \text{as} \; x' \rightarrow -1.
\end{equation*}
This correspondence between times of the hierarchy and meromorphic
differentials on $\Sigma$
\begin{equation*}
t_{n,\pm} \mapsto d \Omega_{n,\pm}
\end{equation*}
is a very general feature of finite-gap integration. We say that the
differential \textit{couples} to the time for obvious reasons from
\eqref{time-differential coupling}. As we saw in the previous
sections, every Hamlitonian corresponds to a Lax matrix which
is responsible for generating the corresponding time in the Lax
formalism. Here we see that every Hamiltonian also corresponds to a
meromorphic differential on $\Sigma$ responsible for generating the
corresponding time in the finite-gap language. Notice the splitting
between differentials singular at $x' = +1$ and those singular at $x'
= -1$. These are related to left and right movers of the string.
For instance, at the zeroth level $n=0$ we have $\sigma^{\pm} \equiv
\frac{\tau \pm \sigma}{2} = - t_{0,\pm}$ and $dq_{\pm} \equiv dq \pm
dp = -2 \pi d \Omega_{0,\pm}$, so in particular
\begin{equation*}
t_{0,+} d\Omega_{0,+} + t_{0,-} d\Omega_{0,-} = \frac{1}{2 \pi} (
\sigma dp + \tau dq),
\end{equation*}
which is the usual $d \mathcal{Q}$ defined in \cite{Paper1, Paper2,
Paper3} where all the higher times are set to zero.

In the next section we will be perturbing finite-gap solutions and so
we give here the explicit formulae for the generic finite-gap solution
in terms of Riemann $\theta$-functions on $\Sigma$. Details can be
found in \cite{Paper1, Paper2, Paper3}. Of particular interest for
constructing the embedding $g$ of the string in $SU(2)$
\begin{equation*}
g = \left( \begin{array}{cc} Z_1&Z_2\\ -\bar{Z}_2&\bar{Z}_1
\end{array} \right) \in SU(2),
\end{equation*}
is the dual Baker-Akhiezer vector which is defined relative to the
conjugate divisor $\hat{\tau} \hat{\gamma}_0$ and opposite
singular parts $- s_{n,\pm}$. The components are explicitly
constructed as \cite{Paper2, Paper3}
\begin{subequations} \label{reconstruction formulae}
\begin{equation} \label{reconstruction formula for Z from psi^+}
Z_1 = C \widetilde{\psi}^+_1(0^+), \quad Z_2 =
\frac{C}{\chi(\infty^-)^{\frac{1}{2}}} \widetilde{\psi}^+_2(0^+),
\end{equation}
where $C \in \mathbb{R}$ is a normalisation constant chosen such
that $|Z_1|^2 + |Z_2|^2 = 1$ and $\chi(P)$ is a meromorphic
function on $\Sigma$ with divisor $(\chi) = \hat{\gamma}_0 \cdot
\hat{\tau} \hat{\gamma}_0 \cdot B^{-1}$ and normalised by
$\chi(\infty^+) = 1$ ($B$ is the divisor of branch points of
$\Sigma$). The components of the dual Baker-Akhiezer vector at
$0^+$ are explicitly given by
\begin{equation} \label{reconstruction formula for psi^+_1}
\widetilde{\psi}^+_1(0^+) = h_-(0^+) \frac{\theta \big(\bm{D}; \Pi \big) \theta
\big( 2 \pi \int^{0^+}_{\infty^+} \bm{\omega} - \int_{\bm{b}}
d\mathcal{Q} - \bm{D}; \Pi \big)}{\theta \big(\int_{\bm{b}}
d\mathcal{Q} + \bm{D}; \Pi \big) \theta \big( 2 \pi
\int^{0^+}_{\infty^+} \bm{\omega} - \bm{D}; \Pi \big)} \exp \left( +
\frac{i}{2} \int_{\infty^-}^{\infty^+} d\mathcal{Q} - \frac{i}{2}
\int_{0^-}^{0^+} d\mathcal{Q} \right),
\end{equation}
\begin{equation} \label{reconstruction formula for psi^+_2}
\widetilde{\psi}^+_2(0^+) = h_+(0^+) \frac{\theta \big(\bm{D}; \Pi \big) \theta
\big( 2 \pi \int^{0^+}_{\infty^-} \bm{\omega} - \int_{\bm{b}}
d\mathcal{Q} - \bm{D}; \Pi \big)}{\theta \big(\int_{\bm{b}}
d\mathcal{Q} + \bm{D}; \Pi \big) \theta \big( 2 \pi
\int^{0^+}_{\infty^-} \bm{\omega} - \bm{D}; \Pi \big)} \exp \left( -
\frac{i}{2} \int_{\infty^-}^{\infty^+} d\mathcal{Q} - \frac{i}{2}
\int_{0^-}^{0^+} d\mathcal{Q} \right).
\end{equation}
\end{subequations}

\subsection{Quasi-actions} \label{section: quasi-actions}

Remember that the Lax matrix in \eqref{Lax connection 4} is
responsible for the flow of the Hamiltonian $\text{tr}\,
\Omega(x) = 2 \cos p(x)$. Thus going back to the
corresponding Hamilton equation in Lax form we can rewrite it as
\begin{equation} \label{pre action Lax equation}
2 \pi i \left\{ - \frac{\sqrt{\lambda}}{8 \pi^2 i} \left( 1 -
\frac{1}{x^2} \right) p(x) , J_1(x') \right\} = \left[
\partial_{\sigma} - J_1(x'), \frac{\Psi(x)
\frac{i}{2} \sigma_3 \Psi(x)^{-1}}{x-x'} \right].
\end{equation}
Integrating this equation in $x$ over the different $a$-cycles,
and recalling that the action variables are defined as $S_i = -
\frac{\sqrt{\lambda}}{8 \pi^2 i} \int_{a_i} \left( 1 -
\frac{1}{x^2} \right) p(x) dx$ we find
\begin{subequations} \label{action, R Lax eqs}
\begin{equation} \label{action Lax equation}
\{ S_i , J_1(x') \} = \left[
\partial_{\sigma} - J_1(x'), \frac{1}{4 \pi}
\int_{a_i} \frac{\Psi(x) \sigma_3 \Psi(x)^{-1}}{x-x'} dx \right],
\end{equation}
and similarly integrating around the point $x = \infty$ and
recalling that the global $SU(2)_R$ charge is defined as
$\frac{R}{2} = \frac{\sqrt{\lambda}}{8 \pi^2 i} \oint_{\infty}
\left( 1 - \frac{1}{x^2} \right) p(x) dx$ we find
\begin{equation} \label{R Lax equation}
\frac{1}{2} \{ R , J_1(x') \} = \left[
\partial_{\sigma} - J_1(x'), - \frac{1}{4 \pi}
\int_{\infty} \frac{\Psi(x) \sigma_3 \Psi(x)^{-1}}{x-x'} dx
\right],
\end{equation}
\end{subequations}
Equations \eqref{action, R Lax eqs} simply say that the
Hamiltonian flow of the action variables $S_i$ and $R$ are
generated by the following respective Lax matrices
\begin{equation} \label{action Lax}
\begin{split}
S_i \quad &\longleftrightarrow \quad A_i(x') = \frac{1}{4 \pi}
\int_{a_i} \frac{\Psi(x) \sigma_3 \Psi(x)^{-1}}{x-x'}
dx,\\
\frac{R}{2} \quad &\longleftrightarrow \quad - \frac{1}{4 \pi}
\oint_{\infty} \frac{\Psi(x) \sigma_3 \Psi(x)^{-1}}{x-x'} dx.
\end{split}
\end{equation}

Because any integral of motion can be expressed in terms of the
action variables $S_i$, one ought to be able to use equation
\eqref{action Lax} to derive the Lax matrix for any other
integral of motion. Indeed, for instance we know that
\begin{equation} \label{dP and dE}
\begin{split}
\delta \mathcal{P} &= \sum_{i=1}^g \left( \int_{b_i} \frac{dp}{2
\pi} \right) \delta S_i + \bigg( \int_{\infty^-}^{\infty^+}
\frac{dp}{2 \pi} \bigg) \; \frac{1}{2} \delta R,\\
\delta \mathcal{E} &= \sum_{i=1}^g \left( \int_{b_i} \frac{dq}{2
\pi} \right) \delta S_i + \bigg( \int_{\infty^-}^{\infty^+}
\frac{dq}{2 \pi} \bigg) \; \frac{1}{2} \delta R
\end{split}
\end{equation}
and so this means one can write
\begin{equation*}
\{ \mathcal{E} \pm \mathcal{P}, \cdot \} = \sum_{i=1}^g \left(
\int_{b_i} \frac{dq_{\pm}}{2 \pi} \right) \{ S_i, \cdot \} +
\bigg( \int_{\infty^-}^{\infty^+} \frac{dq_{\pm}}{2 \pi} \bigg)
\frac{1}{2} \left\{ R, \cdot \right\},
\end{equation*}
where $dq_{\pm} = dq \pm dp$. Making use of the Lax matrix for the
action variables \eqref{action Lax} and the fact that the
differentials $dq_{\pm}$ are normalised as $\int_{a_i} dq_{\pm} =
0$ means we can write the Lax matrix for $\mathcal{E} \pm
\mathcal{P}$ as follows
\begin{multline*}
\mathcal{E} \pm \mathcal{P} \quad \longleftrightarrow \quad
\frac{1}{4 \pi} \sum_{i=1}^g \left[ \int_{a_i} \frac{\Psi(x)
\sigma_3 \Psi(x)^{-1}}{x-x'} dx \int_{b_i} \frac{dq_{\pm}}{2 \pi}
- \int_{b_i} \frac{\Psi(x) \sigma_3 \Psi(x)^{-1}}{x-x'} dx
\int_{a_i} \frac{dq_{\pm}}{2 \pi} \right]\\ - \frac{1}{4 \pi}
\oint_{\infty} \frac{\Psi(x) \sigma_3 \Psi(x)^{-1}}{x-x'} dx
\int_{\infty^-}^{\infty^+} \frac{dq_{\pm}}{2 \pi}.
\end{multline*}
Written in this form we can apply the Riemann bilinear identity to
obtain
\begin{equation} \label{energy momentum Lax}
\mathcal{E} \pm \mathcal{P} \quad \longleftrightarrow \quad - i
\Big( \text{res}_{x = 1} + \text{res}_{x = - 1} \Big)
\frac{\Psi(x) \sigma_3 \Psi(x)^{-1}}{x-x'} \frac{q_{\pm}(x)}{2
\pi} dx,
\end{equation}
where an overall factor of two came from the fact that we get
equivalent contributions from both sheets, namely at $x^{\pm} =
(+1)^{\pm}$ and $x^{\pm} = (-1)^{\pm}$. Note also importantly that
there is no contribution from the apparent pole at $x = x'$. This
is because $x = x'$ is not actually a pole of the Lax equation
itself, as can be seen from \eqref{pre action Lax equation} which
is perfectly regular as $x$ approaches $x'$ since
$[\partial_{t_{n,\pm}} - J_{n,\pm}(x'), \Psi(x') \sigma_3
\Psi(x')^{-1}] = 0$ which follows from \eqref{Psi evolution 2} and
the trivial fact that diagonal matrices commute. An equation such
as \eqref{energy momentum Lax} relating an integral of motion to a
Lax matrix should really always be understood as a relation
between two ingredients of a Lax equation. To evaluate the
residues in \eqref{energy momentum Lax} we note that the Abelian
integrals $q_{\pm}(x)$ have poles at $x = \pm 1$ with the
following asymptotics
\begin{equation*}
q_+(x) \sim_{x \rightarrow +1} - \frac{2 \pi \kappa_+}{x - 1},
\qquad q_-(x) \sim_{x \rightarrow -1} \frac{2 \pi \kappa_-}{x + 1}.
\end{equation*}
It follows now using the identity \eqref{res singular part
identity} that
\begin{equation*}
\mathcal{E} \pm \mathcal{P} \quad \longleftrightarrow \quad \left(
\Psi(x') i \sigma_3 \Psi(x')^{-1} \frac{q_{\pm}(x')}{2 \pi}
\right)_{\pm 1} = \frac{i \kappa_{\pm}}{1 \mp x'} \Psi(\pm 1)
\sigma_3 \Psi(\pm 1)^{-1} = J_{\pm}(x'),
\end{equation*}
and we recover exactly the same expression as before
\eqref{energy-momentum}. It is important to note that it was the
multivaluedness of the Abelian integral $q_{\pm}(P) = \int^P
dq_{\pm}$ (or equivalently the fact that $dq_{\pm}$ had some
non-trivial periods) which resulted in a non-zero answer for the
corresponding Lax matrix. Indeed, the Lax matrix obtained by this
argument clearly depends only on the cohomology class $[dq_{\pm}]
\in H^1(\Sigma, \infty^{\pm})$ of the Abelian differential
$dq_{\pm}$ one starts off with on the singular algebraic curve
$\Sigma / \{ \infty^{\pm} \}$. One can see this explicitly
from the equation preceding \eqref{energy momentum Lax} or
otherwise from \eqref{energy momentum Lax} itself: suppose
$dq_{\pm}, dq'_{\pm}$ are two representatives of the same
cohomology class, then $dq_{\pm} - dq'_{\pm} = df$ is exact and
the corresponding difference of the expressions in \eqref{energy
momentum Lax} is
\begin{equation*}
- \frac{i}{2} \sum_{P = (\pm 1)^{\pm}} \text{res}_P \frac{\Psi(P)
\sigma_3 \Psi(P)^{-1}}{x(P) - x'} \frac{f(P)}{2 \pi} dx,
\end{equation*}
where $\Psi(P) = (\bm{h}(P),\bm{h}(\hat{\sigma} P))$. But this is the
sum over the residues of a well defined meromorphic differential on
$\Sigma / \{ \infty^{\pm} \}$ (since $f(P)$ is single-valued) and so
is zero.

\begin{figure}[h]
\centering \psfrag{BQ}{\small \red $\mathcal{B}_Q$}
\psfrag{Qp}{\small $Q^+$} \psfrag{Qm}{\small $Q^-$}
\includegraphics[height=20mm]{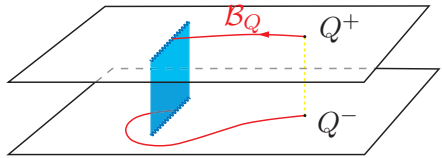}
\caption{Definition of the cycle $\mathcal{B}_Q$ for a given
singular point $Q$.} \label{figure: singular points}
\end{figure}

One could use the same trick as above to compute more explicitly
the Lax matrix for the action variables \eqref{action Lax}. To
simplify the notation we first combine $S_i$ and $\frac{1}{2} R$
into the $g+1$ filling fractions
\begin{equation*}
S_I = - \frac{\sqrt{\lambda}}{8 \pi^2 i} \int_{\mathcal{A}_I}
\left( 1 - \frac{1}{x^2} \right) p(x) dx
\end{equation*}
where $\mathcal{A}_I$ is the cycle going around the
$I^{\text{th}}$ cut counterclockwise on the top sheet. They
satisfy $\sum_{I = 1}^{g+1} S_I = \frac{1}{2}(L - R)$ where $L$ is
the global $SU(2)_L$ charge. So to apply the previous reasoning we
could write
\begin{equation*}
\delta S_I = \sum_{J=1}^{g+1} \delta_{IJ} \delta S_J.
\end{equation*}
For the same argument to follow through we must introduce Abelian
differentials $dq^{(J)}$ of the second kind (so $dq^{(J)}$ has no
residues) such that
\begin{equation} \label{quasi-actions}
\int_{\mathcal{A}_I} dq^{(J)} = 0, \qquad \int_{\mathcal{B}_I}
dq^{(J)} = \delta_{IJ},
\end{equation}
where $\mathcal{B}_I$ is the contour going from $\infty^+$ to
$\infty^-$ through the $I^{\text{th}}$ cut. Such differentials
exist: consider $g + 1$ independent differentials from the
hierarchy, and call them $d \Omega_J$. Then the $(g+1) \times
(g+1)$ matrix $A_{IJ} = \int_{\mathcal{B}_I} d \Omega_J$ is
invertible, and so $dq^{(J)} = A^{-1}_{KJ} d\Omega_K$ have the
desired property. Yet since the conditions \eqref{quasi-actions}
on the differentials $dq^{(J)}$ uniquely specify their cohomology
class in $H^1(\Sigma, \infty^{\pm})$, by the preceding remark they
are also sufficient to uniquely fix the resulting Lax matrix. By
the procedure of section \ref{section: B-A and lin} these Lax
matrices yield unique normalised Abelian differentials which
satisfy \eqref{quasi-actions}, which we still denote $dq^{(J)}$ by
abuse of notation. Since the operations of constructing a Lax
matrix from a given integral of motion and that of constructing an
Abelian differential from a given Lax matrix are both linear, it
follows that the equation for $\mathcal{E}$ in \eqref{dP and dE}
translates into an equation in terms of differential forms on
$\Sigma / \{ \infty^{\pm} \}$. Rewrite this equation as
\begin{equation} \label{dE/dS_I}
\delta \mathcal{E} = \sum_{i=1}^g \bigg( \int_{b_i} \frac{dq}{2
\pi} - \int_{\infty^-}^{\infty^+} \frac{dq}{2 \pi} \bigg) \delta
S_i + \bigg( \int_{\infty^-}^{\infty^+} \frac{dq}{2 \pi} \bigg) \;
\delta \left( \frac{1}{2} R + \sum_{i=1}^g S_i \right) =
\sum_{I=1}^{g+1} \bigg( \int_{\mathcal{B}_I} \frac{dq}{2 \pi}
\bigg) \delta S_I,
\end{equation}
it follows that
\begin{equation} \label{important formula}
dq = \sum_{I=1}^{g+1} \left( \int_{\mathcal{B}_I} dq \right)
dq^{(I)},
\end{equation}
and in particular this leads to the following equation which will
be important later
\begin{equation*}
\int_{\mathcal{B}_Q} dq = \sum_{I=1}^{g+1} \left(
\int_{\mathcal{B}_I} dq \right) \int_{\mathcal{B}_Q} dq^{(I)}
\end{equation*}
where $Q \in \Gamma$ is a singular point on $\Gamma$ which is
blown up on the desingularised curve $\Sigma$ to two points
$Q^{\pm} \in \Sigma$ and $\mathcal{B}_Q$ is a curve joining
$Q^{\pm}$ having zero intersection number with any of the $a$- or
$b$-cycles, see Figure \ref{figure: singular points}.

\section{Perturbations of finite-gap strings} \label{section: perturbations}

\begin{figure}[h] \centering
\begin{tabular}{ccccc}
\psfrag{S}{\footnotesize $\Sigma^{\epsilon}$} \psfrag{a}{\tiny
\red $a_0$} \psfrag{b}{\tiny \lightblue $b_0$}
\includegraphics[height=25mm]{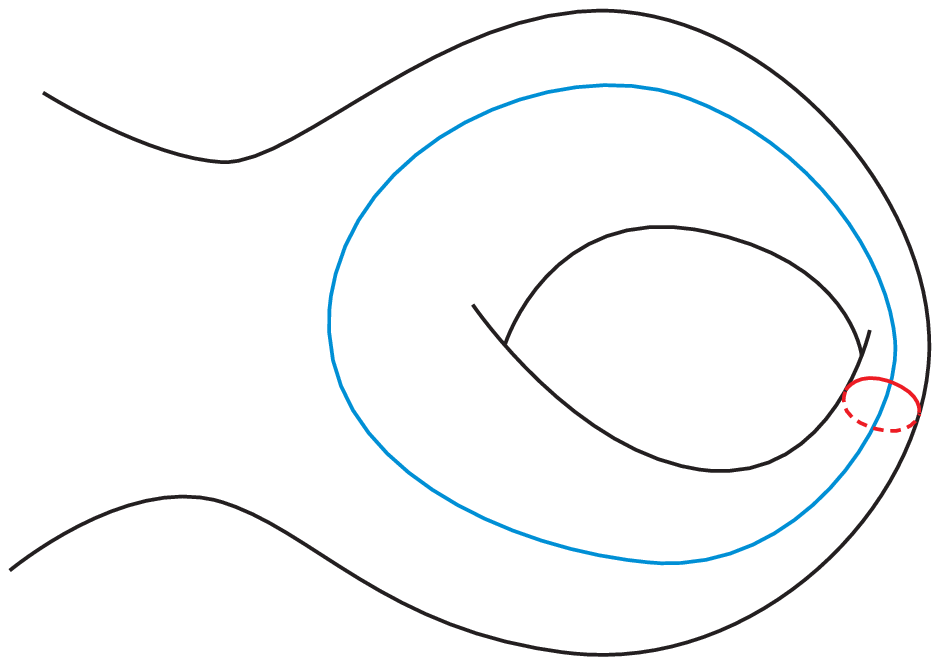} & $\quad$ &
\raisebox{10mm}{$\red \underset{\black \epsilon \rightarrow
0}\longrightarrow$} & $\quad$ & \psfrag{S}{\footnotesize $\Sigma$}
\psfrag{b}{\tiny \lightblue $b_0$}
\includegraphics[height=25mm]{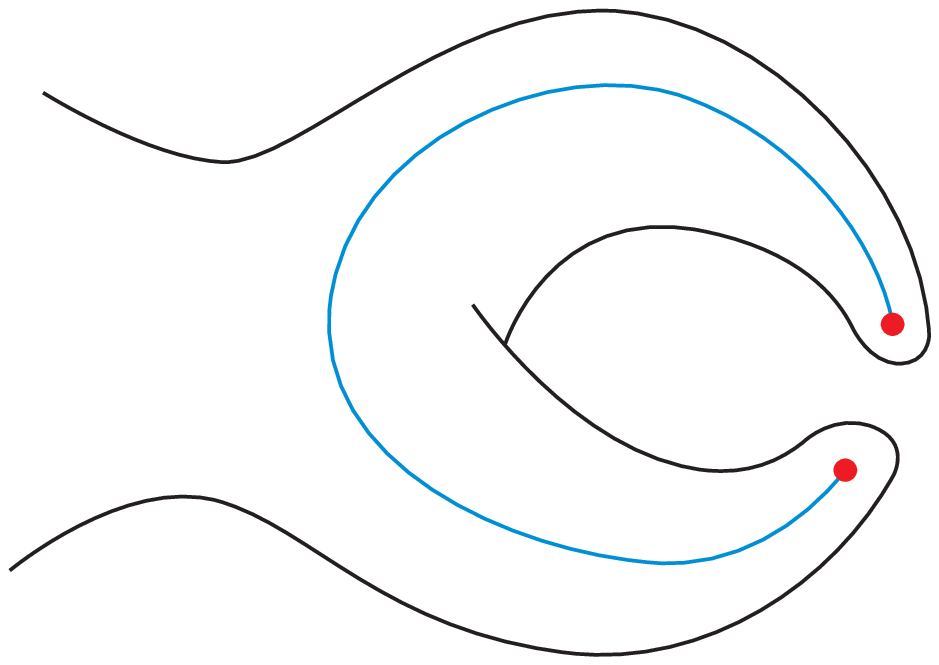}\\
\end{tabular}
\caption{Pinching an $a$-cycle.} \label{Figure: degeneration}
\end{figure}

Given a finite-gap solution $Z_i$ with underlying algebraic curve
$\Sigma$ of genus $g$, one can obtain its stability angles by
considering nearby solutions $Z_i + \delta Z_i$ with algebraic
curves $\Sigma^{\epsilon}$ of genus $g + 1$. In other words,
perturbations of a given finite-gap solution $Z_i$ correspond to
degenerations of a genus $g+1$ algebraic curve $\Sigma^{\epsilon}$
into the genus $g$ curve $\Sigma$ of the solution $Z_i$, see
Figure \ref{Figure: degeneration}.
Now since we are concerned with real finite-gap solutions,
constructed from real algebraic curves $\Sigma$ (see \cite{Paper1}
for a discussion of reality conditions), the degeneration process
in Figure \ref{Figure: degeneration} describing the perturbation
should respect this reality condition. This forces us to consider
degenerations through the pinching of imaginary cycles, which we
can choose to call the $a$-cycles as in \cite{Paper1}. We discuss
the pinching of $a$-cycles in appendix \ref{section: pinching
a-period}.

As discussed in section \ref{section: B-A and lin} the dependence
of the general finite-gap solution on the hierarchy of times $\{ t
\}$ is entirely encoded in the normalised Abelian differential of
the second kind $d\mathcal{Q} = \sum_N t_N d\Omega_N$ defined in
\eqref{time-differential coupling} which enters the reconstruction
formula as follows
\begin{equation} \label{reconstruction formula for Z_i}
Z_i = C_i \; \frac{\theta \big( 2 \pi \int^{0^+}_{P_i} \bm{\omega}
- \int_{\bm{b}} d\mathcal{Q} - \bm{D} ; \Pi \big)}{\theta
\big(\int_{\bm{b}} d\mathcal{Q} + \bm{D} ; \Pi \big)} \; \exp
\left( - i \int^{0^+}_{P_i} d\mathcal{Q} \right),
\end{equation}
where $P_1 = \infty^+$ and $P_2 = \infty^-$. In this expression we
have hidden all the time independent part into the overall
constants $C_i$ whose specific forms can be retrieved from the
complete reconstruction formulae \eqref{reconstruction formulae}.
A nearby solution $Z_i + \delta Z_i$ is constructed with the same
formulae but from slightly deformed data (which includes a
deformed curve $\Sigma^{\epsilon}$)
\begin{equation} \label{reconstruction formula for Z_i + dZ_i}
Z_i + \delta Z_i = C^{\epsilon}_i \; \frac{\theta \big( 2 \pi
\int^{0^+}_{P_i} \vec{\omega}^{\epsilon} -
\int_{\vec{b}^{\epsilon}} d\mathcal{Q}^{\epsilon} - \vec{D} ;
\tilde{\Pi}^{\epsilon} \big)}{\theta
\big(\int_{\vec{b}^{\epsilon}} d\mathcal{Q}^{\epsilon} + \vec{D} ;
\tilde{\Pi}^{\epsilon} \big)} \; \exp \left( - i \int^{0^+}_{P_i}
d\mathcal{Q}^{\epsilon} \right).
\end{equation}
The ingredients of the deformed solution are as follows. First of
all, since the underlying curve $\Sigma^{\epsilon}$ has genus
$g+1$, the arguments of the $\theta$-functions for this curve are
$(g+1)$-component vectors, namely
\begin{equation*}
\vec{D} = \left( \begin{array}{c} \!\!\! D_0 \!\!\! \\ \bm{D}
\end{array} \right) \in \mathbb{C}^{g+1}, \quad \vec{b}^{\epsilon}
= \left( \begin{array}{c} \!\!\! b_0^{\epsilon} \!\!\! \\
\bm{b}^{\epsilon} \end{array} \right) \in H^1(\Sigma^{\epsilon}),
\quad \vec{\omega}^{\epsilon} = \left(
\begin{array}{c} \!\!\! \omega_0^{\epsilon} \!\!\! \\
\bm{\omega}^{\epsilon} \end{array} \right).
\end{equation*}
In the singular limit $\epsilon \rightarrow 0$ one has
$\bm{b}^{\epsilon} \rightarrow \bm{b}$ and $\bm{\omega}^{\epsilon}
\rightarrow \bm{\omega}$ which are the $\bm{b}$-cycles and the $g$
holomorphic differentials on $\Sigma$ respectively. The extra
$b$-cycle $b_0^{\epsilon}$ becomes a degenerate cycle on the curve
$\Sigma$, see Figure \ref{Figure: degeneration}. In appendix
\ref{section: pinching a-period} we show that the extra
holomorphic differential $\omega_0^{\epsilon}$ on
$\Sigma^{\epsilon}$ acquires a simple pole at the singular point
and so becomes a normalised Abelian differential of the third
kind. The Abelian differential $d\mathcal{Q}^{\epsilon}$ on
$\Sigma^{\epsilon}$ is defined by the same singular parts
\eqref{singular parts} as $d\mathcal{Q}$ at $x = \pm 1$ but could
potentially acquire an extra simple pole at the singular point.
However, because $d\mathcal{Q}^{\epsilon}$ is normalised on
$\Sigma^{\epsilon}$, its residue there would vanish in the
$\epsilon \rightarrow 0$ limit, so that in fact
$d\mathcal{Q}^{\epsilon} \rightarrow d\mathcal{Q}$. One can also show
that $C^{\epsilon}_i \rightarrow C_i$ as $\epsilon \rightarrow 0$.

The important object in \eqref{reconstruction formula for Z_i +
dZ_i} when considering the singular limit $\epsilon \rightarrow 0$
is the period matrix which can be broken down into blocks in a
natural way
\begin{equation*}
\tilde{\Pi}^{\epsilon} = \int_{\vec{b}^{\epsilon}}
\vec{\omega}^{\epsilon} = \left(\begin{array}{cc}
\Pi_{00}^{\epsilon} & {\bm{\Pi}_0^{\epsilon}}^{\sf{T}}\\
\bm{\Pi}_0^{\epsilon} & \Pi^{\epsilon}
\end{array}\right).
\end{equation*}
The singular limits of each block follow from the above
considerations of $\vec{b}^{\epsilon}, \vec{\omega}^{\epsilon}$ in
the limit (see appendix \ref{section: pinching a-period} for
details). In particular, $\Pi^{\epsilon} \rightarrow \Pi$ as
$\epsilon \rightarrow 0$ which is simply the period matrix of
$\Sigma$. The vectors $\bm{\Pi}_0^{\epsilon}$ also stay finite in
the limit. The top left component $\Pi_{00}^{\epsilon}$ on the
other hand diverges in this limit, leading to a simplification of
the Riemann $\theta$-function $\theta( \cdot ;
\tilde{\Pi}^{\epsilon} )$ as $\epsilon \rightarrow 0$ which
becomes expressible in terms of the Riemann $\theta$-function
$\theta( \cdot ; \Pi )$ of $\Sigma$ as in \eqref{theta reduction}.

Now taking into account all the above limits and working to first
order in $\epsilon$, a direct computation shows that the
difference $\delta Z_i$ between expressions \eqref{reconstruction
formula for Z_i + dZ_i} and \eqref{reconstruction formula for Z_i}
contains three types of contribution
\begin{multline} \label{variation}
\delta Z_i = \left( \{\text{periodic}\} + \{\text{periodic}\}
\times e^{i \int_{b_0} d\mathcal{Q}} + \{\text{periodic}\} \times
e^{- i \int_{b_0} d\mathcal{Q}} \right) \times e^{\pi i
\Pi_{00}^{\epsilon}},
\end{multline}
where ``$\left\{ \text{periodic} \right\}$'' denotes functions
periodic in all the angle variables $\varphi_I$ of the underlying
finite-gap solution \eqref{reconstruction formula for Z_i}, namely
invariant under $\varphi_I \rightarrow \varphi_I + 2 \pi$ for each
$I = 1, \ldots, g+1$. The three contributions in \eqref{variation}
correspond to three different stability angles of the underlying
solution \eqref{reconstruction formula for Z_i} which can be read
off directly
\begin{equation} \label{stability angles 1}
\nu^{(I)}_0 = 0, \qquad \nu^{(I)}_{\pm} = \pm 2 \pi \int_{b_0}
dq^{(I)}, \quad I = 1, \ldots, g+1.
\end{equation}
The zero stability angles $\nu^{(I)}_0$ are related to the
$\varphi_I$-translation invariance of the equations of motion
which is explicitly broken by the finite-gap solution
\eqref{reconstruction formula for Z_i}.

\begin{figure}[h]
\begin{tabular}{ccc}
\psfrag{a0}{\tiny \red $a_0$} \psfrag{a1}{\tiny \red $a_1$}
\psfrag{ag}{\tiny \red $a_g$} \psfrag{b0}{\tiny \green $b_0$}
\psfrag{b1}{\tiny \green $b_1$} \psfrag{bg}{\tiny \green $b_g$}
\psfrag{d}{\tiny $\cdots$} \psfrag{ip}{\tiny $\infty^+$}
\psfrag{im}{\tiny $\infty^-$}
\includegraphics[height=40mm]{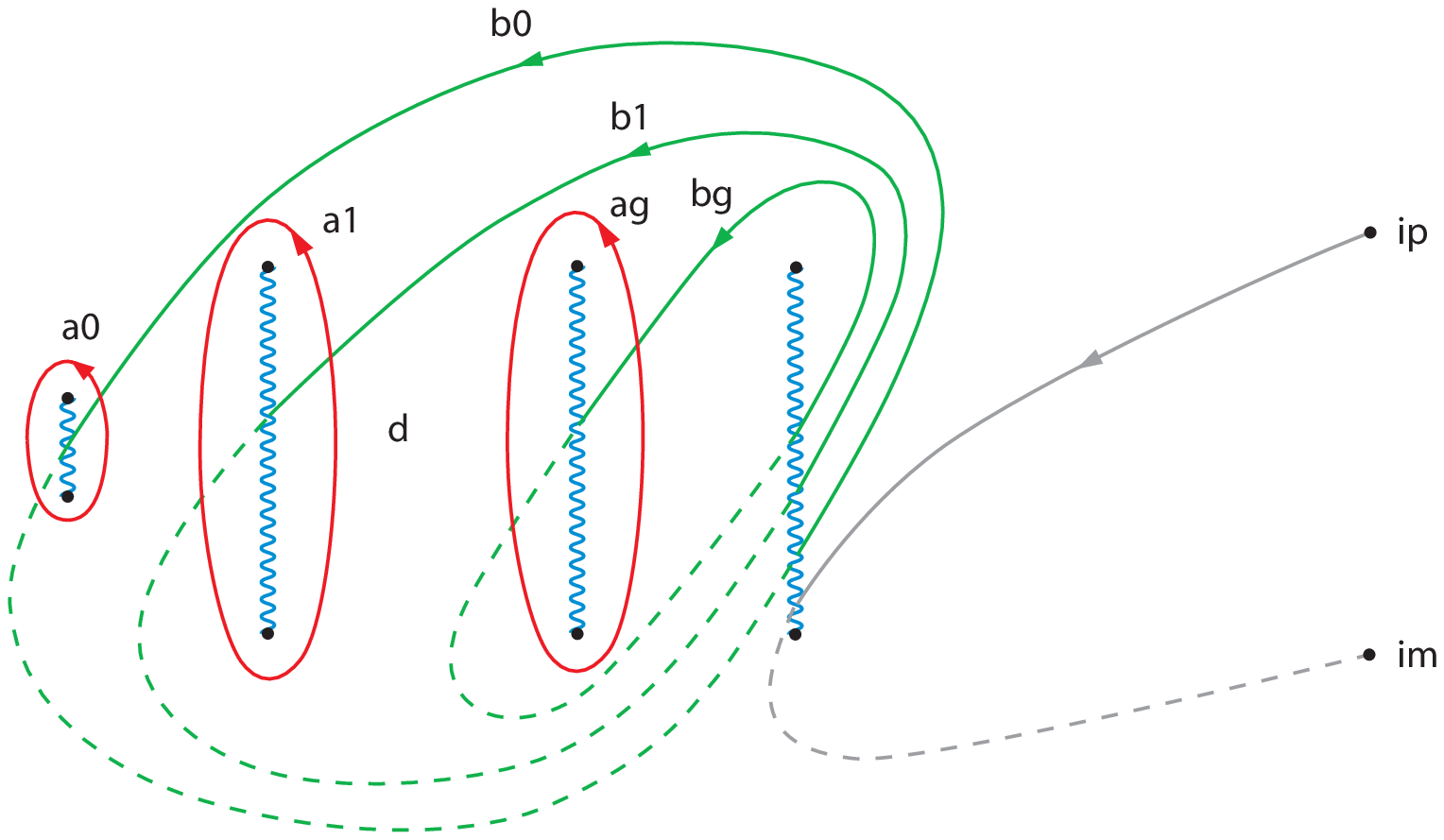} & \qquad \qquad &
\psfrag{P0}{\tiny $P_0$} \psfrag{a1}{\tiny \red $a_1$}
\psfrag{ag}{\tiny \red $a_g$} \psfrag{b0}{\tiny \green
$\mathcal{B}_0$} \psfrag{b1}{\tiny \green $b_1$} \psfrag{bg}{\tiny
\green $b_g$} \psfrag{d}{\tiny $\cdots$} \psfrag{ip}{\tiny
$\infty^+$} \psfrag{im}{\tiny $\infty^-$}
\includegraphics[height=40mm]{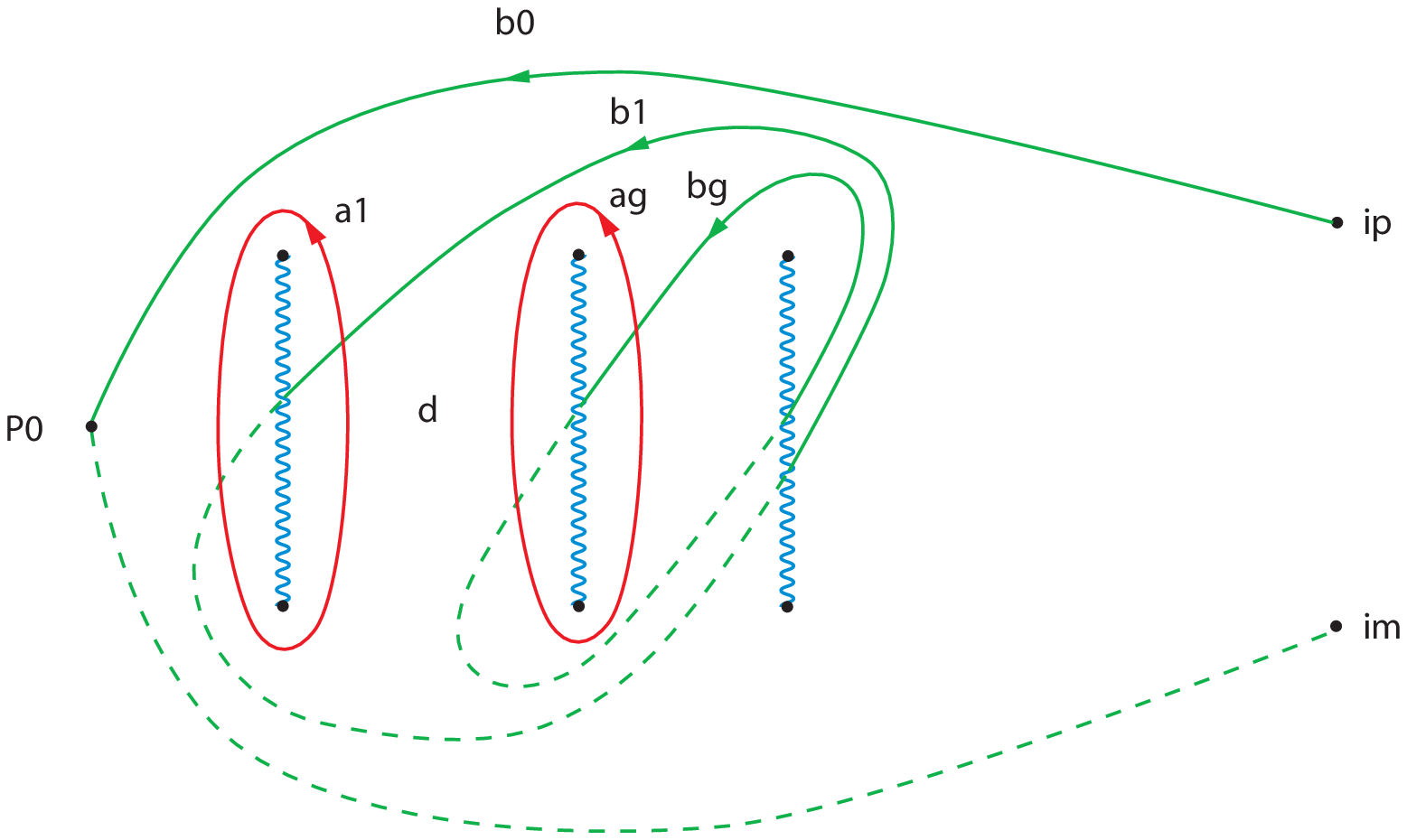}\\
& & \\
$(a)$ & & $(b)$
\end{tabular}
\caption{The canonical cycles before $(a)$ and after $(b)$
shrinking of the $0^{\text{th}}$ cut. Note that it doesn't matter
where the shrinking cut lies with respect to the other cuts, but
for the sake of clarity of the figure we chose it to be the
furthest to the left.} \label{Figure: cycles deg}
\end{figure}

Now stability angles are defined modulo $2 \pi$ but for the
underlying solution \eqref{reconstruction formula for Z_i} to be
periodic requires that
\begin{equation*}
2 \pi \int_{\infty^-}^{\infty^+} dq^{(I)} \in 2 \pi \mathbb{Z},
\qquad I = 1, \ldots, g+1,
\end{equation*}
therefore we can redefine the stability angles $\nu^{(I)}_{\pm}$
as
\begin{equation} \label{stability angles 2}
\nu^{(I)}_{\pm} = \pm 2 \pi \left( \int_{b_0} dq^{(I)} +
\int_{\infty^+}^{\infty^-} dq^{(I)} \right) = \pm 2 \pi
\int_{\mathcal{B}_0} dq^{(I)},
\end{equation}
where the contour $\mathcal{B}_0$ runs from $\infty^+$ on the top
sheet to $\infty^-$ on the bottom sheet, by going through the
$0^{\text{th}}$ cut, see Figure \ref{Figure: cycles deg}. In the
singular limit $\epsilon \rightarrow 0$ the $0^{\text{th}}$ cut
shrinks to a point, say $P_0$ and so \eqref{stability angles 2}
yields
\begin{equation} \label{stability angles 3}
\begin{split}
\nu^{(I)}_{\pm} &= \pm 2 \pi \left( \int_{\infty^+}^{P_0} dq^{(I)}
+ \int_{P_0}^{\infty^-} dq^{(I)} \right) = \pm 2 \pi \left(
\int_{\infty^+}^{P_0} dq^{(I)} -
\int_{P_0}^{\infty^-} \hat{\sigma}^{\ast} dq^{(I)} \right)\\
&= \pm 2 \pi \left( \int_{\infty^+}^{P_0} dq^{(I)} -
\int_{\hat{\sigma} P_0}^{\infty^+} dq^{(I)} \right) = \pm 2 \pi
\left( \int_{\infty^+}^{P_0} dq^{(I)} + \int^{\hat{\sigma}
P_0}_{\infty^+} dq^{(I)} \right)\\ &\equiv \pm 2 \pi \left(
q^{(I)}(P_0) + q^{(I)}(\hat{\sigma}P_0) \right) = \pm 4 \pi
q^{(I)}(P_0),
\end{split}
\end{equation}
where $q^{(I)}(P) \equiv \int_{\infty^+}^P dq^{(I)}$ with the
integral running along the top sheet (the precise choice of
contour then doesn't matter since $dq^{(I)}$ is normalised) and
the last equality follows from $P_0 = \hat{\sigma} P_0$ by virtue
of $P_0$ being a singular point.

By repeating the calculation in \eqref{stability angles 3} but for
the $\mathcal{B}_0$-period of $dp$ (the integrality of the
$b$-periods of $dp$ follows from the closed string requirement,
namely that the finite-gap solution be periodic under $\sigma
\rightarrow \sigma + 2 \pi$),
\begin{equation*}
\int_{\mathcal{B}_0} dp = 2 \pi n_0, \quad n_0 \in \mathbb{Z},
\end{equation*}
the details of which can be found in \cite{Paper1}, we arrive at
an equation for the location of the singular point $P_0$, namely
\begin{equation*}
p(P_0) = n_0 \pi.
\end{equation*}
The above analysis shows that to this singular point $P_0$ there
corresponds two stability angles for each of the $g+1$ cuts
determined by the $\mathcal{B}_0$-period of corresponding
quasi-action $dq^{(I)}$ or
\begin{equation} \label{stability angles 4}
\nu^{(I)}_{\pm} = \pm 4 \pi q^{(I)}(P_0).
\end{equation}

\section{Semi-classical energy spectrum} \label{section: semiclassical 2}

As we recalled in section \ref{section: B-A and lin}, every
finite-gap solution to the equations of motion of a bosonic string
on $\mathbb{R} \times S^3$ is constructed from a finite-genus
algebraic curve $\Sigma$ equipped with an additional set
$\hat{\gamma}_0$ of $g+1$ points on it called a divisor (of degree
$\text{deg } \hat{\gamma}_0 = g+1$). This algebro-geometric data
can be identified with a bundle
$\mathcal{M}_{\mathbb{C}}^{(2g+2)}$ over the moduli space
$\mathcal{L}$ of the algebraic curve $\Sigma$, of dimension
$\text{dim}_{\mathbb{C}} \; \mathcal{L} = g+1$,
\begin{equation*}
S^{g+1}(\Sigma) \rightarrow \mathcal{M}_{\mathbb{C}}^{(2g+2)}
\rightarrow \mathcal{L},
\end{equation*}
whose fibre over every point of the base, specifying a curve
$\Sigma$, is the $(g+1)$-st symmetric product $S^{g+1}(\Sigma) =
\Sigma^{g+1}/S_{g+1}$ of $\Sigma$ (see \cite{Paper1} for moer
details). The finite-gap construction of \cite{Paper1} defines an
injective \textit{geometric map} \cite{Krichever0} from this
algebro-geometric data $\mathcal{M}_{\mathbb{C}}^{(2g+2)}$ into
the space $\mathcal{S}^V_{\mathbb{C}}$ of complexified solutions
$j \in \mathfrak{sl}(2,\mathbb{C})$ to the equations of motion of
a string moving on $\mathbb{R} \times S^3$ which also satisfy the
Virasoro and static gauge conditions,
\begin{equation} \label{geometric map}
\mathcal{G} : \mathcal{M}_{\mathbb{C}}^{(2g+2)} \hookrightarrow
\mathcal{S}^V_{\mathbb{C}}.
\end{equation}
Since a general point in phase-space is the restriction to the
hypersurface $\tau = 0$ of the general solution we can identify
the space $\mathcal{S}_{\mathbb{C}}$ of (complexified) solutions
with (complexified) phase-space
$\mathcal{P}_{\mathbb{C}}^{\infty}$. Furthermore, the subset
$\mathcal{S}^V_{\mathbb{C}} \subset \mathcal{S}_{\mathbb{C}}$ of
solutions satisfying Virasoro and static gauge conditions can be
identified with the second class constraint surface
$\mathcal{P}_{\mathbb{C}}^V \subset
\mathcal{P}_{\mathbb{C}}^{\infty}$ defined by these conditions. We
can describe the map \eqref{geometric map} as an embedding
$\mathcal{M}_{\mathbb{C}}^{(2g+2)} \hookrightarrow
\mathcal{P}_{\mathbb{C}}^V$. If we further impose reality
conditions by restricting the algebro-geometric data
$\mathcal{M}_{\mathbb{C}}^{(2g+2)}$ to \textit{real}
algebro-geometric data (see \cite{Paper1} for a detailed
discussion of reality conditions) then finite-gap integration
describes an injective map \cite{Krichever0}
\begin{equation} \label{reconstruction map}
\mathcal{G}_{\mathbb{R}} : \mathcal{M}_{\mathbb{R}}^{(2g+2)}
\hookrightarrow \mathcal{P}^V_{\mathbb{R}},
\end{equation}
from the $(g+1)$-dimensional toric fibration $\mathbb{T}^{g+1}
\rightarrow \mathcal{M}_{\mathbb{R}}^{(2g+2)} \rightarrow
\mathcal{L}_{\mathbb{R}}$, with $\text{dim}_{\mathbb{R}} \;
\mathcal{L}_{\mathbb{R}} = g+1$, into the (real) phase-space
$\mathcal{P}^V_{\mathbb{R}}$ of strings on $\mathbb{R} \times S^3$
satisfying the Virasoro and static gauge constraints. Introducing
the inclusion $\iota_V : \mathcal{P}_{\mathbb{R}}^V
\hookrightarrow \mathcal{P}_{\mathbb{R}}^{\infty}$ of the second
class constraint surface $\mathcal{P}_{\mathbb{R}}^V \subset
\mathcal{P}_{\mathbb{R}}^{\infty}$, the Dirac bracket on
$\mathcal{P}_{\mathbb{R}}^V$ is the pull-back of the symplectic
structure $\omega$ on $\mathcal{P}_{\mathbb{R}}^{\infty}$. As was
show in \cite{Paper2}, the pull-back to
$\mathcal{M}_{\mathbb{R}}^{(2g+2)}$ of this symplectic structure
$\iota_V^{\ast} \omega$ on $\mathcal{P}^V_{\mathbb{R}}$ takes the
simple form
\begin{equation} \label{symplectic form}
\hat{\omega}_{2g+2} \equiv \mathcal{G}_{\mathbb{R}}^{\ast} \,
\iota_V^{\ast} \omega = \sum_{I=1}^{g+1} \delta S_I \wedge \delta
\varphi_I.
\end{equation}
The different variables in \eqref{symplectic form} are defined as
follows \cite{Paper2}:
\begin{itemize}
\item The \textit{action variables} $S_I$ are given by the filling
fractions
\begin{equation*}
S_I = \frac{1}{2 \pi i} \int_{\mathcal{A}_I} \alpha, \quad I = 1,
\ldots, g+1,
\end{equation*}
where $\mathcal{A}_I$ is the cycle encircling the $I^{\text{th}}$
cut $\mathcal{C}_I$ on the physical sheet of $\Sigma$ represented
as a hyperelliptic curve and $\alpha = \frac{\sqrt{\lambda}}{4
\pi} z dp$ is a special $1$-form on $\Sigma$, with $z \equiv x +
\frac{1}{x}$ and $p(x)$ being the quasi-momentum. \item The
\textit{angle variables} $\varphi_I$ are specified by the image of
the divisor $\hat{\gamma}_0$ on the generalised Jacobian
$J(\Sigma,\infty^{\pm})$ of the curve $\Sigma$ under the extended
Abel map $\vec{\mathcal{A}} : S^{g+1}(\Sigma) \rightarrow
J(\Sigma,\infty^{\pm})$, or more precisely
\begin{equation*}
\varphi_i = \mathcal{A}_i(\hat{\gamma}_0) -
\mathcal{A}_{g+1}(\hat{\gamma}_0), \;\; i = 1,\ldots,g, \quad
\qquad \varphi_{g+1} = - \mathcal{A}_{g+1}(\hat{\gamma}_0).
\end{equation*}
\end{itemize}
The injective map \eqref{reconstruction map} can thus be thought
of as an embedding in phase-space of a $(g+1)$-parameter family of
isotropic $(g+1)$-torii parametrised by $\{S_I\}_{I = 1}^{g+1}$
since the pull-back \eqref{symplectic form} of the symplectic form
$\omega$ to these torii is identically zero. This is the necessary
set-up to apply the Bohr-Sommerfeld conditions \eqref{BS6} for the
quantisation of a $p$-torus in an $n$-dimensional phase-space,
where here the total phase-space is infinite dimensional so that
$n = \infty$ and $p = g+1$. The condition \eqref{BS6} also
involves the stability angles of perturbations around the
$p$-torus which we computed in the section \ref{section:
perturbations}. So applying \eqref{BS6} to the finite-gap string
we can write down the Bohr-Sommerfeld quantisation conditions for
the action variables of the string as follows
\begin{equation} \label{BS7}
\frac{S_I}{\hbar} = N_I + \frac{1}{2} + \sum_{\alpha =
g+2}^{\infty} \left( n_{\alpha} + \frac{1}{2} \right)
\frac{\nu_{\alpha}^{(I)}}{2 \pi} + O(\hbar),
\end{equation}
where we have used the fact that the Maslov index for the
$\mathcal{A}_I$-cycle ($I = 1, \ldots, g+1$) in the generalised
Jacobian $J(\Sigma,\infty^{\pm})$ is simply $\mu_I = 2$. We
emphasise that \eqref{BS7} is only valid in the harmonic
oscillator approximation $N_I \gg n_{\alpha}$ where the
perturbations are much smaller than the background filling
fractions.

\subsection{The main result} \label{section: main result}

In the semiclassical regime, the Hamiltonian is defined by the
same classical function of the actions
$E_{\text{cl}}[S_1,\ldots,S_{g+1}]$ but evaluated on the action
operators since by semiclassical integrability we have that
$[\hat{S}_i, \hat{S}_j] = O(\hbar^3)$, so
\begin{equation*}
\hat{\mathcal{H}}_{\text{string}} =
E_{\text{cl}}[\hat{S}_1,\ldots,\hat{S}_{g+1}] + O(\hbar^2).
\end{equation*}
It follows that the energy spectrum is simply the classical energy
$E_{\text{cl}}$ evaluated on the eigenvalues of the action
variables \eqref{BS7} namely
\begin{multline*}
E = E_{\text{cl}} \left[ N_1 \hbar + \frac{\hbar}{2} +
\sum_{\alpha = g+2}^{\infty} \left( n_{\alpha} + \frac{1}{2}
\right) \frac{\nu_{\alpha}^{(1)}}{2 \pi} \hbar,\ldots, \right. \\
\left. N_{g+1} \hbar + \frac{\hbar}{2} + \sum_{\alpha =
g+2}^{\infty} \left( n_{\alpha} + \frac{1}{2} \right)
\frac{\nu_{\alpha}^{(g+1)}}{2 \pi} \hbar \right] + O(\hbar^2).
\end{multline*}
We now Taylor expand this using the fact that $N_I \gg n_{\alpha}$
to obtain
\begin{equation*}
E = E_{\text{cl}} \left[\left(N_1 + \frac{1}{2}\right)
\hbar,\ldots, \left(N_{g+1} + \frac{1}{2}\right)\hbar \right] +
\sum_{I= 1}^{g+1} \sum_{\alpha = g+2}^{\infty} \left( n_{\alpha}
+ \frac{1}{2} \right) \frac{\partial E_{\text{cl}}}{\partial S_I}
\frac{\nu^{(I)}_{\alpha}}{2 \pi} \hbar.
\end{equation*}
Using equations \eqref{dE/dS_I} and \eqref{stability angles 2} to
express $\partial E_{\text{cl}}/\partial S_I$ and
$\nu^{(I)}_{\alpha}$ respectively as $\mathcal{B}$-periods,
\begin{equation*}
E = E_{\text{cl}} \left[\left(N_1 + \frac{1}{2}\right)
\hbar,\ldots, \left(N_{g+1} + \frac{1}{2}\right)\hbar \right] +
\sum_{I= 1}^{g+1} \sum_{\alpha = g+2}^{\infty} \left( n_{\alpha}
+ \frac{1}{2} \right) \int_{\mathcal{B}_I} \frac{dq}{2 \pi}
\int_{\mathcal{B}_{\alpha}} dq^{(I)} \hbar.
\end{equation*}
where $\mathcal{B}_{\alpha}$ is the contour running from
$\infty^+$ to the singular point labelled $\alpha$ on the top
sheet, and back on the bottom sheet to $\infty^-$. The sum over
$I$ can now be performed using equation \eqref{important formula}
which yields
\begin{equation*}
E = E_{\text{cl}} \left[\left(N_1 + \frac{1}{2}\right)
\hbar,\ldots, \left(N_{g+1} + \frac{1}{2}\right)\hbar \right] +
\sum_{\alpha = g+2}^{\infty} \left( n_{\alpha} + \frac{1}{2}
\right) \int_{\mathcal{B}_{\alpha}} \frac{dq}{2 \pi} \hbar.
\end{equation*}
If we now formally think of the function $E_{\text{cl}}$ as
depending on the infinite set of filling fractions $\{ S_I
\}_{I=1}^{g+1}$, $\{ S_{\alpha} \}_{\alpha = g+2}^{\infty}$ (all
but finitely many of which are turned off for the classical
finite-gap solutions) then we can interpret the
$\mathcal{B}_{\alpha}$-period of $dq/2 \pi$ as $\partial
E_{\text{cl}}/\partial S_{\alpha}$ using a formal analogue of
\eqref{dE/dS_I} for an infinite gap solution. One can then resum
the resulting Taylor expansion to obtain the following formal
expression for the semiclassical energy spectrum
\begin{equation} \label{main result}
E = E_{\text{cl}} \left[\left(N_1 + \frac{1}{2}\right)
\hbar,\ldots, \left(N_{g+1} + \frac{1}{2}\right)\hbar,
\left(n_{g+2} + \frac{1}{2}\right) \hbar, \ldots \right].
\end{equation}
We stress that this is only a formal derivation as rigorously one
would have to regularise the divergent infinite sum over stability
angles at the intermediate steps as well as subtract off the energy of
the vacuum (\textit{i.e.} the zero cut finite-gap solution). But
formally at least the result of the above derivation is the following:
\begin{itemize}
  \item The semiclassical energy spectrum is obtained by evaluating
the classical energy function of an infinite-gap solution on filling
fractions quantised to half-integer multiples of $\hbar$.
  \item The infinite number of singular points of the spectral curve
$\text{det}\, (\Omega(x) - y {\bf 1}) = 0$ which accumulate at $x
= \pm 1$ must be filled with half a unit of $\hbar$ in their
ground state with an additional integer multiple of $\hbar$ for
excitations.
\end{itemize}

\subsection{Comparison with alternative approach} \label{section: algebraic curve approach}

In \cite{Gromov+Vieira1} an alternative method was proposed for
extracting the semiclassical energy spacing around any given
classical solution from the algebraic curve $\Sigma$ itself, without
making use of the divisor $\hat{\gamma}_0$ on $\Sigma$, and which the
subsequent papers \cite{Gromov+Vieira2, Gromov+Vieira3} built
upon. The heart of the method resides in the assumption that the
filling fractions $S_I$ become quantised in integer units at least in
a semiclassical approximation. This assumption seems natural because
the filling fractions constitute the action variables of the theory
(a fact proved only in the $\mathbb{R} \times S^3$ subsector
\cite{Paper2}) and we expect that after Bohr-Sommerfeld quantisation
the action variables become half-integer multiples of $\hbar$. Yet we
see from \eqref{BS7} that this is not the case. As we have argued at
the start of this section, a finite-gap solution can be pictured as a
degenerate isotropic $(g+1)$-torus within the full infinite
dimensional phase-space. And although the Bohr-Sommerfeld conditions
\eqref{BS3} for an integrable system do imply that the action
variables become half-integer multiples of $\hbar$, we saw in section
\ref{section: BS} and appendix \ref{section: BS isolated} that these
conditions receive 1-loop corrections, when applied to a degenerate isotropic
torus, from fluctuations transverse to the torus in the form of
stability angles. Only after the calculation in the previous
subsection can one conclude that the semiclassical spectrum of a
finite-gap solution is given by the classical energy of an
infinite-gap solution whose fillings are half-integrer multiples of
$\hbar$.

Now the algebraic curve $\Sigma$ is characterised by the
quasi-momentum $p(x)$ used to define the filling fractions $S_I$ as
\begin{equation} \label{filling fractions}
S_I = \frac{1}{2\pi i} \frac{\sqrt{\lambda}}{4 \pi} \int_{\mathcal{A}_I}
\left(x+\frac{1}{x}\right)dp = - \frac{1}{2\pi i}
\frac{\sqrt{\lambda}}{4 \pi} \int_{\mathcal{A}_I} \left(1 - \frac{1}{x^2}
\right) p(x) dx, \quad I = 1, \ldots, g+1.
\end{equation}
The integer quantisation of these filling fractions in the
semiclassical limit can be interpreted in the language of the
gauge theory side by attributing to a single Bethe root one unit
of filling fraction. In the semiclassical quantisation of a
solution each cut of its algebraic curve thus turns into a large
clump of Bethe roots with the filling fraction counting the number
of such roots \cite{Gromov+Vieira1}. The idea of
\cite{Gromov+Vieira1} for obtaining the semiclassical energy
spacings is then to compare the energies of two neighbouring
classical solutions differing only by a single Bethe root. If the
underlying solution is characterised by the quasi-momentum $p(x)$
and has $K$ cuts $\mathcal{C}_j$ with mode numbers $n_j, j = 1,
\ldots, K$,
\begin{equation} \label{cuts}
p(x + i0) + p(x - i0) = 2 \pi n_j, \quad x \in \mathcal{C}_j, j =
1, \ldots, K,
\end{equation}
then its perturbation is characterised by a perturbed
quasi-momentum $p(x) + \delta p(x)$ with still the same $K$ cuts
but also with an extra isolated Bethe root at $x_{K+1}$ with mode
number $n_{K+1}$
\begin{subequations} \label{perturbed cuts}
\begin{equation} \label{perturbed cuts 1}
p(x + i0) + \delta p(x + i0) + p(x - i0) + \delta p(x - i0) = 2
\pi n_j, \quad x \in \mathcal{C}_j, j = 1, \ldots, K,
\end{equation}
\begin{equation} \label{perturbed cuts 2}
p(x_{K+1}) + \delta p(x_{K+1}) + p(x_{K+1}) + \delta p(x_{K+1}) =
2 \pi n_{K+1}.
\end{equation}
\end{subequations}
By using \eqref{cuts} we may simplify \eqref{perturbed cuts 1} to
\begin{subequations} \label{perturbed cuts bis}
\begin{equation} \label{perturbed cuts bis 1}
\delta p(x + i0) + \delta p(x - i0) = 0, \quad x \in
\mathcal{C}_j, j = 1, \ldots, K.
\end{equation}
and since $\delta p(x)$ is small, by working to lowest order we
can approximate \eqref{perturbed cuts 2} as
\begin{equation} \label{perturbed cuts bis 2}
p(x_{K+1}) = \pi n_{K+1},
\end{equation}
\end{subequations}
Equations \eqref{perturbed cuts bis} are the starting point in
\cite{Gromov+Vieira1} for obtaining the semiclassical energy
spacings by reading them off from $\delta p(x)$.

Let us now show that the semiclassical energy spacings obtained by
this method agrees with the semiclassical spectrum \eqref{main result}
obtained in the previous subsection. We know from \eqref{dE/dS_I}
that the variation of the energy $E$ of a classical solution as we
vary the moduli $S_I$ is
\begin{equation*}
\delta E = \sum_{I = 1}^{g+1} \left( \int_{\mathcal{B}_I} \frac{dq}{2\pi} \right) \delta S_I.
\end{equation*}
It follows that adding a single Bethe root (which would correspond to
setting $\delta S_J = \hbar$ for some $J$) should increase the energy of
the solution by
\begin{equation} \label{Energy spacing}
\delta E = \int_{\mathcal{B}_J} \frac{dq}{2\pi} \hbar.
\end{equation}
This is exactly what one gets if we set $N_J \rightarrow N_J
+ 1$ in \eqref{main result} and Taylor expand in the $J^{\text{th}}$
entry using $N_J \gg 1$. We easily find that the energy evaluated on
the solution with $S_J = N_J + 1$ is equal to the energy evaluated on
the solution with $S_J = N_J$ plus the perturbation \eqref{Energy
spacing}. Thus \eqref{main result} predicts the same energy spacing
\eqref{Energy spacing} as we would expect if Bethe roots carried
$\hbar$ units of filling fraction.

Note finally that the energy $E_{\text{cl}}$ we have been using is not
the space-time energy of the classical solution but rather the
worldsheet energy or the Hamiltonian of the fields $Z_i$ in the
action. It can however be related to the space-time energy
$\Delta$ by the following simple formula
\begin{equation*}
E = \frac{\Delta^2}{2 \sqrt{\lambda}}.
\end{equation*}

\section{Summary and Outlook}

We have obtained the semiclassical energy spectrum of bosonic
string theory on $\mathbb{R} \times S^3$ as expressed in equation
\eqref{main result} by semiclassically quantising the general
finite-gap solution of this theory. The derivation of \eqref{main
result} can be summarised as follows. We have argued that the
generic finite-gap solution can be thought of as an embedding of a
$(g+1)$-torus $\Sigma_f$ into the full infinite dimensional
phase-space of the theory. Since these torii are
finite-dimensional, they are all degenerate isotropic torii
located on the boundary $\partial \mathcal{S}$ of the infinite
region $\mathcal{S} \equiv \{ S_I \geq 0, \forall I \}$. But a
procedure due to Voros \cite{Voros, Voros1} provides a way of
semiclassically quantising such degenerate torii: the method
consists of studying neighbouring orbits in the small oscillator
approximation, which would live on a neighbouring non-degenerate
torus in the interior of $\mathcal{S}$, and then quantise this
torus in the usual way using Bohr-Sommerfeld-Maslov quantisation
conditions. The computation in section \ref{section: main result}
consisted in formally rewriting the quantised energy of such a
linearised torus in terms of the energy of the infinite-gap
solution it is approximating in the interior of $\mathcal{S}$ (but
still near $\partial \mathcal{S}$). The result is that the
semiclassical energy spectrum can be obtained by evaluating the
classical energy function on points of the following infinite
lattice in $\mathcal{S}$,
\begin{equation} \label{lattice}
S_I \in \hbar \left( \frac{1}{2} + \mathbb{N} \right), \forall I.
\end{equation}
Yet because we computed the semiclassical spectrum around
finite-gap solutions which only describe $\partial \mathcal{S}$,
it follows that \eqref{main result} only describes this lattice
structure near the boundary of $\mathcal{S}$ where the $n$'s are
much smaller than the finitely many $N$'s in \eqref{main result}.
However, since the number of $N$'s is finite but arbitrary, by
formally considering the infinite genus limit of finite-gap
solutions, the complete spectrum is described by $E =
E_{\text{cl}} \left[\left(N_1 + \frac{1}{2}\right) \hbar, \ldots
\right]$ so that \eqref{lattice} should give the correct lattice
structure in the whole bulk of $\mathcal{S}$.

Such a procedure for semiclassically quantising finite-gap
solutions should be sufficiently general to apply with little
modification to more general settings and in particular to the
case of superstrings on $AdS_5 \times S^5$. As stated in the
introduction, it would therefore be very interesting to obtain the
divisor for the full algebraic curve of $AdS_5 \times S^5$ by
constructing the finite-gap solution in full generality on $AdS_5
\times S^5$.

Finally, in view of ultimately obtaining an exact
quantisation of string theory on $AdS_5 \times S^5$ we have argued
that operator ordering issues will be of crucial importance since
they already appear in the semiclassical analysis. In
this paper we assumed for simplicity that the cohomology class of the
subprincipal form vanished since with this assumption we were able to
reproduce the semiclassical spectrum of \cite{Gromov+Vieira1,
Gromov+Vieira2, Gromov+Vieira3} at least for the fluctuations in the
$\mathbb{R} \times S^3$ subspace. This rules out many operator
orderings for the exact quantisation, namely all those for which
the action variables have a subprincipal Weyl symbol.

\section*{Acknowledgements}

I am very grateful to Nick Dorey for many useful insights and
discussions on semiclassical quantisation as well as for following the
details of this work very closely. I would also like to thank Harry
Braden for interesting discussions on finite-gap integration.

\appendix

\section{Symbolic calculus of Pseudo-differential operators} \label{section: PsiDO}

The passage from a classical system on phase-space $T^{\ast} X$ to
its quantum counterpart involves promoting the algebra of
classical observable $C(T^{\ast} X)$ to a noncommutative algebra
$\mathcal{A}$ of operators. Classically, the Poisson algebra of
observables is uniquely specified by the choice of a symplectic
structure $\omega = \sum_i dx_i \wedge d\xi_i$ and the Poisson
bracket of two observables $f,g \in C(T^{\ast} X)$ is then defined
by $\{ f, g\} = \omega(X_f, X_g)$, where $X_H$ denotes the
Hamiltonian vector field associated to any function $H \in
C(T^{\ast} X)$ satisfying $i_{X_H} \omega = dH$. To pass to
quantum mechanics, the prescription of \textit{canonical
quantisation} is to promote the special functions $x_i, \xi_i \in
C(T^{\ast} X)$ to operators $\hat{x}_i,\hat{\xi}_i$ and the
symplectic structure $\omega = \sum_i dx_i \wedge d\xi_i$ to the
Weyl algebra $[ \hat{x}_i, \hat{\xi}_j] = i \hbar \delta_{ij}$
which admits the unique representation $\hat{x}_i = x_i,
\hat{\xi}_i = -i \hbar \partial/\partial x_i \equiv -i \hbar
\partial_i$ in terms of differential operators on $L^2(X)$. The
problem that remains after canonical quantisation is to associate
with any other given observable $f \in C(T^{\ast} X)$ (which is a
function of $x_i, \xi_i$) a (pseudo-)differential operator
$\hat{f}$ on $L^2(X)$, and it is immediately obvious that this is
by no means unique. Many different operators correspond to the
same classical function: for instance, given any $t \in
\mathbb{R}$, the differential operator $t x_1 \partial_1 + (1-t)
\partial_1 \cdot x_1$ is a possible candidate for the quantisation
of the function $x_1 \xi_1$. In other words, it is not possible to
specify the operator ordering in an operator $\hat{f}$ starting
from just single function $f \in C(T^{\ast} X)$. However, with an
infinite set of functions $f_k \in C(T^{\ast} X)$ it turns out to
be possible to associate a unique operator $\hat{f}$ by canonical
quantisation. Such a set defines a function of $\hbar$ through the
asymptotic expansion
\begin{equation} \label{classical Weyl symbol}
f(x,\xi;\hbar) \underset{\hbar \rightarrow 0}\sim \sum_{k \geq 0}
f_k(x,\xi) \hbar^k.
\end{equation}
We refer to such a $\hbar$-dependent function $f(\hbar) \in
C(T^{\ast} X)$ as a \textit{classical (Weyl) symbol}, which is
technically required to satisfy certain estimates, such as all its
partial derivatives being uniformly bounded by some order
function.

Without going into details of the construction, we now state the
map from symbols to \textit{pseudo-differential
operators}\footnote{When the symbol $f(x, \xi; \hbar)$ is a
polynomial in $x, \xi$ the associated operator is an ordinary
partial differential operator. To include the more general case when
$f(x, \xi; \hbar)$ might not be a polynomial we talk about
pseudo-differential operators.} ($\Psi$DO for short). Given a symbol
$f(\hbar)$, we define the corresponding $\Psi$DO by specifying its
action on $u \in L^2(X)$ using the \textit{Weyl quantisation} formula
\cite{Martinez}
\begin{equation*}
\left( \text{Op}_{\hbar}^W (f(\hbar)) u \right) (x) = \frac{1}{(2 \pi
\hbar)^{n}} \int_{\mathbb{R}^{2n}} e^{\frac{i}{\hbar}(x-y) \cdot \xi}
f\left( \frac{x+y}{2}, \xi ; \hbar \right) u(y) dy d\xi.
\end{equation*}
It is important to note here that the choice of Weyl quantisation
in the definition of the $\Psi$DO from its symbol does not limit
us to having only Weyl ordered $\Psi$DOs. Indeed, the operator
$\text{Op}_{\hbar}^W (f(\hbar))$ is Weyl ordered only when the
corresponding Weyl symbol is $\hbar$-independent. So it is
precisely the subleading terms in the asymptotic expansion
\eqref{classical Weyl symbol} of the symbol $f(x, \xi; \hbar)$
which account for the different possible choices of orderings in
the definition of the $\Psi$DO. For example, the Weyl ordered
operator of the classical observable $x_1 \xi_1$ is given simply
by the Weyl symbol $x_1 \xi_1$, namely
\begin{equation*}
\text{Op}_{\hbar}^W (x_1 \xi_1) = \frac{-i \hbar}{2}\left( x_1
\partial_1 + \partial_1 \cdot x_1 \right),
\end{equation*}
whereas the left ordered operator $-i \hbar x_1 \partial_1$ which
corresponds to the same classical observable $x_1 \xi_1$ as
$\text{Op}_{\hbar}^W (x_1 \xi_1)$ is given by a Weyl symbol with a
subleading term in $\hbar$ since
\begin{equation*}
\text{Op}_{\hbar}^W \left(x_1 \xi_1 + \frac{i \hbar}{2}\right) = -i
\hbar x_1 \partial_1.
\end{equation*}
Naturally the right ordered operator $-i \hbar \partial_1 \cdot
x_1$ has Weyl symbol $x_1 \xi_1 - \frac{i \hbar}{2}$. A general
$\Psi$DO $A$ always has a unique Weyl symbol, which is a
$\hbar$-dependent function $f(x,\xi;\hbar)$ denoted $\sigma^W(A)$.
The leading non-zero term in the asymptotic expansion
\eqref{classical Weyl symbol} of this Weyl symbol is called the
\textit{principal symbol}, denoted $\sigma_0^W(A)$, and the
subleading term is called the \textit{subprincipal symbol},
denoted $\sigma_{\text{sub}}^W(A)$. For instance, if
$f_0(x,\xi)\neq 0$ then $\sigma_0^W(A) = f_0(x,\xi)$ and
$\sigma_{\text{sub}}^W(A) = f_1(x,\xi) \hbar$.

An important object for the study of quantum integrability is the
commutator $[A,B]$ of two operators $A$ and $B$. In the present
context of $\Psi$DOs one can show that if $A,B$ are $\Psi$DOs then
their commutator $[A,B]$ is also a $\Psi$DO with principal symbol
\begin{equation*}
\sigma_0^W([A,B]) = -i \hbar \left\{ \sigma_0^W(A), \sigma_0^W(B)
\right\},
\end{equation*}
(so that $-i \hbar \sigma_0^W$ is a Lie algebra homomorphism) and
subprincipal symbol
\begin{equation*}
\sigma_{\text{sub}}^W([A,B]) = -i \hbar \left\{ \sigma_0^W(A),
\sigma_{\text{sub}}^W(B) \right\} - i \hbar \left\{
\sigma_{\text{sub}}^W(A), \sigma_0^W(B) \right\}.
\end{equation*}

\section{Bohr-Sommerfeld for isolated periodic orbit} \label{section: BS isolated}

Let $\gamma$ be a given periodic orbit of energy $E$, i.e. $\gamma
\subset \Sigma_E$. We henceforth assume that $E$ is a regular
value of $H$ so that $\Sigma_E$ is a smooth codimension one
submanifold of $T^{\ast} X$. Given a point $p_0 \in \gamma$, we
call a section of $\gamma$ at $p_0$ a smooth codimension one
\begin{figure}[h]
\begin{center}
\psfrag{g}{\tiny $\gamma$} \psfrag{p0}{\tiny $p_0$}
\psfrag{p}{\tiny $p'$} \psfrag{p2}{\tiny $p$} \psfrag{S}{\tiny
$S$}
\includegraphics[height=45mm]{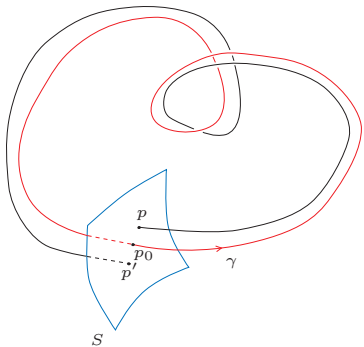}
\caption{Poincar\'e map: global perturbations of a periodic orbit
$\gamma$ can be studied locally in terms of a map $\psi : S
\rightarrow S$ defined by the flow of the Hamiltonian vector field
$X_H$.} \label{Poinc}
\end{center}
\end{figure}
surface $S \subset \Sigma_E$ transverse to $\gamma$ and
intersecting it at $p_0$. We then define the local map $\psi : S
\rightarrow S$ near $p_0$ by letting $p' = \psi(p)$ be the unique
point obtained by following $p \in S$ around the Hamiltonian flow
$X_H$ for a time close to the period $T_\gamma$ of $\gamma$ (see
Figure \ref{Poinc}). Note that fixed points $p = \psi(p)$
(respectively periodic points $p = \psi^k(p), k \geq 2$) of $\psi$
correspond to periodic orbits of the Hamiltonian flow $X_H$ of
period close to $T_\gamma$ (respectively close to $k T_\gamma$).
In particular, since $p_0 = \psi(p_0)$ we define the Poincar\'e
map as the differential of $\psi$ at $p_0$ \cite{Moser}
\begin{equation*}
P = d\psi_{p_0} : T_{p_0} S \rightarrow T_{p_0} S.
\end{equation*}
We say that the periodic orbit $\gamma$ is non-degenerate if and
only if $1$ is not an eigenvalue of the Poincar\'e map. This is a
way of saying that $\gamma$ is isolated on $\Sigma_E$ in the sense
that there are no periodic orbits on $\Sigma_E$ arbitrarily close
to it. However, although $\gamma$ is isolated on $\Sigma_E$, it
belongs to a continuous $1$-parameter family $\gamma_E$ of
periodic orbits intersecting $\Sigma_E$ at $\gamma$(see Figure
\ref{figure: cylinder}). This is the content of the ``cylinder
theorem'' (see for instance \cite{Moser, Marsden}).

If $\gamma$ is stable, then the eigenvalues of the Poincar\'e map
defined at a point $p_0 \in \gamma$ come in complex conjugate
pairs of the form $(e^{i \nu_{\alpha}}, e^{- i \nu_{\alpha}}),
\nu_{\alpha} \in \mathbb{R}$ and hence the Poincar\'e map is
merely a product of rotations by angles $\nu_{\alpha}$ in $n-1$
disjoint planes $\mathbb{R}^2_{\alpha} \subset T_{p_0} S$.
\begin{figure}[h]
\begin{center}
\psfrag{g}{\tiny $\gamma$} \psfrag{R}{\tiny $T_{p_0} S =
\mathbb{R}^2_{\alpha}$} \psfrag{p}{\tiny $p_0$} \psfrag{S}{\tiny
$S^1_{F_{\alpha}}$}
\includegraphics[height=40mm]{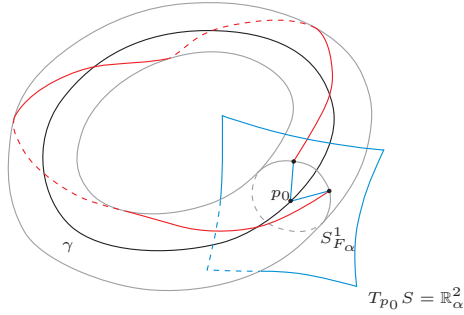}
\caption{The infinitesimal torus around a stable isolated periodic
orbit $\gamma$ ($p=1$) illustrated in the case $n=2$ where there is
only one stability angle $\nu_{\alpha}$ and $T_{p_0} S =
\mathbb{R}^2_{\alpha}$.} \label{local_torus}
\end{center}
\end{figure}
In other words, every point $p_0 \in \gamma$ of the stable isolated
periodic orbit $\gamma$ is surrounded by an infinitesimal torus
$S^1_{F_2} \times \ldots \times S^1_{F_n}$, where
$S^1_{F_{\alpha}} = \{ x_{\alpha} \in \mathbb{R}^2_{\alpha} \; |
\; ||x_{\alpha}||^2 = F_{\alpha} \} \subset
\mathbb{R}^2_{\alpha}$, which is preserved by the Poincar\'e map
to first approximation in $F_{\alpha} \ll 1$. By the cylinder
theorem the periodic orbit $\gamma$ belongs to a continuous family
$\gamma_E$ parametrised by the energy $E$, and so one could now
apply the Bohr-Sommerfeld-Maslov quantisation conditions to the
family of torii $\Lambda \equiv \gamma_E \times S^1_{F_2} \times
\ldots \times S^1_{F_n}$ just constructed (see Figure
\ref{local_torus})
\begin{align*}
\int_{S^1_{F_{\alpha}}} \alpha &= 2 \pi \left( n_{\alpha} +
\frac{1}{2} \right) \hbar + O(\hbar^2), \quad \alpha = 2,\ldots,n\\
\int_{\tilde{\gamma}} \alpha &= 2 \pi \left( N +
\frac{\mu_{\gamma}}{4} \right) \hbar + O(\hbar^2),
\end{align*}
where $\tilde{\gamma}$ is the closed path on $\Lambda$ consisting of a
classical path going from $T_{p_0} S$ once around $\Lambda$ back to
$T_{p_0} S$ and the set of arcs of angles $- \nu_{\alpha}$ on $T_{p_0} S$
to close off this classical path (see red curve in Figure
\ref{local_torus}).

Consider the 2-dimensional surface $\Gamma$ bounded by the
periodic orbit $\gamma$ and the closed curve $\tilde{\gamma}$,
constructed in the obvious way: at any point $t \neq 0$ along the
curve $\gamma(t)$, $\Gamma$ looks locally like $\{ \gamma(t) +
\tau y(t) | 0 < t < T, 0\leq \tau \leq 1\}$ where $y(t)$ is the
transversal vector to $\gamma$ joining the points $\gamma(t)$ and
$\tilde{\gamma}(t)$. At $t=0$ we complete the surface by adding
the sections of the disc of angle $- \nu_{\alpha}$ on $T_{p_0} S$.
Then by Stokes's theorem we have
\begin{equation*}
\left( \int_{\tilde{\gamma}} - \int_{\gamma} \right) \alpha =
\int_{\partial \Gamma} \alpha = \int_{\Gamma} \omega.
\end{equation*}
On the part of $\Gamma$ corresponding to $t \neq 0$ we have
$\omega|_{\Gamma} = 0$ since the tangent space to $\Gamma$ is
spanned by $X_H$ and the transversal vector $y$ ($i_y i_{X_H}
\omega = i_y dH = y(H) = 0$ since $y$ lies in the energy surface
$\Sigma_E$). And since $\Gamma_{t=0}$ looks like sections of angle
$-\nu_{\alpha}$ of the disc of radius $\sqrt{F_{\alpha}}$ it
follows that
\begin{equation*}
\left( \int_{\tilde{\gamma}} - \int_{\gamma} \right) \alpha =
\int_{\Gamma_{t=0}} \omega = - \sum_{\alpha = 2}^n \nu_{\alpha}
F_{\alpha}.
\end{equation*}
On the other hand we have that
\begin{equation*}
\int_{S^1_{F_{\alpha}}} \alpha = \int_{D^1_{F_{\alpha}}} \omega =
2 \pi F_{\alpha},
\end{equation*}
where $D^1_{F_{\alpha}}$ is the disc in $\mathbb{R}^2_{\alpha}$
bounded by the circle $S^1_{F_{\alpha}}$. The last equality
follows by a direct computation, in analogy with the harmonic
oscillator. Finally, by combining all the above we obtain
\cite{Voros, Voros1}
\begin{equation} \label{BS4}
\int_{\gamma} \alpha = \left[ 2 \pi \left( N +
\frac{\mu_{\gamma}}{4}\right) + \sum_{\alpha = 2}^n \left(
n_{\alpha} + \frac{1}{2} \right) \nu_{\alpha} \right] \hbar +
O(\hbar^2),
\end{equation}
which is just the Bohr-Sommerfeld condition for an isolated
periodic orbit.

\section{Dirac brackets} \label{section: Dirac brackets}

Just as in \cite{Paper2}, in this paper we work in conformal static
gauge in order to isolate the physical degrees of freedom of the
string. This is done by imposing the Virasoro constraints and static
gauge fixing condition. However, these constraints together form a
set of second class constraints and so to consistently impose these
constraints from the outset one must work with Dirac brackets instead
of Poisson brackets. In this section we show that the for the type of
brackets $\{ \text{tr}\, \Omega(x), \cdot \}$ considered in section
\ref{section: hierarchy} this distinction does not matter since
\begin{equation*}
\{ \text{tr}\, \Omega(x), f \}_{\text{D.B.}} = \{ \text{tr}\,
\Omega(x), f \}_{\text{P.B.}}
\end{equation*}
for an arbitrary function $f$ of the principal chiral model fields $j
= -g^{-1}dg$ and so by abuse of notation we drop the suffices on both
brackets and write $\{ \cdot, \cdot \}$ throughout section
\ref{section: hierarchy}.

We start with the Poisson bracket \eqref{PB T,J}. To compute
Poisson brackets on the circle we shall work on the universal
cover $\mathbb{R}$. So let $\sigma_1 = \sigma + 2 \pi$, $\sigma_2
= \sigma$ and $\sigma_3 = \sigma'$ in \eqref{PB T,J} to obtain the
Poisson bracket $\{ \Omega(\sigma,x) \overset{\otimes},
J_1(\sigma',x') \}$. This easily leads to the Poisson brackets $\{
\Omega(\sigma,x) \overset{\otimes}, j_{\pm}(\sigma') \}$ after
noting from the definition of $J_1(x)$ that $J_1(0) =
\frac{1}{2}(j_+ - j_-)$ and $ \lim_{x \rightarrow \infty} (-x)
J_1(x) = \frac{1}{2}(j_+ + j_-)$, in particular
\begin{multline*}
\{ \Omega(\sigma, x) \overset{\otimes}, j_{\pm}(\sigma')
\}_{\text{P.B.}} = (T(\sigma + 2 \pi,\sigma',x) \otimes {\bf 1})
\times \left( (\delta(\sigma' - \sigma - 2 \pi) - \delta(\sigma' -
\sigma)) \frac{4 \pi}{\sqrt{\lambda}} \frac{1 \pm x}{1 - x^2} \eta
\right.\\ \left. +\chi(\sigma'; \sigma + 2 \pi,\sigma) \left[ -
\frac{2 \pi}{\sqrt{\lambda}} \frac{2 x}{1 - x^2} \eta, (x \pm 1)
J_1(\sigma', x) \otimes {\bf 1} \pm {\bf 1} \otimes \frac{1}{2}
(j_+(\sigma') - j_-(\sigma')) \right] \right) \\ \times
(T(\sigma',\sigma,x) \otimes {\bf 1}),
\end{multline*}
where we have used the definitions of the $r,s$-matrices \cite{Paper2}
which involve the tensor product $\eta = \frac{1}{2} \sigma_a \otimes
\sigma_a$. Using the identity $\text{tr}_2 (\eta {\bf 1} \otimes A) =
A$ for any matrix $A \in \mathfrak{su}(2)$ one can show that after
multiplying the above equation by ${\bf 1} \otimes j_{\pm}(\sigma')$
and taking the trace $\text{tr}_2$ over the second tensor factor the
commutator disappears and we are left with
\begin{multline*}
\left\{ \Omega(\sigma, x), \frac{1}{2} \text{tr}\, j^2_{\pm}(\sigma')
\right\}_{\text{P.B.}} = \frac{4 \pi}{\sqrt{\lambda}} (\delta(\sigma'
- \sigma - 2 \pi) - \delta(\sigma' - \sigma)) T(\sigma + 2
\pi,\sigma',x) J_{\pm}(\sigma',x) T(\sigma',\sigma,x),
\end{multline*}
where $J_{\pm}(\sigma',x) = j_{\pm}(\sigma')/(1 \mp x)$. Next we
multiply both sides by $e^{\pm i n \sigma'}$ and integrate over $\sigma'$
from $0$ to $2 \pi$. However, since we are on the universal cover
$\mathbb{R}$ of $S^1$ we get two non-zero contributions, namely from
the integrations over the two lifts $[0, 2\pi]$ and $[2\pi, 4\pi]$
(assuming $\sigma \in (0, 2\pi)$). Definition the Virasoro generators
\begin{equation*}
L_n = \frac{\sqrt{\lambda}}{8 \pi} \int_0^{2 \pi} d\sigma' e^{i n
\sigma'} \frac{1}{2} j^2_+(\sigma'), \quad \tilde{L}_n =
\frac{\sqrt{\lambda}}{8 \pi} \int_0^{2 \pi} d\sigma' e^{-i n \sigma'}
\frac{1}{2} j^2_-(\sigma'),
\end{equation*}
we can write the result as follows
\begin{equation*}
\{ \Omega(\sigma, x), L_n \}_{\text{P.B.}} = \frac{1}{2} e^{i n
\sigma} [J_+(\sigma,x), \Omega(\sigma,x)], \quad \{ \Omega(\sigma,
x), \tilde{L}_n \}_{\text{P.B.}} = \frac{1}{2} e^{- i n \sigma}
[J_-(\sigma,x), \Omega(\sigma,x)].
\end{equation*}
Note that in the above calculation it is because of the presence of
the $s$-matrix, which arises from non-ultralocality of the Poisson
brackets of the model, that we end up with the correct transformation
property for $\Omega(x)$ under conformal transformations. Finally,
since the right hand sides are commutators, taking the trace shows
that $\text{tr}\, \Omega(x)$ is invariant under conformal
transformations generated by $L_n, \tilde{L}_n$, namely
\begin{equation*}
\{ \text{tr}\, \Omega(x), L_n \}_{\text{P.B.}} = \{ \text{tr}\,
\Omega(x), \tilde{L}_n \}_{\text{P.B.}} = 0.
\end{equation*}
The assertion that the Dirac and Poisson brackets involving the
quantity $\text{tr}\, \Omega(x)$ are equal now follows from the
definition of the Dirac bracket which in the present case reads,
\begin{multline*}
\{ \text{tr}\, \Omega(x), f \}_{\text{D.B.}} = \{ \text{tr}\,
\Omega(x), f \}_{\text{P.B.}} - \{ \text{tr}\, \Omega(x), L_n
\}_{\text{P.B.}} \{ L_n , L_m\}_{\text{P.B.}}^{-1} \{ L_m, f
\}_{\text{P.B.}}\\ - \{ \text{tr}\, \Omega(x), \tilde{L}_n
\}_{\text{P.B.}} \{ \tilde{L}_n , \tilde{L}_m\}_{\text{P.B.}}^{-1} \{
\tilde{L}_m, f \}_{\text{P.B.}},
\end{multline*}
for any function $f$ of the principal chiral model fields $j = -
g^{-1} dg$.

\section{Pinching an $a$-period} \label{section: pinching a-period}

In this appendix we determine the behaviour of the Riemann
$\theta$-function when the underlying algebraic curve $\Sigma$ becomes
singular \cite{Fay, McKean}. To determine the effect of degenerating an
$a$-cycle on the algebraic curve $\Sigma$, let us consider a family
$\Sigma^{\epsilon}$ of Riemann surfaces ($\epsilon > 0$) of genus
$g+1$ with homology basis $\left\{ a_i^{\epsilon},b_i^{\epsilon}
\right\}_{i=0}^g$ of $H_1(\Sigma^{\epsilon},\mathbb{R})$. Let
$\left\{ \omega_i^{\epsilon} \right\}_{i=0}^g$ be a dual basis of
holomorphic 1-forms canonically normalised as
\begin{equation} \label{g+1 a-period}
\int_{a_j^{\epsilon}} \omega_k^{\epsilon} = \delta_{jk}.
\end{equation}
We model the pinching of an $a$-cycle of the algebraic curve
$\Sigma$ by choosing a family $\{ \Sigma^{\epsilon} \}_{\epsilon >
0}$ for which a particular marked cycle $\tilde{a}_0^{\epsilon}$
on $\Sigma^{\epsilon}$ homotopic to $a_0^{\epsilon}$ shrinks to a
point $P_0$ in the singular limit $\epsilon \rightarrow 0$. The
resulting surface $\Sigma^0$ is singular at $P_0$, and we denote
by $\Sigma'$ its desingularisation.
\begin{center}
\begin{tabular}{cccc}
$\;$ & \psfrag{a1}{\green $a_0^{\epsilon}$} \psfrag{a2}{\red
${a'}_0^{\epsilon}$} \psfrag{a3}{\green $\tilde{a}_0^{\epsilon}$}
\psfrag{M}{$\Sigma^{\epsilon}$}
\includegraphics[height=30mm]{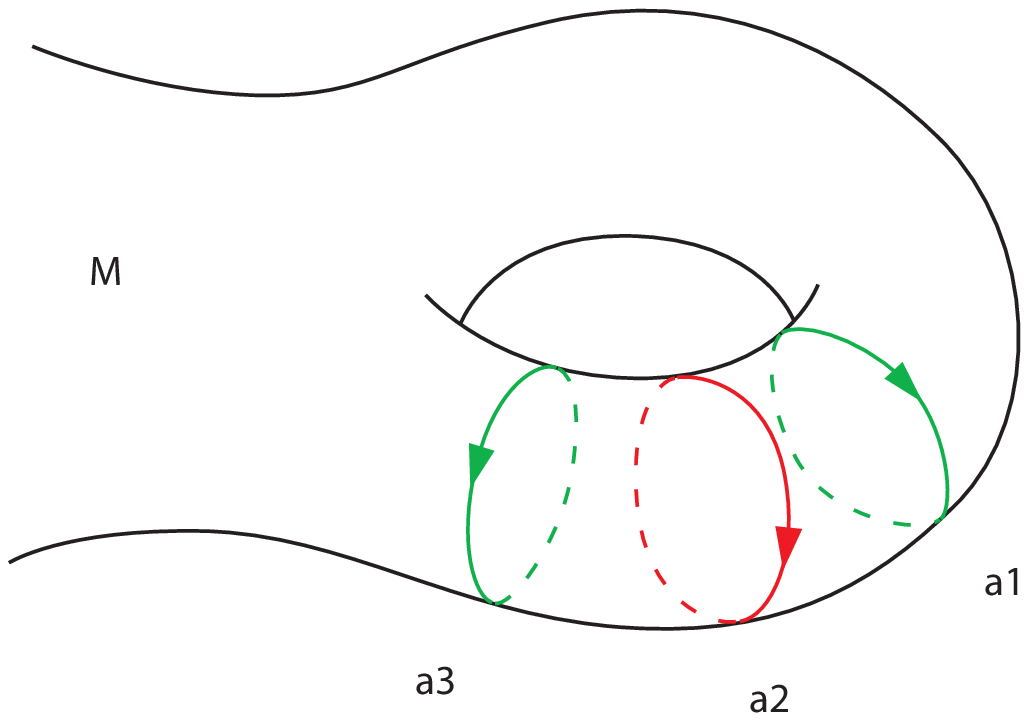} &
\raisebox{16mm}{\blue $\qquad \rightsquigarrow \qquad$} &
\psfrag{a1}{\green $a_0$} \psfrag{a2}{\green ${a'}_0$}
\psfrag{ap}{\footnotesize \red $P_0^-$} \psfrag{am}{\footnotesize
\red $P_0^+$} \psfrag{M}{$\Sigma'$}
\raisebox{7mm}{\includegraphics[height=20mm]{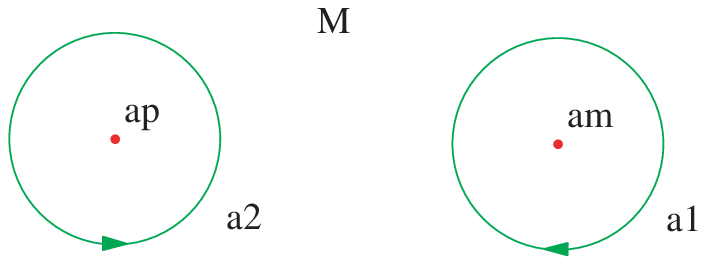}}
\end{tabular}
\end{center}
In the limit $\epsilon \rightarrow 0$ where the cycle
$\tilde{a}_0^{\epsilon}$ shrinks to a single point $P_0 \in
\Sigma^0$, the cycles $a_0$ and ${a'}_0$ are homotopic to the
punctures $P_0^+$ and $P_0^-$ on $\Sigma'$ corresponding to the
desingularisation of $P_0$ on $\Sigma^0$. It follows from
\eqref{g+1 a-period} that in the limit $\epsilon \rightarrow 0$
the 1-form $\omega_0^{\epsilon}$ acquires simple poles at the pair
of points $P_0^{\pm}$ with residues
\begin{equation*}
\text{res}_{P_0^+} \omega_0 = \frac{1}{2 \pi i}\int_{a_0} \omega_0
= \frac{1}{2 \pi i}, \qquad \text{res}_{P_0^-} \omega_0 =
\frac{1}{2 \pi i}\int_{{a'}_0} \omega_0 = - \frac{1}{2 \pi i}.
\end{equation*}
Since $\omega_0$ has no further poles it is a normalised
($\int_{a_i} \omega_0 = 0$, $i = 1,\ldots,g$) Abelian differential
of the third kind on $\Sigma'$. Moreover, $\left\{ \omega_i
\right\}_{i=1}^g$ is a basis of holomorphic 1-forms on $\Sigma'$
dual to the homology basis $\left\{ a_i,b_i \right\}_{i=1}^g$ for
$\Sigma'$. Since the curve $b_0$ starts and ends at $P_0^{\pm}$,
the component $\Pi_{00}^{\epsilon} =
\int_{b_0^{\epsilon}}\omega_0^{\epsilon}$ of the period matrix
will blow up as $\epsilon \rightarrow 0$. All other components of
the period matrix $\Pi_{ij}^{\epsilon} =
\int_{b_i^{\epsilon}}\omega_j^{\epsilon}$ and $\Pi_{0j}^{\epsilon}
= \int_{b_0^{\epsilon}}\omega_j^{\epsilon}$ stay finite in the
limit $\epsilon \rightarrow 0$. The behaviour of the Riemann
$\theta$-function associated with $\Sigma^{\epsilon}$
\begin{equation} \label{theta g+1}
\theta(\vec{z};\tilde{\Pi}^{\epsilon}) = \sum_{\vec{m} \in
\mathbb{Z}^{g+1}} \text{exp}\, \left\{ i \langle \vec{m},\vec{z}
\rangle + \pi i \langle \tilde{\Pi}^{\epsilon} \vec{m},\vec{m}
\rangle \right\}
\end{equation}
can now be analysed in the limit $\epsilon \rightarrow 0$. Using
the fact that the imaginary part $\text{Im}\,\Pi$ of the period
matrix $\Pi$ is positive definite we have $\text{Im}\,\Pi_{00} =
\text{Im}\,\langle \Pi e^{(0)},e^{(0)} \rangle > 0$. It follows
that the quantity $e^{\pi i \Pi_{00}^{\epsilon}}$ is vanishingly
small in the limit $\epsilon \rightarrow 0$ and one finds that
\eqref{theta g+1} can be expanded as follows
\begin{equation} \label{theta reduction}
\theta(\vec{z};\tilde{\Pi}^{\epsilon}) =
\theta(\bm{z};\Pi^{\epsilon}) + \left[ \theta(\bm{z} +
\bm{\Pi}_0^{\epsilon};\Pi^{\epsilon}) e^{i z_0} + \theta(\bm{z} -
\bm{\Pi}_0^{\epsilon};\Pi^{\epsilon}) e^{-i z_0} \right] e^{\pi i
\Pi_{00}^{\epsilon}} + O\left(e^{2 \pi i
\Pi_{00}^{\epsilon}}\right)
\end{equation}
where
\begin{equation*}
\vec{z} = \left(\begin{array}{c}z_0 \\ \bm{z}\end{array}\right)
\in \mathbb{C}^{g+1}, \quad \tilde{\Pi}^{\epsilon} =
\left(\begin{array}{cc}\Pi_{00}^{\epsilon} &
{\bm{\Pi}_0^{\epsilon}}^{\sf T}\\ \bm{\Pi}_0^{\epsilon} &
\Pi^{\epsilon}\end{array}\right).
\end{equation*}

\end{document}